%% file: main.tex
\documentclass[10pt]{article}
\usepackage[margin=1in]{geometry}
\usepackage[affil-it]{authblk}
\usepackage[title]{appendix}
\usepackage[font=footnotesize,labelfont=bf]{caption}
\usepackage{xcolor}
\usepackage[nocompress]{cite}
\input{config.tex}

\usepackage[normalem]{ulem}
\usepackage{tikz}

\usepackage{graphicx}
\usepackage[utf8]{inputenc}
\usepackage{pgf}

\newtheorem{definition}{Definition}[section]

\newtheorem{theorem}{Theorem}[section]

\theoremstyle{definition}
\newtheorem{remark}{Remark}[section]

\newcommand{\apdxref}[1]{\hyperref[#1]{Appendix~\ref{#1}}}

\title{\Large\textbf{Kernel manifolds: nonlinear-augmentation dimensionality reduction using reproducing kernel Hilbert spaces}}
\author{Alejandro~N.~Diaz\thanks{Corresponding author. E-mail: \href{mailto:andiaz@sandia.gov}{andiaz@sandia.gov}.}}
\author{Jacob~T.~Needels}
\author{Irina K. Tezaur}
\author{Patrick~J.~Blonigan}
\affil{\normalsize Sandia National Laboratories}

\date{}

\begin{document}

\thispagestyle{empty}

\maketitle

\input{abstract.tex}

\vspace{2mm}

\input{introduction.tex}
\input{feature_map_manifolds.tex}
\input{kernel_manifolds.tex}
\input{numerics_intro.tex}
\input{numerics_adv_diff.tex}
\input{numerics_pseudo_hifire.tex}
\input{numerics_bracket.tex}
\input{numerics_doublemach.tex}

\input{conclusion.tex}

\section*{Acknowledgements}

Support for this work was received through Sandia National Laboratories’ Laboratory Directed
Research and Development (LDRD) program and through the U.S. Department of Energy, Office of
Science, Office of Advanced Scientific Computing Research, Mathematical Multifaceted Integrated
Capability Centers (MMICCs) program, under Field Work Proposal 22025291 and the Multifaceted
Mathematics for Predictive Digital Twins (M2dt) project. 
A.N.\ Diaz acknowledges support from Sandia National Laboratories' S. Scott Collis Fellowship in Data Science. 
Additionally, the writing of this manuscript
was funded in part by I.K. Tezaur's Presidential Early Career Award for
Scientists and Engineers (PECASE).
The authors wish to thank Alejandro Mota for creating the {\tt Norma.jl} finite element code and for helping to set up the 3D flexible bracket problem within this code.

Sandia National Laboratories is a multi-mission laboratory managed and operated by National Technology \& Engineering Solutions of Sandia, LLC (NTESS), a wholly owned subsidiary of Honeywell International Inc., for the U.S.~Department of Energy's National Nuclear Security Administration~(DOE/NNSA) under contract DE-NA0003525. 
This paper describes objective technical results and analysis. Any subjective views or opinions that might be expressed in the paper do not necessarily represent the views of the U.S. Department of Energy or the United States Government.
SAND2025-10914O

\clearpage

\begin{appendices}
\input{appendix}
\end{appendices}

\bibliographystyle{abbrv}
\bibliography{my_refs}

\end{document}

%% file: config.tex
\usepackage{amsmath, amsfonts, amssymb, amsthm}
\usepackage{color}
\usepackage{caption}
\usepackage{enumerate}
\usepackage{fancyhdr}   % Headers and footers
\usepackage{float}
\usepackage{graphicx}
\usepackage{hyperref}
\usepackage{mathtools}
\usepackage{multirow}
\usepackage{subcaption}
\usepackage{soul}
\usepackage{tikz}
\usepackage[nameinlink]{cleveref}

\hypersetup{
    colorlinks=true,
    linkcolor=blue,
    urlcolor=blue,
    citecolor=blue,
    pdftitle={Kernel manifolds: nonlinear augmentation dimensionality reduction using reproducing kernel Hilbert spaces}
    pdfauthor={Alejandro N. Diaz, Jacob Needels, Irina K. Tezaur, Patrick J. Blonigan},
}

\usetikzlibrary{decorations.pathreplacing, arrows.meta, calligraphy}
\crefformat{equation}{#2(#1)#3}
\numberwithin{equation}{section}

\newcommand{\trp}{^{\mathsf{T}}}  % transpose

\newcommand{\set}[1]{\left\{ #1 \right\} }
\newcommand{\norm}[1]{\left\| #1 \right\|}
\newcommand{\innerprod}[1]{\left\langle #1 \right\rangle}

\newcommand{\real}{\mathbb{R}}

\newcommand{\cD}{{\cal D}}

\newcommand{\cH}{{\cal H}}

\newcommand{\cO}{{\cal O}}

\newcommand{\cQ}{{\cal Q}}

\newcommand{\cV}{{\cal V}}

\newcommand{\bzero}{\mathbf{0}}

\newcommand{\be}{\mathbf{e}}
\newcommand{\bff}{\mathbf{f}}
\newcommand{\bg}{\mathbf{g}}
\newcommand{\bh}{\mathbf{h}}

\newcommand{\bm}{\mathbf{m}}
\newcommand{\bn}{\mathbf{n}}

\newcommand{\bq}{\mathbf{q}}

\newcommand{\bs}{\mathbf{s}}

\newcommand{\bv}{\mathbf{v}}
\newcommand{\bw}{\mathbf{w}}
\newcommand{\bx}{\mathbf{x}}
\newcommand{\by}{\mathbf{y}}

\newcommand{\bA}{\mathbf{A}}

\newcommand{\bG}{\mathbf{G}}

\newcommand{\bI}{\mathbf{I}}

\newcommand{\bM}{\mathbf{M}}

\newcommand{\bQ}{\mathbf{Q}}

\newcommand{\bU}{\mathbf{U}}
\newcommand{\bV}{\mathbf{V}}

\newcommand{\bX}{\mathbf{X}}
\newcommand{\bY}{\mathbf{Y}}

\newcommand{\tbQ}{\widetilde{\mathbf{Q}}}

\newcommand{\hbp}{\widehat{\mathbf{p}}}
\newcommand{\hbq}{\widehat{\mathbf{q}}}

\newcommand{\obq}{\bar{\mathbf{q}}}

\newcommand{\obV}{\overline{\mathbf{V}}}

\newcommand{\hbQ}{\widehat{\mathbf{Q}}}

\newcommand{\hbP}{\widehat{\mathbf{P}}}

\newcommand{\hcQ}{{\widehat{\cal Q}}}

\newcommand{\bOmega}{{\boldsymbol{\Omega}}}

\newcommand{\bbeta}{{\boldsymbol{\beta}}}

\newcommand{\bmu}{{\boldsymbol{\mu}}}
\newcommand{\bnu}{{\boldsymbol{\nu}}}

\newcommand{\bphi}{{\boldsymbol{\phi}}}

\newcommand{\bpsi}{{\boldsymbol{\psi}}}

\newcommand{\bSigma}{{\boldsymbol{\Sigma}}}

\newcommand{\bXi}{{\boldsymbol{\Xi}}}

\newcommand{\pdt}{\frac{\partial}{\partial t}}
\newcommand{\pdx}{\frac{\partial}{\partial x}}

\newcommand{\dt}{\frac{\textrm{d}}{\textrm{d}t}}

%% file: abstract.tex
%!TEX root = main.tex

\begin{abstract}
\noindent
This paper generalizes recent advances on quadratic manifold (QM) dimensionality reduction by developing kernel methods-based nonlinear-augmentation dimensionality reduction. 
QMs, and more generally feature map-based nonlinear corrections, 
augment linear dimensionality reduction with a nonlinear correction term in the reconstruction map to overcome approximation accuracy limitations of purely linear approaches.
While feature map-based approaches typically learn a least-squares optimal polynomial correction term, we generalize this approach by learning an optimal nonlinear correction from a user-defined reproducing kernel Hilbert space. 
Our approach allows one to impose arbitrary nonlinear structure on the correction term, including polynomial structure, and includes feature map and radial basis function-based corrections as special cases.
Furthermore, our method has relatively low training cost and has monotonically decreasing error as the latent space dimension increases. 
We compare our approach to proper orthogonal decomposition and several recent QM approaches on data from several example problems.
\end{abstract}

%% file: introduction.tex
%!TEX root = main.tex

\section{Introduction}\label{sec:introduction}
In recent years, data-driven surrogate modeling of complex physical systems has been an active research area in scientific machine learning (SciML). 
A common challenge of constructing surrogate models is that data from simulations of partial differential equations (PDEs) are often high dimensional, with the dimensionality dictated by the discretization methods used and the fidelity of the corresponding mesh. 
Consequently, surrogate modeling techniques typically involve a dimensionality reduction component to 
extract low-dimensional features from high-dimensional simulation data, which can then be used to construct a low-dimensional, computationally efficient surrogate model. 
Crucially, the dimensionality reduction technique used fundamentally limits the achievable accuracy of the resulting surrogate model. 

Classical linear dimensionality reduction techniques, such as proper orthogonal decomposition (POD) \cite{MGubisch_SVolkwein_2017a, MHinze_SVolkwein_2005a}, seek to identify a low-dimensional affine subspace that best captures the high-dimensional data. 
POD identifies this subspace as the span of the leading left singular vectors of a representative set of high-dimensional snapshot data. 
The POD basis induces an affine \emph{encoder} map that projects high-dimensional snapshots onto generalized coordinates that lie in a low-dimensional \emph{latent space}, which can then be mapped to an approximate reconstruction of the high-dimensional snapshot via an affine \emph{decoder} map. 
While POD is effective in numerous applications, it is also well-known to poorly approximate data from problems with slowly decaying Kolmogorov $n$-width  \cite{MOhlberger_SRave_2016a}, such as transport-dominated problems or problems with sharp gradients. 
In these cases, a large POD basis is required to accurately approximate the given data. 
However, despite lying in a high-dimensional affine subspace, these data may still possess an inherent low-dimensional nonlinear structure. This assumption motivates using \emph{nonlinear} dimensionality reduction for problems with slowly decaying Kolmogorov $n$-width.

A number of nonlinear dimensionality reduction approaches have recently been proposed to circumvent the Kolmogorov $n$-width barrier. 
These techniques are often collectively referred to as \emph{nonlinear manifolds} to distinguish them from linear dimensionality reduction, 
and include autoencoders (AEs) \cite{KLee_KTCarlberg_2020a, YKim_YChoi_DWidemann_TZohdi_2022a, FRomor_GStabile_GRozza_2023a, JCocola_JTencer_FRizzi_EParish_PBlonigan_2023a, ANDiaz_YChoi_MHeinkenschloss_2024a, goyal2024generalized, CBonneville_etal_2024a, CBonneville_YChoi_DGhosh_JLBelof_2024a, WDFries_XHe_YChoi_2022a}, 
quadratic manifolds (QMs) \cite{JBarnett_CFarhat_2022a, benner2023quadratic,  RGeelen_SWright_KWillcox_2023a, RGeelen_LBalzano_SWright_KWillcox_2024a, PSchwerdtner_BPeherstorfer_2024a, schwerdtner2025online, jain2017quadratic, rutzmoser2017generalization},
the projection-based reduced-order model + artificial neural network (PROM-ANN) technique \cite{JBarnett_CFarhat_YMaday_2023a}, and the recent Gaussian process (GP) and radial basis function (RBF) interpolation-based nonlinear closure approach \cite{deParga:2025}.
AEs consist of fully nonlinear encoder and decoder maps, which can offer substantial accuracy improvements over linear dimensionality reduction. 
However, they can be difficult to train due to their dependence on the neural network architecture and large data requirements. 
Furthermore, they often lack interpretable structure and can be slow to evaluate, which is detrimental in use cases requiring fast online evaluations, such as reduced-order modeling.
To address these issues, the QM, PROM-ANN, and GP/RBF-interpolation approaches instead augment linear dimensionality reduction with a nonlinear correction term in the decoder.
We characterize this class of approaches as \emph{nonlinear-augmentation manifolds}.
QMs model the nonlinear augmentation term as a quadratic polynomial, while PROM-ANN models it as an ANN and the GP/RBF-interpolation approach models it as the mean function of a GP or as an RBF interpolant. 
These approaches have interpretable structure and, with the exception of PROM-ANN, comparatively low training cost relative to AEs, while still providing improved accuracy over linear dimensionality reduction approaches. 
However, QMs, and more generally, feature-map based nonlinear-augmentation manifolds, are typically limited to polynomial corrections, which can restrict the achievable accuracy of these methods, while becoming increasingly difficult to train as the polynomial order increases. 
Although using GPs and RBF interpolants as in \cite{deParga:2025} can address these difficulties, this approach can be further generalized to improve the flexibility of nonlinear-augmentation manifolds.
In this paper we develop a nonlinear-augmentation dimensionality reduction technique using regularized kernel interpolation, which generalizes recent work on QMs and GP/RBF interpolation-based nonlinear closures, yielding greater expressivity in the nonlinear approximation, reduced training cost compared to QMs, and improved accuracy on out-of-sample data. 

Regularized kernel interpolation is a generalization of RBF interpolation relying on mature mathematical theory from reproducing kernel Hilbert spaces (RKHS) \cite{GSantin_2018a, GSantin_BHaasdonk_2021a, CAMicchelli_MPontil_2005a, wendland2004scattered}. 
While kernel methods are a common tool in classical data science and machine learning, they have become increasingly popular in SciML in recent years, with notable applications in operator and equation learning \cite{batlle2024kernel, lowery2024kernel, jalalian2025data}, tensor decomposition \cite{larsen2024tensor}, and reduced-order and surrogate modeling \cite{Diaz:2025, PJBaddoo_BHermann_BJMcKeon_SLBrunton_2022a}. 
In our dimensionality reduction method, which we refer to as \emph{kernel manifold} dimensionality reduction, one can model the nonlinear augmentation term with arbitrary nonlinearities by choosing an appropriate kernel, and includes QMs and GP/RBF-based closures as special cases. 
Furthermore, unlike feature map-based nonlinear augmentations, regularized kernel interpolation allows one to model the nonlinear term using polynomials of arbitrarily high order without increasing the number of degrees of freedom in the learning problem. 
We also note that our approach is distinct from kernel principal component analysis (KPCA) \cite{scholkopf1997kernel} and kernel POD (KPOD) \cite{salvador2021non} in that it augments classical POD with a nonlinear correction term, whereas KPCA/KPOD returns a nonlinear kernel-based encoder without an explicit decoder, thus requiring one to learn an approximate decoder function through, e.g., kernel interpolation or neural networks. 
While we purely focus on dimensionality reduction in this paper, 
we emphasize that our method is general-purpose, and can be used for standalone dimensionality reduction, intrusive model reduction as in \cite{deParga:2025}, or non-intrusive model reduction through integration with operator inference (OpInf) \cite{ghattas2021acta,kramer2024survey,BPeherstorfer_KWillcox_2016a, EQian_BKramer_BPeherstorfer_KWillcox_2020a, RGeelen_LBalzano_SWright_KWillcox_2024a} or the recent kernel methods-based approach of \cite{Diaz:2025}. 

The outline of this paper is as follows.
We first describe the problem setup and review POD and feature map-based nonlinear-augmentation dimensionality reduction in \Cref{sec:feature_map_manifolds}. 
Next, we describe our new kernel manifold approach in \Cref{sec:kernel_manifolds}, beginning with a brief review of regularized kernel interpolation in \Cref{sec:kernel_interpolation}. 
We then compare our method with POD and the recently proposed Alternating QM \cite{RGeelen_LBalzano_SWright_KWillcox_2024a} and Greedy QM \cite{PSchwerdtner_BPeherstorfer_2024a} approaches on data from several example problems in \Cref{sec:numerics}. 
The results show that our approach can yield improved accuracy over POD and QM in the small latent dimension size regime in several cases while requiring reduced training cost compared to Greedy QM. 
Finally, \Cref{sec:conclusion} provides several concluding remarks and directions for future work. 

%% file: feature_map_manifolds.tex
\section{Review of feature map-based nonlinear-augmentation manifolds}\label{sec:feature_map_manifolds}
This section reviews a class of nonlinear manifold dimensionality reduction that augments linear dimensionality reduction with a nonlinear correction term in the decoder map. 
These approaches are typically applied to data from large-scale physics simulations.
We first discuss a unified problem setting in \Cref{sec:problem_setting}.
We then briefly review the proper orthogonal decomposition (POD) in \Cref{sec:pod}, since nonlinear-augmentation manifolds approaches use POD as a starting point. Lastly, we review the feature map-based nonlinear augmentation approach in \Cref{sec:feature_map_reduction}.

\subsection{Problem setting}\label{sec:problem_setting}

Let $\cD\subset \real^p$ denote a set of parameters for a given physical system and let $\cO : \cD\to \real^N$ denote an operator that maps a parameter $\bmu\in \cD$ to a discretized state of the physical system on a mesh of size $N$:
\begin{equation}
    \cO : \bmu \mapsto \bq(\bmu) \in \real^N.
\end{equation}
For example, $\cO$ can represent the solution operator of a discretized PDE.
We seek a low-dimensional representation of the solution space of the operator $\cO$,
\begin{equation}\label{eq:solution_manifold}
    \cQ = \cO(\cD)\subset \real^N.
\end{equation}
In particular, we wish to construct an \emph{encoder} map 
    $\bh:\cQ \to \hcQ\subset\real^r$
and a \emph{decoder} map
    $\bg:\hcQ\to \cQ,$
where $r\ll N$,
such that 
\begin{equation}\label{eq:approximation_problem}
    \bg(\bh(\bq))\approx \bq \quad \forall \quad \bq \in \cQ.
\end{equation}
In this setting, $\cQ\subset \real^N$ is referred to as the \emph{full space} and $\hcQ\subset\real^n$ is referred to as the \emph{latent space}.
We will commonly use the \emph{projection error} 
\begin{equation}\label{eq:projection_error}
    \be(\bq)=\bq - \bg(\bh(\bq))
\end{equation}
to assess the quality of the encoder and decoder.
In the following sections, we review methods for constructing $\bh$ and $\bg$ using data samples from the full space $\cQ$.

\subsection{Proper orthogonal decomposition}\label{sec:pod}
POD computes an affine encoder 
$\bh$ and affine decoder $\bg$
optimal affine subspace $\obq+\cV\subset \real^N$ that best approximates $\cQ$,
where $\obq\in \real^N$ is a fixed offset vector and $\cV\subset \real^N$ is a linear subspace of dimension $r\ll N$. 
We review the essential details here and refer the reader to, e.g., ~\cite{MHinze_SVolkwein_2005a,MGubisch_SVolkwein_2017a} for a more complete overview. 

Let
$\bQ = \begin{bmatrix}
    \bq_1, \dots, \bq_M
\end{bmatrix}\in \real^{N\times M}$
be a data matrix consisting of columns $\set{\bq_j}_{j=1}^M\subset \cQ$.
We then shift the data $\bQ$ by the offset vector $\obq$,
\begin{equation}\label{eq:shifted_snapshot_matrix}
    \tbQ = \bQ-\obq \mathbf{1}\trp,
\end{equation}
where $\mathbf{1}\in \real^N$ is the vector of all ones. 
The offset $\obq$ can be chosen by the user, but is often taken to be the mean of the snapshots:
\begin{equation}\label{eq:snapshot_mean}
    \obq = \frac{1}{M}\sum_{j=1}^{M} \bq_j,
\end{equation}
or set to zero: $\obq=\bzero$.  It can also be used to strongly enforce Dirichlet boundary conditions in a projection-based reduced-order model (ROM) \cite{Gunzburger:2007}.
We then compute the singular value decomposition (SVD) of $\tbQ$,
\begin{equation*}
    \bU\bSigma\bY\trp = \tbQ,
\end{equation*}
and define the POD basis matrix $\bV\in \real^{N\times r}$ to be the first $r$ leading left singular vectors of $\tbQ$, 
\begin{equation*}
    \bV = \bU(\,:\,, 1:r).
\end{equation*}
The subspace $\cV$ is taken to be the span of the columns of $\bV$, $\cV={\rm span}(\bV)$.
Furthermore, the basis matrix $\bV$ minimizes 
\begin{equation}\label{eq:pod_basis_optimality}
    \min_{\bV \in \real^{N\times r}, \bV\trp\bV = \bI_r} \; \norm{\tbQ - \bV\bV\trp \tbQ}_F^2,
\end{equation}
and hence $\bV$ optimally reconstructs the shifted data $\tbQ$.
Lastly, the encoder $\bh$ and decoder $\bg$ are defined as 
\begin{align}\label{eq:affine_encoder_decoder}
    \bh(\bq) &= \bV\trp(\bq-\obq), &
    \bg(\hbq) &= \obq+\bV\hbq.
\end{align}

While POD can successfully reduce the dimensionality of a given dataset $\bQ$ in a number of applications with small projection errors,
it is well-known that it struggles to approximate data coming from transport-dominated problems or from problems with sharp gradients. 
In these cases, a large basis size $r$ is needed to reconstruct the data to achieve a desired accuracy. 
This issue is known as the Kolmogorov $n$-width barrier \cite{MOhlberger_SRave_2016a}. 
To overcome this barrier, a number of recent works have proposed that augment the affine decoder $\bg$ from equation \cref{eq:affine_encoder_decoder} with a nonlinear correction term. 
In particular, these include 
quadratic manifolds (QMs)~\cite{JBarnett_CFarhat_2022a, benner2023quadratic,  RGeelen_SWright_KWillcox_2023a, RGeelen_LBalzano_SWright_KWillcox_2024a, PSchwerdtner_BPeherstorfer_2024a, schwerdtner2025online, jain2017quadratic, rutzmoser2017generalization},
the projection-based ROM + artificial neural network (PROM-ANN) approach~\cite{JBarnett_CFarhat_YMaday_2023a}, 
and recent work using Gaussian processes and radial basis function (RBF) interpolation to fit nonlinear closure terms in the latent space \cite{deParga:2025}.
We review a class of these approaches in the following section.

\subsection{Feature map-based nonlinear-augmentation manifolds}\label{sec:feature_map_reduction}
The feature map-based nonlinear-augmentation manifold approach, which we will abbreviate to ``feature map manifold'' (FM), takes an affine encoder $\bh$ of the form \cref{eq:affine_encoder_decoder} and augments the affine decoder map $\bg$ with a nonlinear correction term, resulting in a decoder of the form 
\begin{equation}\label{eq:feature_map_decoder}
    \bg(\hbq) = \obq + \bV \hbq + \obV \bXi \bphi(\hbq), 
\end{equation}
where $\bV \in \real^{N\times r}$ is a \emph{reduced basis} matrix, 
$\obV\in \real^{N\times m}$ is an \emph{augmenting basis} matrix,
$\bphi: \real^r\to \real^{n_\phi}$ is a nonlinear feature map,
and $\bXi \in \real^{m\times n_\phi}$ is a coefficient matrix.
The basis matrices $\bV$ and $\obV$ are typically chosen beforehand, 
while the coefficient matrix $\bXi$ is learned by minimizing the norms of the projection errors $\norm{\bq_j-\bg(\bh(\bq_j))}_2$ of the given data $\set{\bq_j}_{j=1}^M\subset \cQ$:
\begin{equation}\label{eq:feature_map_manifold_min_problem_1}
\min_{\bXi \in \real^{m\times n_\phi}} \; \sum_{j=1}^M \; \norm{\bq_j - \left(\obq + \bV\bV\trp(\bq_j-\obq) + \obV \bXi \bphi(\bV\trp(\bq_j-\obq))\right)}_2^2 + \lambda \norm{\bXi}_F^2,
\end{equation}
where $\lambda \geq 0$ is a regularization parameter. 

If $\bV$ and $\obV$ are chosen to be 
orthonormal matrices satisfying 
\begin{align}\label{eq:orthonormality_assumption}
    \bV\trp\bV &= \bI_r, &
    \obV\trp\obV &= \bI_m, &
    \bV\trp\obV &= \bzero,
\end{align}
and the projections of $\bq_j-\obq$ by $\bV$ and $\obV$, respectively, are denoted as
\begin{align}\label{eq:projected_state_notation}
    \hbq_j &= \bV\trp(\bq_j - \obq),  &
    \hbp_j &= \obV\trp(\bq_j - \obq),
\end{align}
then one can show that \cref{eq:feature_map_manifold_min_problem_1} is equivalent to 
\begin{align}\label{eq:feature_map_manifold_min_problem_2}
\min_{\bXi \in \real^{m\times n_\phi}} \; \sum_{j=1}^M \; \norm{\hbp_j - \bXi \bphi(\hbq_j)}_2^2 + \lambda \norm{\bXi}_F^2.
\end{align}
With the data matrices $\hbQ$ and $\hbP$ defined as
\begin{align}\label{eq:modal_data_matrices}
    \hbQ &= 
    \begin{bmatrix}
        \hbq_1 & \dots & \hbq_M
    \end{bmatrix} \in \real^{r\times M}, &
    \hbP &= 
    \begin{bmatrix}
        \hbp_1 & \dots & \hbp_M
    \end{bmatrix} \in \real^{m\times M},
\end{align}
a minimizer $\bXi\in \real^{m\times n_\phi}$ can be computed by solving the normal equations
\begin{align}\label{eq:feature_map_coef_minimizer_equation}
    \left(\bphi(\hbQ)\bphi(\hbQ)\trp + \lambda \bI_{n_\phi}\right)\bXi\trp = \bphi(\hbQ)\hbP\trp,
\end{align}
where $\bphi$ applied to a matrix is understood to operate column-wise.

A typical choice for $\bV \in \real^{N\times r}$ is the same POD basis from Section \ref{sec:pod} consisting of the first $r$ leading POD modes, 
with the augmenting basis $\obV\in \real^{N\times m}$ chosen to be the subsequent $m$ modes: 
\begin{align}\label{eq:pod_bases_choice}
    \bV &= \bU(\, :\:, 1:r), &
    \obV &= \bU(\, :\:, r+1:r+m),
\end{align}
see, e.g., \cite[Algorithm 1]{RGeelen_LBalzano_SWright_KWillcox_2024a}.
With this choice of $\bV$ and $\obV$, $\hbq_j$ from equation \cref{eq:projected_state_notation} corresponds to the projection of $\bq_j-\obq$ onto the first $r$ leading modes and $\hbp_j$ corresponds to the projection onto the subsequent $m$ modes.
Moreover, solving \cref{eq:feature_map_manifold_min_problem_2} is equivalent to constructing a function 
\begin{align}\label{eq:pod_mode_map_feature_map}
    \bn&:\real^r\to \real^{m}, & \hbq &\mapsto \bXi\bphi(\hbq)\approx \hbp,
\end{align}
that approximates a map from the lower order modal coefficients $\hbq$ to higher order modal coefficients $\hbp$.
We will leverage this interpretation in the kernel-based approach in Section \ref{sec:kernel_manifolds}.

\begin{remark}
The feature map $\bphi:\real^n\to \real^{n_\phi}$ is often chosen to be a quadratic feature map that omits duplicated entries, 
\begin{equation}\label{eq:quadratic_feature_map}
    \bphi(\hbq) = \begin{bmatrix}
        \hbq_1^2 &
        \hbq_2 \hbq_1 &
        \hbq_2^2 &
        \dots &
        \hbq_n \hbq_{n-1}&
        \hbq_n^2
    \end{bmatrix}\trp \in \real^{n(n+1)/2}.
\end{equation}
Using a decoder of the form \cref{eq:feature_map_decoder} with the quadratic feature map \cref{eq:quadratic_feature_map}
is referred to as quadratic manifold (QM) dimensionality reduction.
\end{remark}
\begin{remark}
    The approaches in \cite{JBarnett_CFarhat_2022a, RGeelen_SWright_KWillcox_2023a, PSchwerdtner_BPeherstorfer_2024a} use $\obV=\bI_N$ and construct $\bV$ using POD modes.
    Since $\bV\trp\obV = \bV\trp \neq \bzero$ in this case, the simplified minimization problem defined by Equations \cref{eq:projected_state_notation} and \cref{eq:feature_map_manifold_min_problem_2} no longer holds, and thus \cref{eq:feature_map_manifold_min_problem_1} must be solved directly.
    While \cite{JBarnett_CFarhat_2022a, RGeelen_SWright_KWillcox_2023a} use the $r$ leading POD modes to construct $\bV$, the greedy QM approach in \cite{PSchwerdtner_BPeherstorfer_2024a} greedily selects (potentially high-order) POD modes, which can lead to substantial accuracy gains; see \cite{PSchwerdtner_BPeherstorfer_2024a} for further details.
\end{remark}
\begin{remark}
    The nonlinear correction term $\obV\bXi\bphi(\hbq)$ in the decoder \cref{eq:feature_map_decoder} can also be interpreted as a closure modeling term, since $\obV\bXi\bphi(\hbq)$ approximates the effect of discarded POD modes without introducing additional degrees of freedom.
    However, it is important to note that an exact mapping $\bn_{*}:\hbq\mapsto\hbp$ from low-order modes to high-order modes is not well-defined in general.  
    Indeed, consider the case $r=m=1$ and $\obq=\bzero$. Then $\bV=\bv_1=\bU(:, 1)$ and $\obV=\bv_2=\bU(:,2).$ Suppose that $\bq:=\bv_1\in \cQ$ and $\bq':=\bv_1+\bv_2\in \cQ$.
    Notice that 
    \begin{align*}
    \hbq &=\bV\trp\bq = 1, \; 
    \hbp = \obV\trp\bq = 0 
    \quad \implies \quad
    \hbp=\bn_{*}(\hbq) = \bn_{*}(1) = 0.
    \end{align*}
    However, 
    \begin{align*}
    \hbq' &=\bV\trp\bq' = 1, \;
    \hbp' = \obV\trp\bq' = 1
    \quad \implies \quad
    \hbp'=\bn_{*}(\hbq') = \bn_{*}(1) = 1 \neq 0.
    \end{align*}
    Therefore $\bn_*$ is not well-defined. 
    Nevertheless, fitting an approximate map of the form \cref{eq:pod_mode_map_feature_map} is useful for dimensionality reduction.
\end{remark}

While the FM approach outlined in this section is relatively general, typically $\bphi$ is chosen to be a low-order polynomial.
Extending the approach to higher-order polynomials results in an increasingly large linear system \cref{eq:feature_map_coef_minimizer_equation} that must be solved to compute the coefficients $\bXi$. 
Moreover, using a high-order polynomial for $\bphi$ would increase the evaluation cost of the decoder $\bg$, which can be detrimental in use-cases where $\bg$ is evaluated many times in an online stage, such as reduced-order modeling.
In the next section, we further generalize the FM approach to a kernel-based nonlinear manifold to address these issues.

%% file: kernel_manifolds.tex
%!TEX root = main.tex

\section{Kernel-based nonlinear-augmentation manifolds}\label{sec:kernel_manifolds}
In the kernel-based nonlinear-augmentation manifold approach, which we will abbreviate to ``kernel manifold'' (KM), we again use an affine encoder $\bh$ and augment the affine decoder with a nonlinear correction term.
We express the decoder as 
\begin{equation}\label{eq:kernel_decoder}
    \bg(\hbq) = \obq + \bV \hbq + \obV \bn(\hbq), 
\end{equation}
where $\bV \in \real^{N\times r}$ is a reduced basis matrix and
$\obV\in \real^{N\times m}$ is an augmenting basis matrix as before,
and $\bn:\real^r\to \real^m$ is an \emph{optimal} nonlinear function that is learned through regularized kernel interpolation.
We briefly review regularized kernel interpolation in Section \ref{sec:kernel_interpolation} and present the details of our new approach in Section \ref{sec:application_to_kernel_manifolds}.

\subsection{Review of regularized kernel interpolation}\label{sec:kernel_interpolation}

\begin{definition}[Positive-definite kernels]
    \label{def:pd_kernels}
    A function $K:\real^d \times \real^d\to \real$ is a (real-valued) \emph{kernel function} if it is symmetric, i.e., $K(\bx,\bx') = K(\bx',\bx)$ for all $\bx,\bx'\in\real^d$. A kernel function $K$ is said to be \emph{positive-definite} if for any $\bX=[~\bx_1~~\cdots~~\bx_n~]\in \real^{n}$ with pairwise distinct columns, the kernel matrix $K(\bX, \bX)\in \real^{n\times n}$ with entries $K(\bX, \bX)_{ij}=K(\bx_i, \bx_j)$ is positive semi-definite.
    Moreover, if $K(\bX, \bX)$ is strictly positive-definite, then $K$ is also said to be \emph{strictly positive-definite}.
\end{definition}

\begin{definition}[Reproducing kernel Hilbert space (RKHS)]
Consider the pre-Hilbert space 
\begin{align*}
    \cH_K^0(\real^d)
    = \set{
        f:\real^{d}\to\real
        ~\bigg|~
        \exists\,n\in\mathbb{N},\,
        \bw\in\real^{n},\,
        \set{\bx_j}_{j=1}^{n}\subset\real^{d}
        ~\textup{such that}~
        f(\bx) = \sum_{j=1}^n w_j K(\bx_j, \bx)
    }
\end{align*}
where $K:\real^d\times\real^d\to\real$ is a positive definite kernel.
The \emph{reproducing kernel Hilbert space (RKHS)} $\cH_K(\real^{d})$ induced by $K$ is the (unique) completion of $\cH_K^0(\real^d)$ with respect to the norm $\norm{\cdot}_{\cH_K(\real^d)} \coloneqq \innerprod{\cdot,\cdot}_{\cH_K(\real^{d})}^{1/2}$ induced by the inner product
\begin{align*}
    \innerprod{f, f'}_{\cH_K(\real^d)}
    \coloneqq \sum_{j=1}^{n} \sum_{k=1}^{n'} w_j w_k' K(\bx_j, \bx_k'),
\end{align*}
where $f(\bx) = \sum_{j=1}^n w_j K(\bx_j, \bx)$ and $f'(\bx) = \sum_{j=1}^{n'} w_k' K(\bx_k', \bx)$.
\end{definition}

For a set of pairwise-distinct vectors $\{\bx_j\}_{j=1}^n \subset \real^d$, we use the notation $K(\bX,\bX)\in \real^{n\times n}$ to denote the \emph{kernel matrix} of \Cref{def:pd_kernels} and define the vector $K(\bX,\bx) = [~K(\bx_1,\bx)~~\cdots~~K(\bx_n,\bx)~]\trp\in\real^n$.
To simplify notation, we write $\cH_K$ instead of $\cH_K(\real^d)$ when it is clear that $K$ is defined over $\real^d\times \real^d$.

While the RKHS $\cH_K$ is a Hilbert space of scalar-valued functions, one can easily construct an RKHS of vector-valued functions by taking the $p$-fold Cartesian product of $\cH_K$, i.e. $\cH_K^p=\cH_K\times \dots \times \cH_K$, with inner product 
$\innerprod{\bff, \bff'}_{\cH_K^p} = \sum_{i=1}^p\innerprod{f_i, f_i'}_{\cH_K}$ for all $\bff = (f_1,\ldots,f_p), \bff' = (f_1',\ldots,f_p') \in \cH_K^p$.
We now state the Representer Theorem, a key result from the RKHS literature underlying our methodology.

\begin{definition}[Regularized kernel interpolant]
    \label{def:kernel_interpolant}
    Let $\bff:\real^d\to \real^p$, $\set{\bx_j}_{j=1}^n\subset \real^d$ be pairwise distinct, and denote $\by_j=\bff(\bx_j)$.
    For an RKHS $\cH_K^p$ and regularization parameter $\lambda \geq 0$, a \emph{regularized interpolant} $\bs_{\bff}^\lambda \in \cH_K^p$ of $\bff$ solves the minimization problem
\begin{align}\label{eq:kernel_optimization_problem}
    \min_{\bs\in \cH_K^p} \; \sum_{j=1}^n \norm{\by_j - \bs(\bx_j)}_2^2 + \lambda\norm{\bs}_{\cH_K^p}^2.
\end{align}
\end{definition}

\begin{theorem}[Representer Theorem]
    \label{thm:representer}
    The minimization problem \cref{eq:kernel_optimization_problem} has a solution of the form
    \begin{equation}\label{eq:vv_kernel_interpolant}
        \bs_\bff^\lambda(\bx)
        = \bOmega\trp K(\bX, \bx),
    \end{equation}
    where the coefficient matrix $\bOmega\in \real^{n\times p}$ solves the linear system
    \begin{equation}\label{eq:coefficient_equation}
        (K(\bX, \bX) + \lambda \bI)
        \bOmega
        =
        \bY\trp.
    \end{equation}
    Moreover, if $K$ is strictly positive definite, $s_\bff^\lambda$ is the unique minimizer. 
\end{theorem}
This result is a straightforward extension of the Representer Theorem for scalar-valued functions (see, e.g., \cite[Theorem 9.3]{GSantin_BHaasdonk_2021a}).
A key observation from \Cref{thm:representer} is that a solution to the infinite-dimensional minimization problem \cref{eq:kernel_optimization_problem} can be obtained by solving the finite-dimensional linear system \cref{eq:coefficient_equation}.

\subsection{Learning kernel-based nonlinear augmentations}\label{sec:application_to_kernel_manifolds}
For our KM approach, we form the decoder \cref{eq:kernel_decoder} where the nonlinear function $\bn:\real^r\to \real^m$ belongs to the RKHS $\cH_K^m=\cH_K^m(\real^r)$ for a given positive-definite kernel $K:\real^r\times \real^r\to \real$.
In particular, we use the optimal $\bn\in \cH_K^m$ that minimizes the norms of the projection errors $\norm{\bq_j-\bg(\bh(\bq_j))}_2$ of the given data $\set{\bq_j}_{j=1}^M\subset \cQ$:
\begin{equation}\label{eq:kernel_manifold_min_problem_1}
    \min_{\bn\in \cH_K^m} \; \sum_{j=1}^M \; \norm{\bq_j - \left(\obq + \bV\bV\trp(\bq_j-\obq) + \obV \bn(\bV\trp(\bq_j-\obq))\right)}_2^2 + \lambda \norm{\bn}_{\cH_K^m}^2,
\end{equation}
where $\lambda \geq 0$ is a regularization parameter.

For our approach, we select 
$\bV \in \real^{N\times r}$ to be the POD basis from Section \ref{sec:pod} consisting of the first $r$ leading POD modes and
the augmenting basis $\obV\in \real^{N\times m}$ to be the subsequent $m$ modes as in \cref{eq:pod_bases_choice}.
Thus $\bV$ and $\obV$ satisfy \cref{eq:orthonormality_assumption}.
As in \cref{eq:projected_state_notation}, again let 
\begin{align}\label{eq:projected_state_notation_again}
    \hbq_j &= \bV\trp(\bq_j - \obq),  &
    \hbp_j &= \obV\trp(\bq_j - \obq),  
\end{align}
i.e., $\hbq_j\in \real^r$ denotes the first $r$ leading POD modal coefficients and $\hbp_j\in \real^m$ the subsequent $m$ modal coefficients. 
One can again show that \cref{eq:kernel_manifold_min_problem_1} is equivalent to 
\begin{align}\label{eq:kernel_manifold_min_problem_2}
    \min_{\bn\in \cH_K^m} \; \sum_{j=1}^M \; \norm{\hbp_j - \bn(\hbq_j)}_2^2 + \lambda \norm{\bn}_{\cH_K^m}^2.
\end{align}
With the data matrices $\hbQ$ and $\hbP$ from \cref{eq:modal_data_matrices},
we can apply \Cref{thm:representer} to \cref{eq:kernel_manifold_min_problem_2} and obtain the solution 
\begin{equation}\label{eq:kernel_optimal_nonlinear_function}
    \bn(\hbq) = \bOmega\trp K(\hbQ, \hbq),
\end{equation}
where $\bOmega\in \real^{M\times m}$ solves 
\begin{equation}\label{eq:kernel_optimal_coefficients}
    (K(\hbQ, \hbQ) + \lambda \bI)
    \bOmega
    =
    \hbP\trp.
\end{equation}
We then obtain the desired nonlinear decoder 
\begin{equation}\label{eq:kernel_decoder_again}
    \bg(\hbq) = \obq + \bV \hbq + \obV \bn(\hbq), 
\end{equation}
with $\bn$ given by \cref{eq:kernel_optimal_nonlinear_function}.

An important distinction between our KM approach and the FM approach in Section \ref{sec:feature_map_reduction} is that the nonlinear function $\bn$ is optimal in the RKHS $\cH_K^m$, whereas the coefficient matrix $\bXi\in \real^{m\times n_\phi}$ from Section \ref{sec:feature_map_reduction} is optimal in the corresponding vector space of matrices. 
Secondly, the linear system \cref{eq:kernel_optimal_coefficients} scales with the amount of data $M$, whereas the linear system \cref{eq:feature_map_coef_minimizer_equation} scales with the dimension of the feature map $\bphi$.
This can be problematic when using high order polynomial feature maps $\bphi$, where $n_\phi$ can be much larger than the amount of data $M$. 
Moreover, in the FM approach, the structure of the feature map $\bphi$ must be explicitly defined, whereas the nonlinear term $\bn$ can have implicit structure enforced by different types of kernels. 
Lastly, our KM allows one to construct a feature map-based reconstruction of the form \cref{eq:feature_map_decoder} as a special case.

\begin{remark}
    The projection-based ROM + artificial neural network (PROM-ANN) approach of \cite{JBarnett_CFarhat_YMaday_2023a} also considers decoders of the form \cref{eq:kernel_decoder_again} with $\bV$ and $\obV$ consisting of POD modes, but instead models the nonlinear function $\bn$ as an ANN. 
    Furthermore, \cite{JBarnett_CFarhat_YMaday_2023a} and \cite{deParga:2025} suggest selecting $r$ and $m$ by first identifying $n_{\rm tot}:=r+m$ satisfying the energy criterion
    \begin{equation}\label{eq:energy_criterion}
        n_{\rm tot} := \underset{\tilde{n}}{\arg \min} \left\{1-\frac{\sum_{j=1}^{\tilde{n}}\sigma_j^2}{\sum_{j=1}^{{\rm rank}(\tbQ)}\sigma_j^2} \leq \nu \right\},
    \end{equation}
    for some $\nu>0$ where $\sigma_j$ is the $j$th singular value of $\tbQ$, then setting $r$ to some small fraction of $n_{\rm tot}$ and $m$ to $m=n_{\rm tot}-r$. 
\end{remark}
\subsection{Kernel selection and equivalence to feature map approach}\label{sec:kernel_selection}
By the Moore--Aronszajn Theorem (see, e.g., \cite[Theorem 3.10]{GSantin_2018a}),
a positive-definite kernel $K$ uniquely defines an RKHS $\cH_K$.
Consequently, the choice of kernel determines what form $\bn\in \cH_K^m$ can take and how accurately it can be used to approximate the data $\bQ$.
We summarize several effective choices for kernels and show that specific types of kernels, i.e. feature map kernels, can result in nonlinear augmentations of the same form as \cref{eq:feature_map_decoder}.

An effective general-purpose choice of kernel are radial basis function (RBF) kernels.
These kernels have the form
\begin{equation}\label{eq:rbf_kernel}
    K(\bx, \bx')=\psi(\epsilon \norm{\bx-\bx'}_2)
\end{equation}
where $\psi:\real_{\geq 0}\to\real$ for all $\bx, \bx'\in \real^n$ and $\epsilon>0$ is a \emph{shape parameter}, and hence an 
RBF interpolant of the data \cref{eq:modal_data_matrices} has the form
\begin{align}\label{eq:rbf_interpolant}
    \bn(\hbq) &= \bOmega\trp
    \bpsi_\epsilon(\hbq), 
     &
    \bpsi_\epsilon(\hbq) &=
    \begin{bmatrix}
        \psi(\epsilon\norm{\hbq_1-\hbq}_2) \\
        \vdots \\
        \psi(\epsilon\norm{\hbq_M-\hbq}_2)
    \end{bmatrix}.
\end{align}
\Cref{tbl:rbf_kernels} provides some examples of commonly used RBF kernels.
A thorough discussion of RBFs in kernel interpolation can be found in, e.g., \cite{GBWright_2003a}.
\begin{table}[h!]
    \centering
    \begin{tabular}{r|l}
        Name & $\psi(r)$ \\ \hline
        Gaussian & $\exp(-r^2)$ \\ 
        Basic Mat\'{e}rn & $\exp(-r)$ \\ 
        Linear Mat\'{e}rn & $(1+r)\exp(-r)$ \\ 
        Quadratic Mat\'{e}rn & $(3+3r + r^2)\exp(-r)$ \\ 
        Inverse Quadratic & $(1+r^2)^{-1}$ \\ 
        Inverse Multiquadric & $(1+r^2)^{-1/2}$ \\ 
        Thin Plate Spline & $r^2 \log(r)$ 
    \end{tabular}
    \caption{Examples of RBF kernels.}
    \label{tbl:rbf_kernels}
\end{table}

Another type of kernel is the polynomial kernel $K:\real^d\times \real^d\to \real$
\begin{equation}\label{eq:polynomial_kernel}
    K(\bx, \bx') = \left(c+\rho\bx\trp\bx'\right)^\ell, 
\end{equation}
where $c, \rho \in \real$, $c\geq 0$, $\rho > 0$, are hyper-parameters and $\ell\in \mathbb{N}$ is the order of the polynomial.
Typical values of $c$ and $\rho$ are $c=1$ and $\rho = 1/d$.
Using a polynomial kernel results in a polynomial nonlinear term $\bn$. 
Therefore, the resulting decoder \cref{eq:kernel_decoder_again} defines a \emph{polynomial manifold}.
However, using a polynomial kernel only \emph{implicitly} defines the polynomial structure of \cref{eq:kernel_decoder_again}. 
In some use cases, such as reduced-order modeling, it may be beneficial to know the \emph{explicit} polynomial structure. 
In these cases, using a \emph{feature map kernel} may be preferable. 

A feature map $\bphi:\real^d\to \real^{n_\phi}$ and a strictly positive definite matrix $\bG \in \real^{n_\phi\times n_\phi}$ induce a postive-definite kernel $K:\real^d\times\real^d \to \real$ defined as the $\bG$-weighted inner product of two feature vectors:
\begin{equation}\label{eq:feature_map_kernel}
    K(\bx, \bx') = \bphi(\bx)\trp \bG \bphi(\bx').
\end{equation}
The resulting interpolant $\bn$ can then be expressed as
\begin{align}\label{eq:feature_map_interpolant}
    \bn(\hbq) &= \bOmega\trp K(\hbQ, \hbq)
    = \bOmega\trp \bphi(\hbQ)\trp\bG \bphi(\hbq)
    = \underbrace{\bOmega\trp \bphi(\hbQ)\trp\bG}_{:=\bXi} \bphi(\hbq)
    = \bXi\bphi(\hbq), 
\end{align}
and thus the decoder $\bg$ is of the form 
\begin{equation}\label{eq:feature_map_decoder_2}
    \bg(\hbq) = \obq + \bV \hbq + \obV \bXi \bphi(\hbq).
\end{equation}
Therefore, feature map-based nonlinear decoders of the form \cref{eq:feature_map_decoder} can be obtained as a special case of the KM approach.
In particular, taking $\bphi$ to be the quadratic feature map \cref{eq:quadratic_feature_map} results in a QM.

\begin{remark}
    \label{rmk:relation_to_GPs}
    Although the GP-based closure of \cite[Section 3.2]{deParga:2025} relies on a probabilistic framework, the authors' use of the GP mean for $\bn$ is computationally equivalent to our KM framework with an RBF kernel and regularization $\lambda$ equal to the GP prior covariance $\sigma^2$. 
    Similarly, the RBF interpolation-based closure of \cite[Section 3.3]{deParga:2025} is a special case of our KM approach using an RBF kernel. 
\end{remark}

\begin{remark}[Input normalization]
    \label{rmk:input_normalization}
    Depending on the input data $\bX$ and choice of kernel $K$, the kernel matrix $K(\bX, \bX)$ may have poor conditioning, which can adversely affect the solve for the coefficient matrix $\bOmega.$
    Although increasing the regularization parameter $\lambda$ may ameliorate the conditioning of the linear system \cref{eq:coefficient_equation}, it may also increase the error of the resulting kernel interpolant at the given training data.
    Alternatively, one can improve the conditioning of $K(\bX, \bX)$ by normalizing the inputs $\bX$. 
    This relies on the following observation: for an injective $\bnu:\real^d\to\real^d$, if $K$ is a positive definite kernel, then the function $K_\bnu:\real^d\times\real^d\to \real$ defined by
    \begin{equation}\label{eq:normalized_kernel}
        K_\bnu(\bx, \bx') = K(\bnu(\bx), \bnu(\bx'))
    \end{equation}
    is also a positive-definite kernel \cite[Prop.~2.7]{GSantin_2018a}.
    Therefore, choosing an appropriate $\bnu$ can improve the conditioning of $K_\bnu(\bX,\bX)$ compared to $K(\bX,\bX)$.
    An effective choice is the function $\bnu(\bx) = \bM^{-1}(\bx - \bar{\bx})$, where $\bM = \operatorname{diag}(\bm)\in\real^{d \times d}$ and $\bar{\bx}\in\real^{d}$ are given by
    \begin{align}\label{eq:kernel_normalization_choice}
        m_i = \max_{j}(\bX_{ij}) - \min_{j}(\bX_{ij}),
        \qquad
        \bar{x}_i = \min_{j}(\bX_{ij}),
        \qquad
        i = 1, \dots, d,
    \end{align}
    which maps the components of each row of $\bX$ to the interval $[0, 1]$.
    With this choice of $\bnu$, we set the weighting matrix $\bG$ in a kernel induced by a feature map $\bphi:\real^d\to\real^{n_\phi}$ to $\bG = (1/n_\phi)\bI$.
\end{remark}

%% file: numerics_intro.tex
%!TEX root = main.tex
\section{Numerical results}\label{sec:numerics}
We compare our KM approach with POD and two recent QM approaches: the Greedy QM approach from \cite{PSchwerdtner_BPeherstorfer_2024a} and the Alternating minimization QM approach from \cite{RGeelen_LBalzano_SWright_KWillcox_2024a}.
For the Alternating QM approach, we use the same stopping tolerance of $10^{-3}$ for the alternating minimization procedure as in \cite{RGeelen_LBalzano_SWright_KWillcox_2024a}.
For the KM approach, we compare several RBF kernels with varying shape parameter $\epsilon$ and a feature map kernel induced by the quadratic feature map \cref{eq:quadratic_feature_map}. 
Where appropriate, the RBF KM will be denoted ``Kernel RBF'' and the KM with kernel induced by the quadratic feature map will be denoted ``Kernel QM''.
Recall from \Cref{sec:kernel_selection} that a feature map kernel with a quadratic feature map results in a QM of the form \cref{eq:feature_map_decoder}.

%% file: numerics_adv_diff.tex
%!TEX root = main.tex
\subsection{2D Advection-diffusion-reaction -- boundary layer}\label{sec:numerics_adv_diff}

The first problem we consider is given by state snapshot data from the 2D advection-diffusion-reaction equation with homogeneous Dirichlet boundary conditions, adapted from one of the examples in \cite{Parish:2025}. 
The 2D advection-diffusion-reaction equation on the unit square $\Omega=(0, 1)^2$ is given by
\begin{subequations}\label{eq:advection_diffusion_pde}
    \begin{align}
        \pdt u(t, x, y) - \alpha \Delta u(t, x, y) + \bbeta\cdot \nabla u(t, x, y) +\gamma u(t, x, y) &= f(t, x, y), 
        & (x,y) &\in \Omega, & t &\in (0, T), \\
        u(t, x, y) &= 0, &(x, y) &\in \partial \Omega, & t &\in (0, T), \\
        u(0, x, y) &= 0, & (x,y) & \in \Omega,
    \end{align}
\end{subequations}
where $\alpha>0$ is the diffusion parameter, $\bbeta = [\beta_1, \beta_2] \in \real^2$, $\beta_1, \beta_2 \geq 0$ is the advection parameter, $\gamma\geq 0$ is the reaction parameter, $T>0$ is the final time, and $f:(0, T)\times \Omega \to \real$ is a forcing function.  
We discretize the PDE \cref{eq:advection_diffusion_pde} using a centered finite difference scheme for the second derivatives and a first-order upwind backward difference scheme for the first derivatives, resulting in a linear system of ODEs of the form 
\begin{subequations}\label{eq:advection_diffusion_fom}
    \begin{align}
        \dt \bq(t) &= \bA \bq(t) + \mathbf{F}(t), & t &\in (0, T), \\
        \bq(0) &= \bzero.
    \end{align}
\end{subequations}

\begin{figure}[H]
    \centering
    \includegraphics[width=\textwidth]{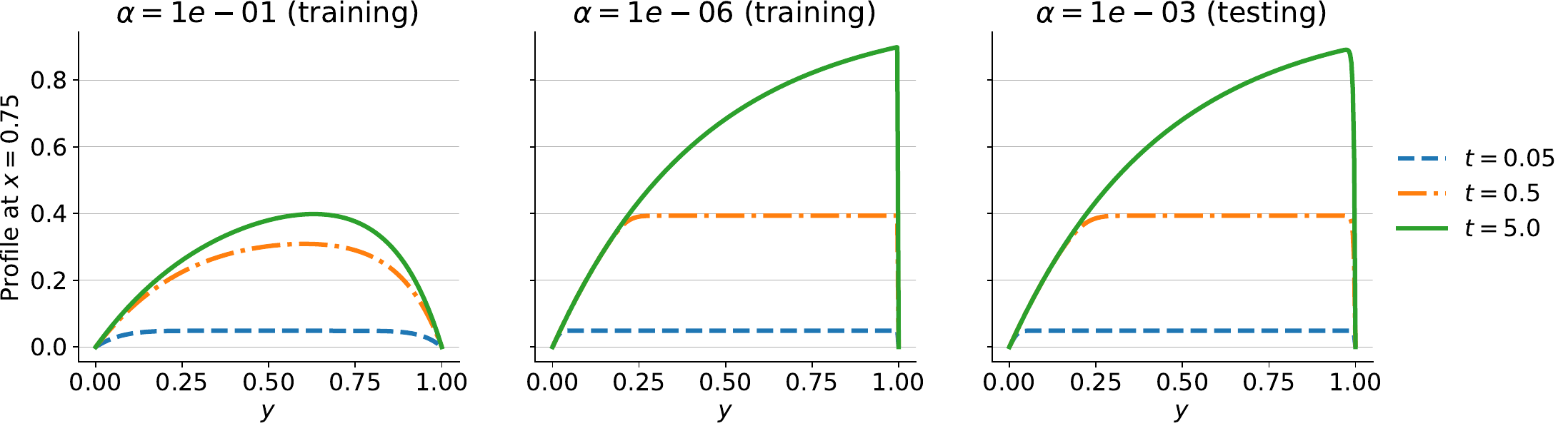}
    \caption{Solution profile along $x=0.75$ for different diffusion values at different timesteps.}
    \label{fig:advdiff_profiles}
\end{figure}

For this experiment, we take $258$ uniformly spaced grid points in both the $x$ and $y$ directions, resulting in a full state dimension of $N=256^2 = 65,536$.
The time stepping is done using \texttt{scipy.interpolate.solve\_ivp()} in Python with a BDF scheme \cite{virtanen2020scipy,LFShampine_MWReichelt_1997a}, time step $\Delta t=0.01$, and final time $T=5$.
We take an advection parameter of $\bbeta = (1/2)[\cos(\pi/3), \sin(\pi/3)]$, a reaction parameter of $\gamma=1$, forcing $f=1$, and vary the diffusion parameter $\alpha$.
This combination of parameters yields a sharp boundary layer for smaller values of $\alpha$. 
A similar test case was used in \cite[Section 7.2]{Parish:2025} using a fixed value of $\alpha=10^{-3}$ and a finite element discretization.
\Cref{fig:advdiff_profiles} plots profiles of the solution at $x=0.75$ for different values of $\alpha$ to illustrate how varying $\alpha$ affects the sharpness of the boundary layer. 

\begin{figure}[H]
    \centering
    \includegraphics[width=0.45\textwidth]{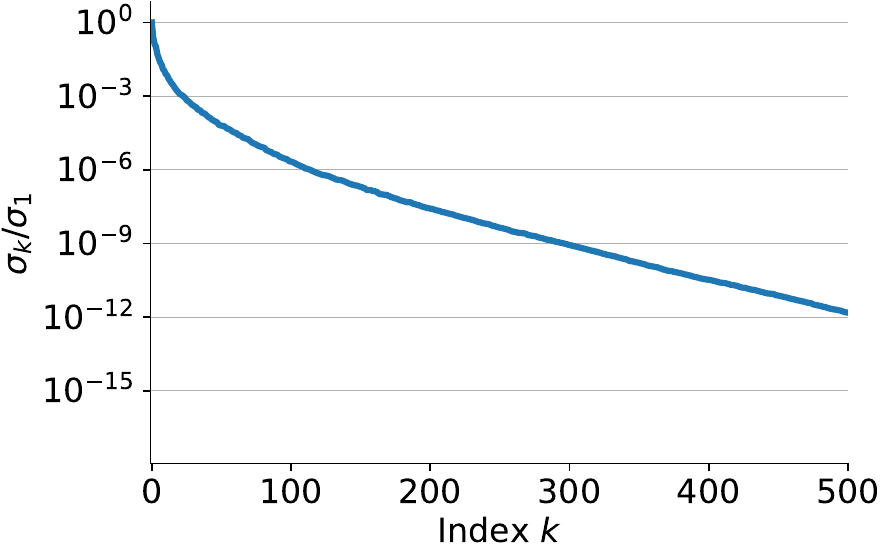}
    \caption{First $500$ normalized singular values for the 2D advection-diffusion-reaction example.}
    \label{fig:advdiff_singvals}
\end{figure}

To collect training data $\bQ$, we simulate \cref{eq:advection_diffusion_pde} for $\ell=10$ logarithmically spaced values of $\alpha\in [10^{-6}, 10^{-1}]$. 
This results in a training dataset 
\begin{equation}\label{eq:advdiff_training_data}
    \bQ = \begin{bmatrix}
        \bq_0^{(1)} & \dots & \bq_{N_t}^{(1)} & \dots &
        \bq_0^{(\ell)} & \dots & \bq_{N_t}^{(\ell)} 
    \end{bmatrix}\in \real^{N\times M},
\end{equation}
where $N = 65,536$ and $M = \ell(N_t+1)=5,010$ and where $\bq_j^{(k)}$ corresponds to the state of \cref{eq:advection_diffusion_fom} at time step $j$ and the $k$th diffusion training diffusion parameter.
The offset $\obq$ is taken to be the mean of the training data \cref{eq:snapshot_mean}.
We then test different dimensionality reduction approaches for $\alpha=10^{-3}$, which is outside of training set.
The singular value decay for the first $500$ normalized singular values is plotted in \Cref{fig:advdiff_singvals}.
The error metric that we use for this example is the relative $\ell^2$ projection error 
\begin{equation}\label{eq:relative_l2_projection_error}
    E=\left(\frac{\sum_{j=0}^{N_t}\norm{\bq_j^*-\bg(\bh(\bq_j^*))}_2^2}{\sum_{j=0}^{N_t}\norm{\bq_j^*}_2^2}\right)^{1/2},
\end{equation}
where $\bq_j^*$ is test data for $\alpha=10^{-3}$ at time step $j$.

For this example, the Kernel RBF approach uses the quadratic Mat\'{e}rn kernel with shape parameter $\epsilon = 10^{-2}.$
We do not use the input normalization discussed in \Cref{rmk:input_normalization} for any of the KMs computed.
The Kernel QM, Kernel RBF, Alternating QM, and Greedy QM approaches use the regularization values in \Cref{tbl:adv_diff_regs}. 
For further details on the selection of the RBF kernel, shape parameter, and regularization values, see \Cref{app:advdiff}. 
\begin{table}[H]
    \centering
    \begin{tabular}{c|c}
        Manifold type & regularization $\lambda$ \\ \hline
        Kernel QM & $10^{-2}$ \\ 
        Kernel RBF & $10^{-11}$ \\
        Alternating QM & $10^0$ \\ 
        Greedy QM & $10^0$ 
    \end{tabular}
    \caption{Fixed regularization values for different dimensionality reduction approaches for the 2D advection-diffusion-reaction example.}
    \label{tbl:adv_diff_regs}
\end{table}

\begin{figure}[H]
    \centering
    \includegraphics[width=0.6\textwidth]{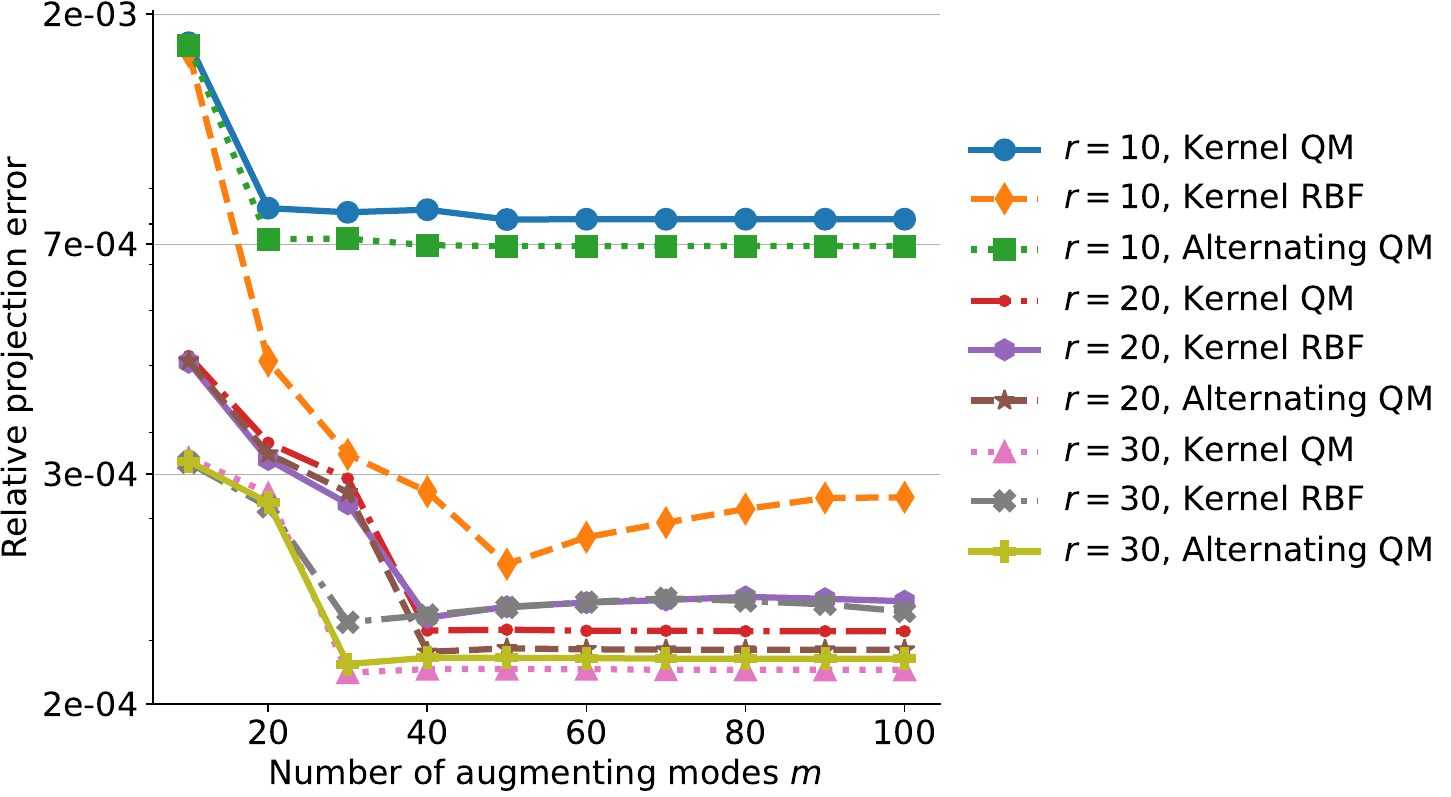}
    \caption{Relative projection error as a function of number of augmenting modes $m$ for several different latent dimension sizes $r$ for the 2D advection-diffusion-reaction example.}
    \label{fig:advdiff_error_vs_augmenting_modes}
\end{figure}

We examine the effect of the number of augmenting modes on the projection error by varying the number of augmenting modes $m \in [10, 100]$ for the Kernel QM, Kernel RBF, and Alternating QM approaches for $r\in \set{10, 20, 30}$.
The Greedy QM approach is excluded from this test because $\obV=\bI$ in the greedy case, and hence does not depend on a choice of number of augmenting modes.
\Cref{fig:advdiff_error_vs_augmenting_modes} indicates that for a fixed value of $r$, the accuracy improvement by increasing the number of augmenting modes $m$ plateaus as $m$ exceeds a small integer multiple of $r$.
For $r=10$, the Kernel QM and Alternating QM errors plateau for $m>2r=20$, while the Kernel RBF error decreases for $m\leq5r=50$, then increases slightly and plateaus for $m>50$. 
For $r=20$, each method plateaus and achieves similar error for $m>2r=40$, while for $r=30$, each method's error plateaus for $m\geq30$. 
These results further indicate that the benefit of nonlinear augmentation is more substantial for smaller values of $r$.
Since Kernel RBF at $r=10$ is the slowest method to plateau for $m\geq 5r$, we fix $m=5r$ for each approach for the remaining tests on the current numerical example.

\begin{figure}[H]
    \centering
    \includegraphics[width=\textwidth]{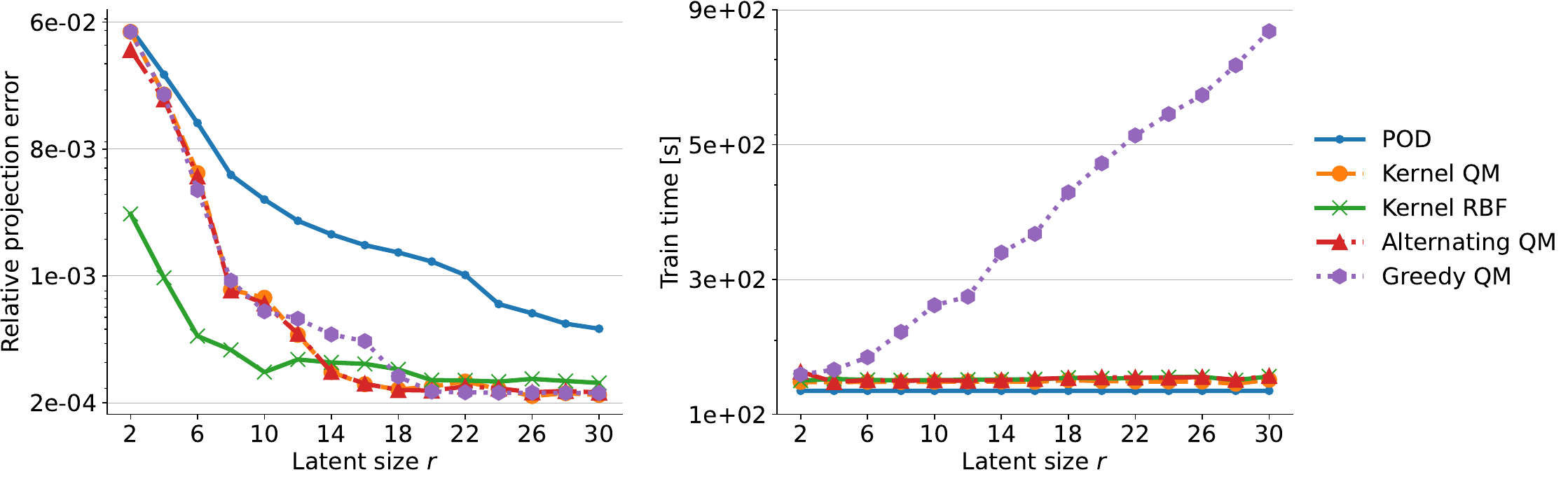}
    \caption{Relative projection error and train time as a function of the latent dimension size $r$ for different dimensionality reduction approaches.}
    \label{fig:advdiff_error_vs_romsize}
\end{figure}

\begin{figure}[H]
    \begin{subfigure}{0.33\columnwidth}
        \centering
        \includegraphics[width=0.9\textwidth]{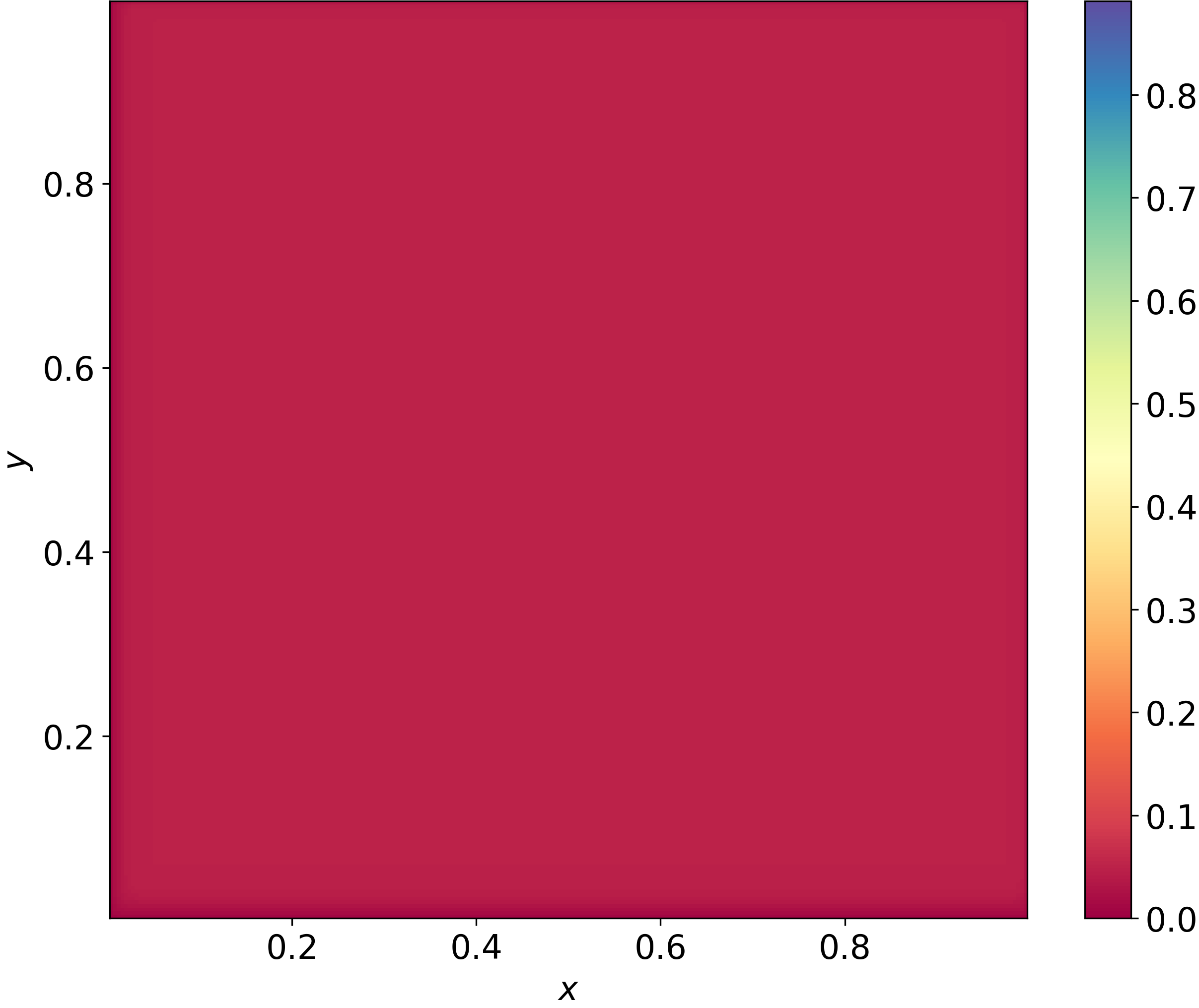}

        \includegraphics[width=0.9\textwidth]{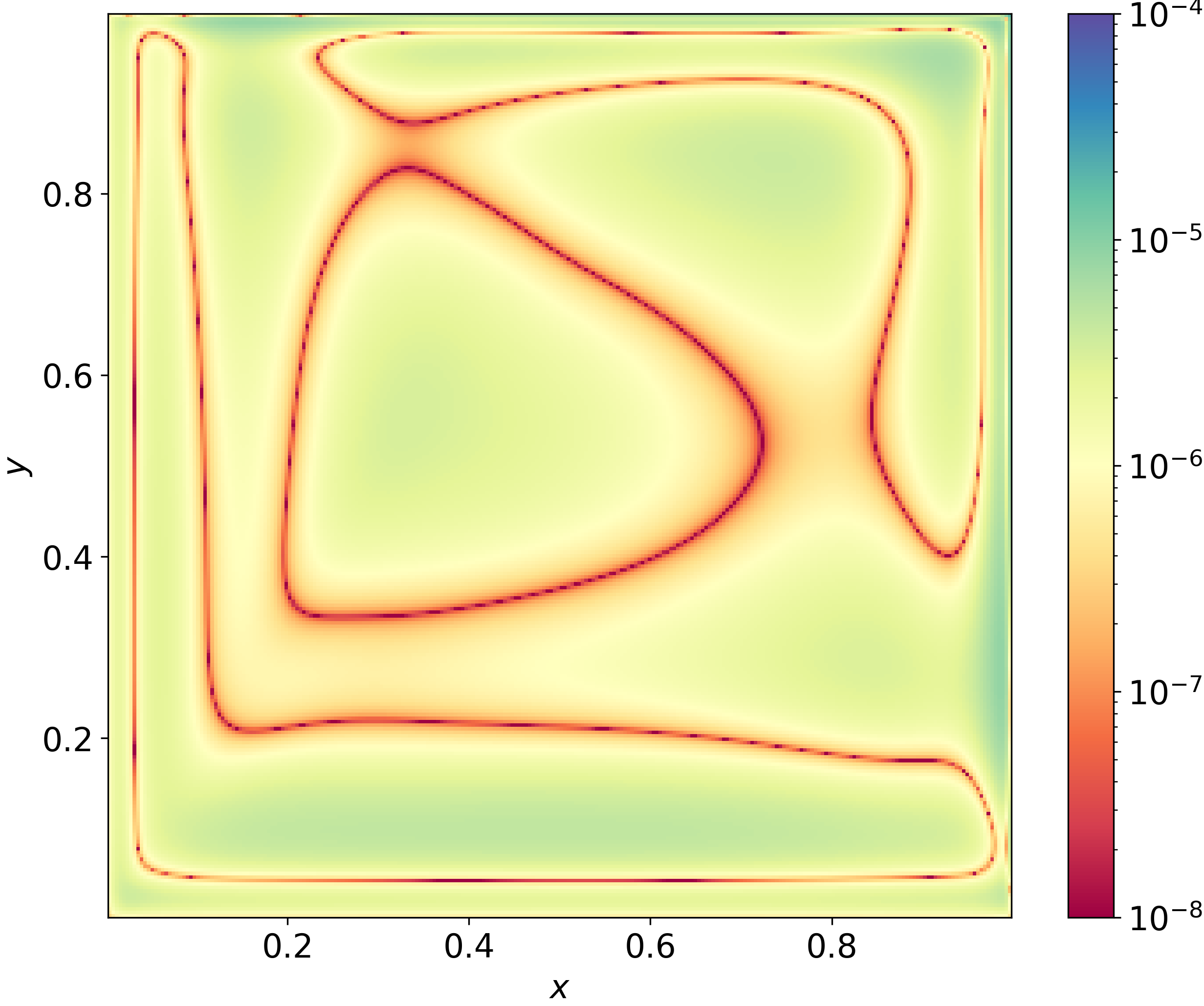}

        \includegraphics[width=0.9\textwidth]{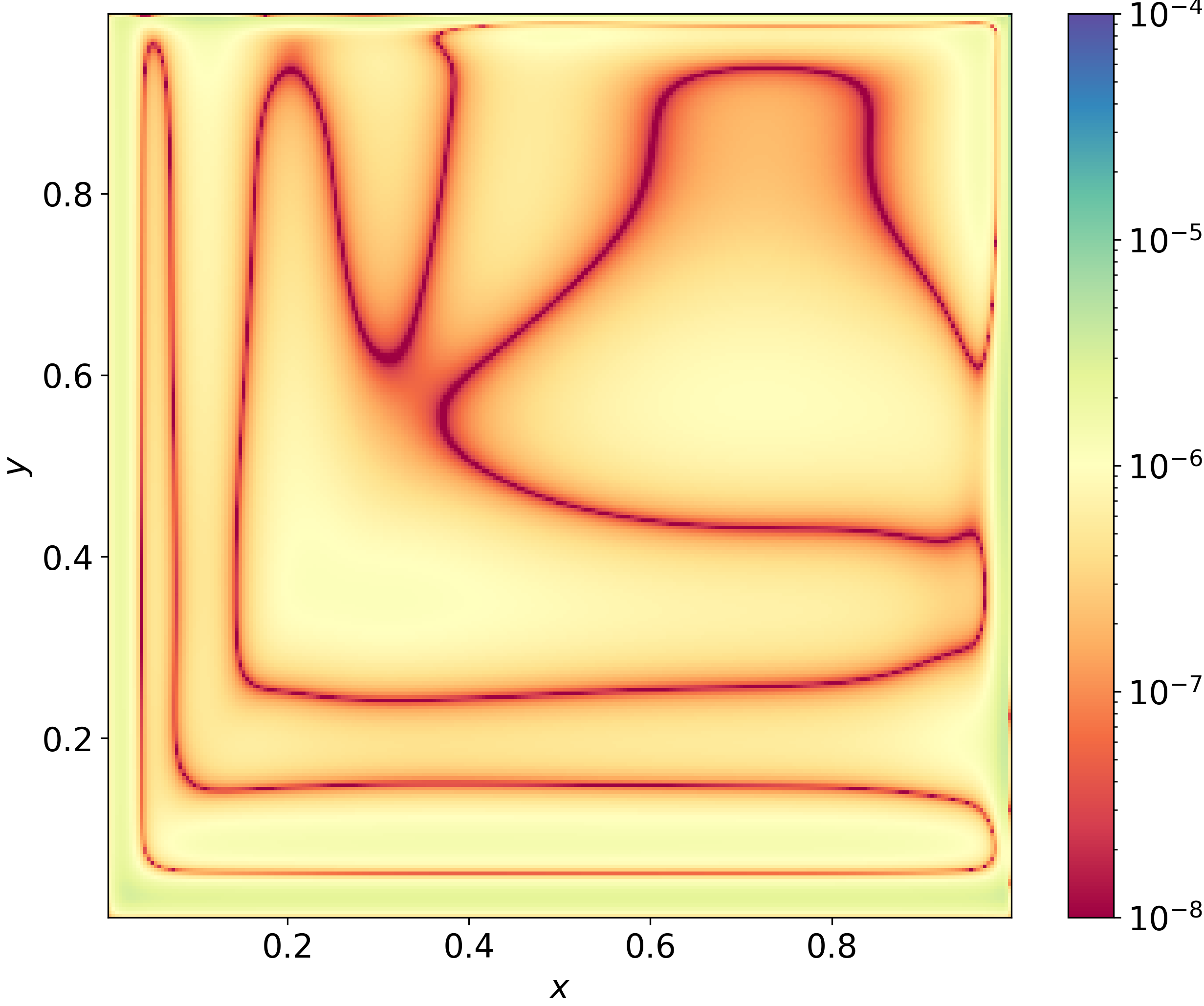}

        \includegraphics[width=0.9\textwidth]{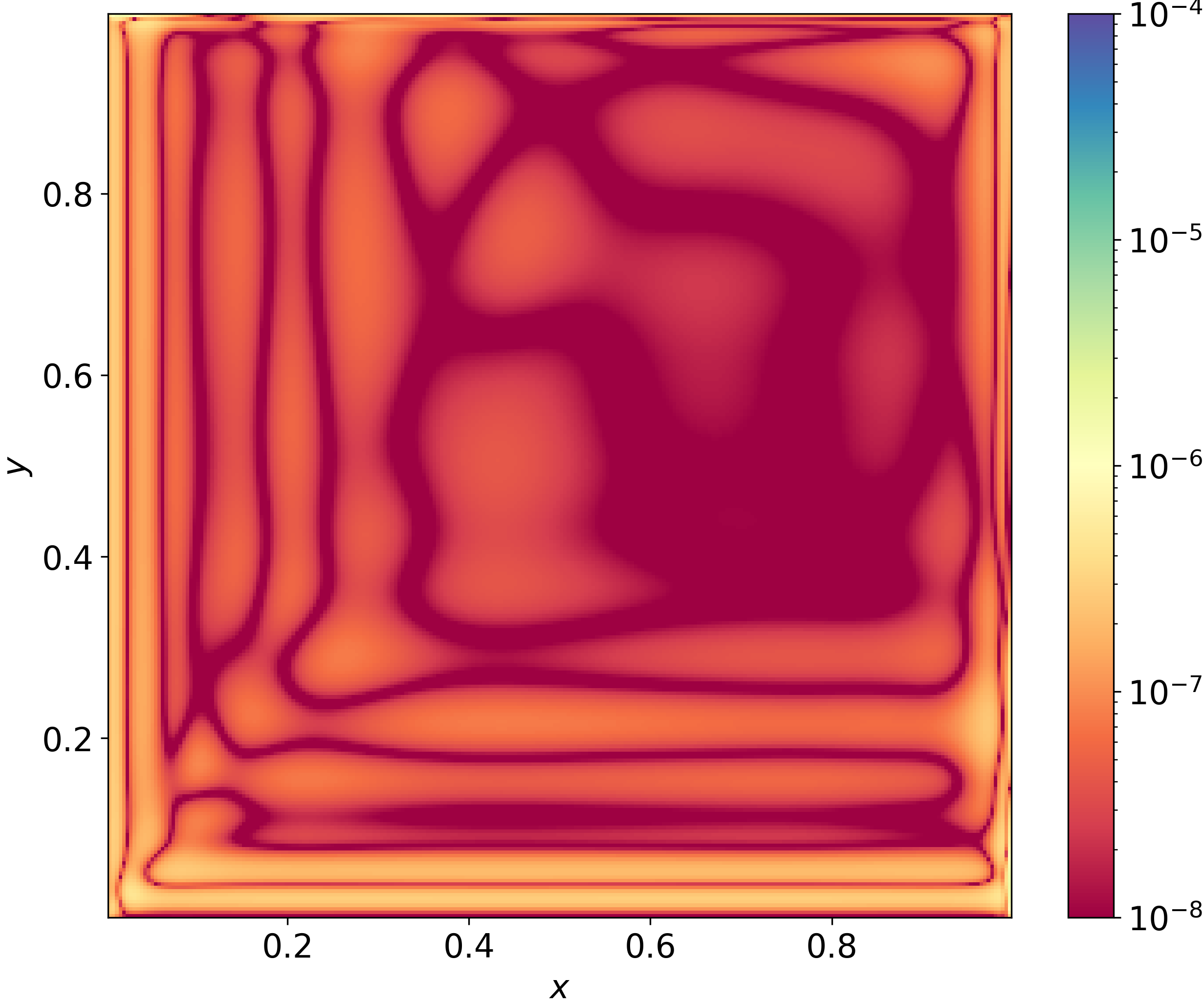}  

        \caption{$t=0.05$}
    \end{subfigure}
    \begin{subfigure}{0.33\columnwidth}
        \centering
        \includegraphics[width=0.9\textwidth]{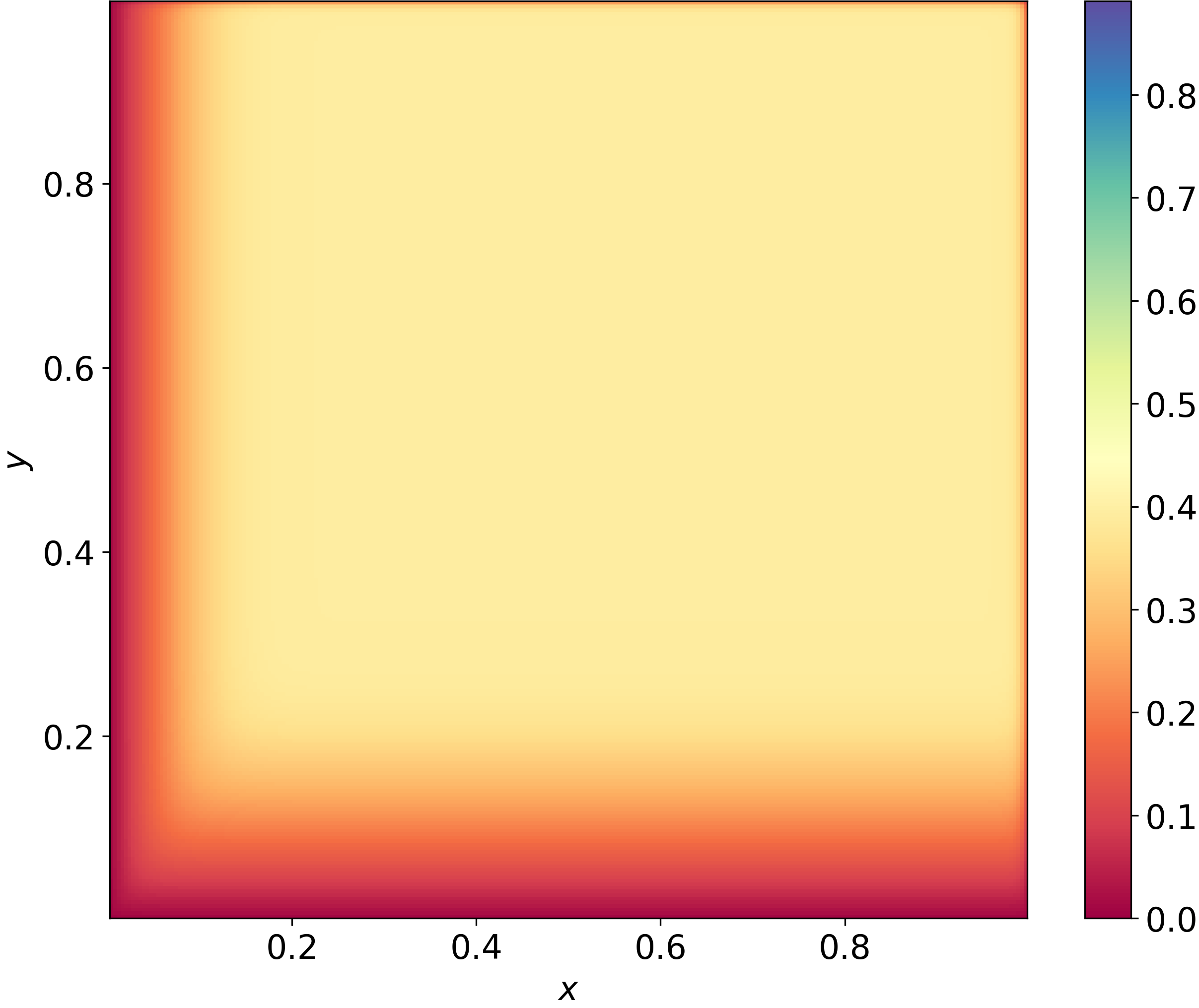}

        \includegraphics[width=0.9\textwidth]{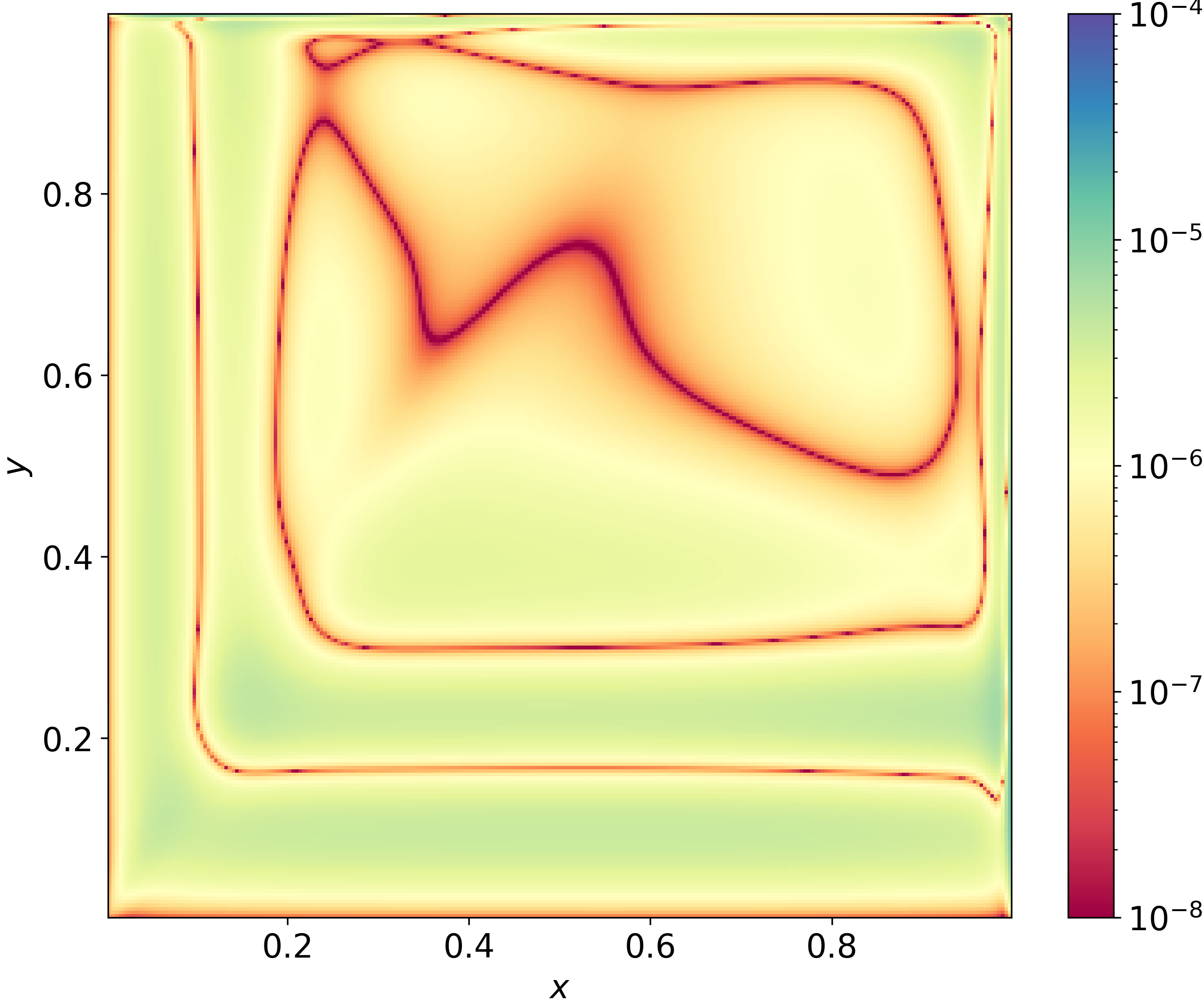}

        \includegraphics[width=0.9\textwidth]{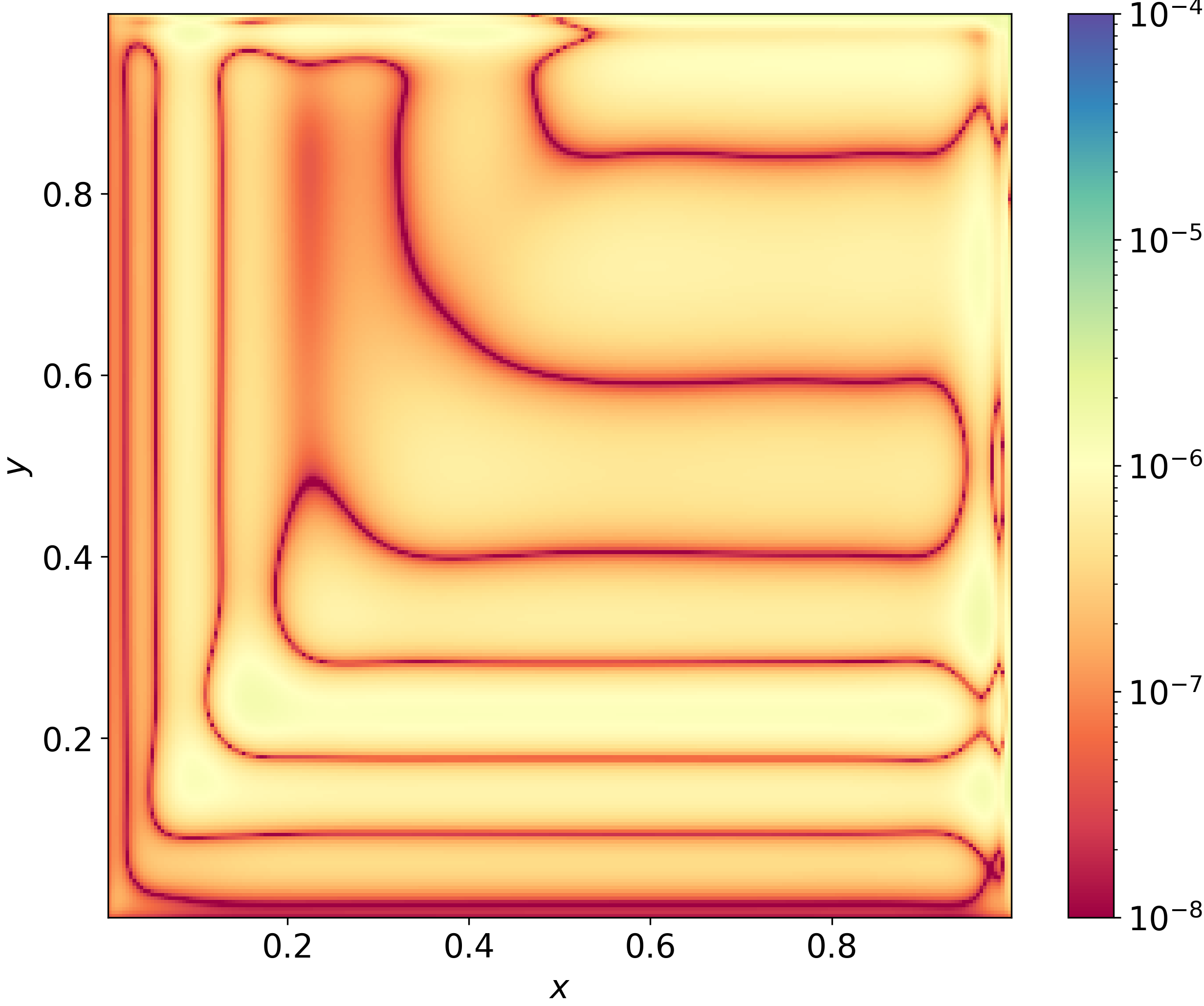}

        \includegraphics[width=0.9\textwidth]{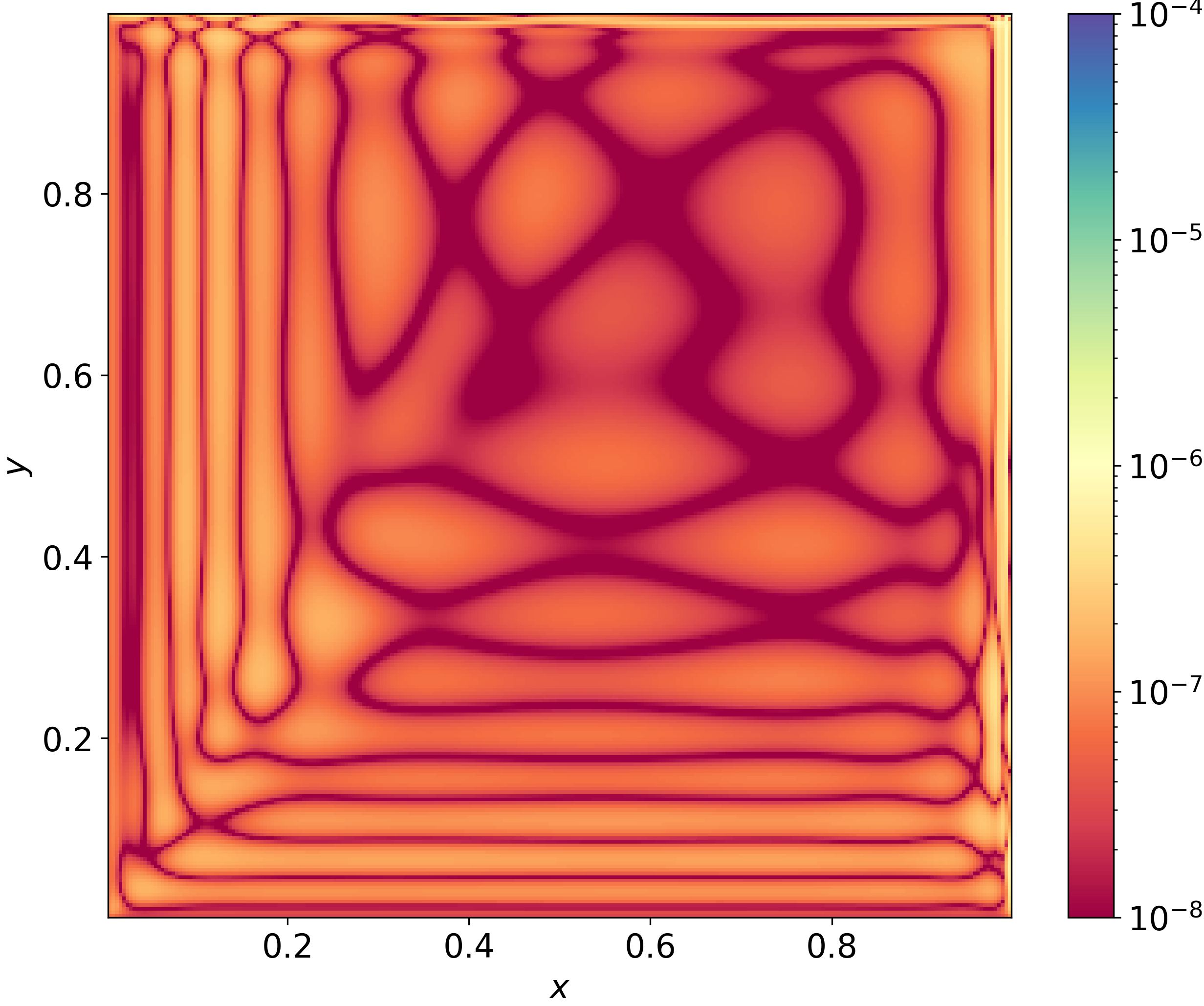}

        \caption{$t=0.5$}
    \end{subfigure}
    \begin{subfigure}{0.33\columnwidth}
        \centering
        \includegraphics[width=0.9\textwidth]{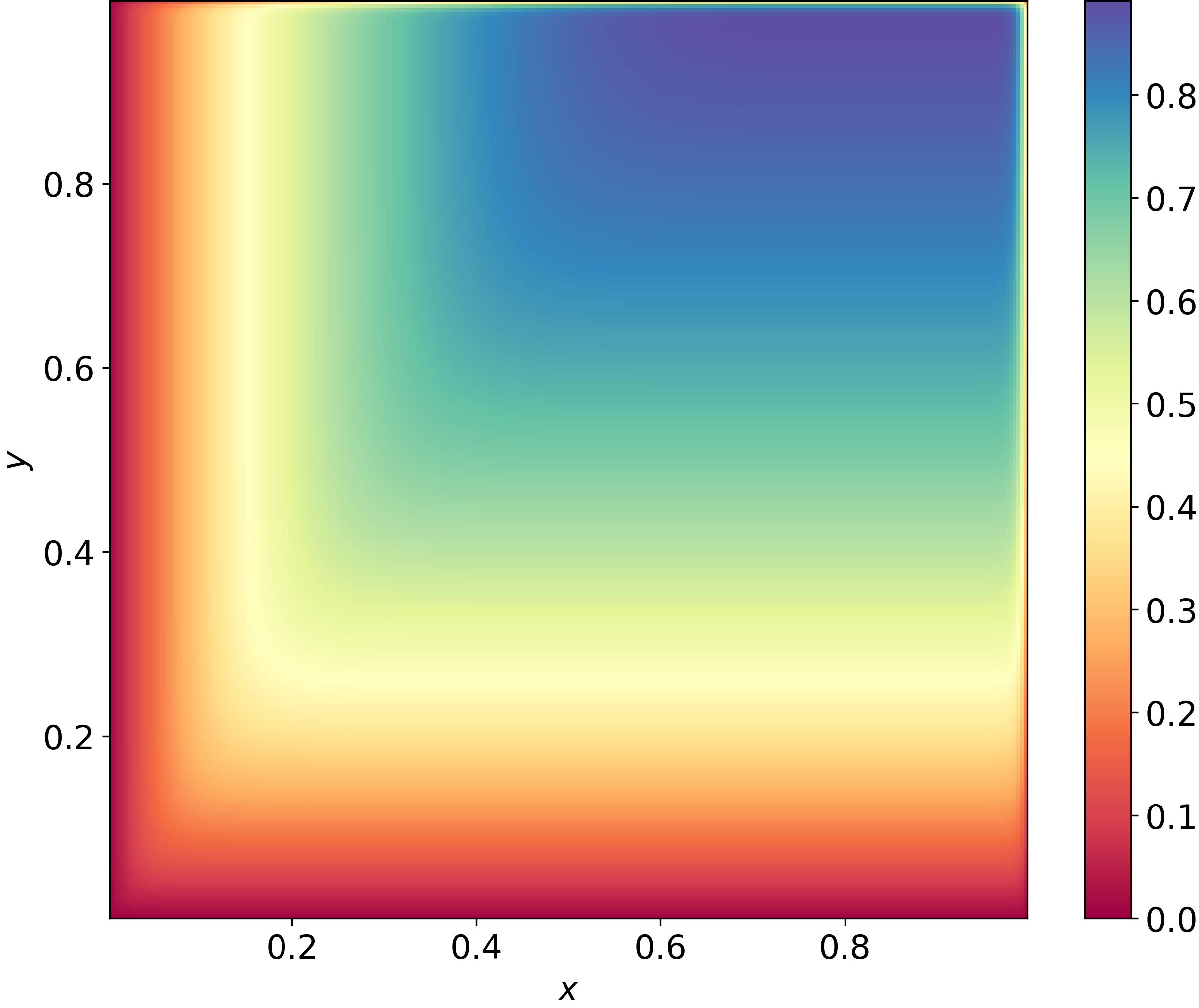}

        \includegraphics[width=0.9\textwidth]{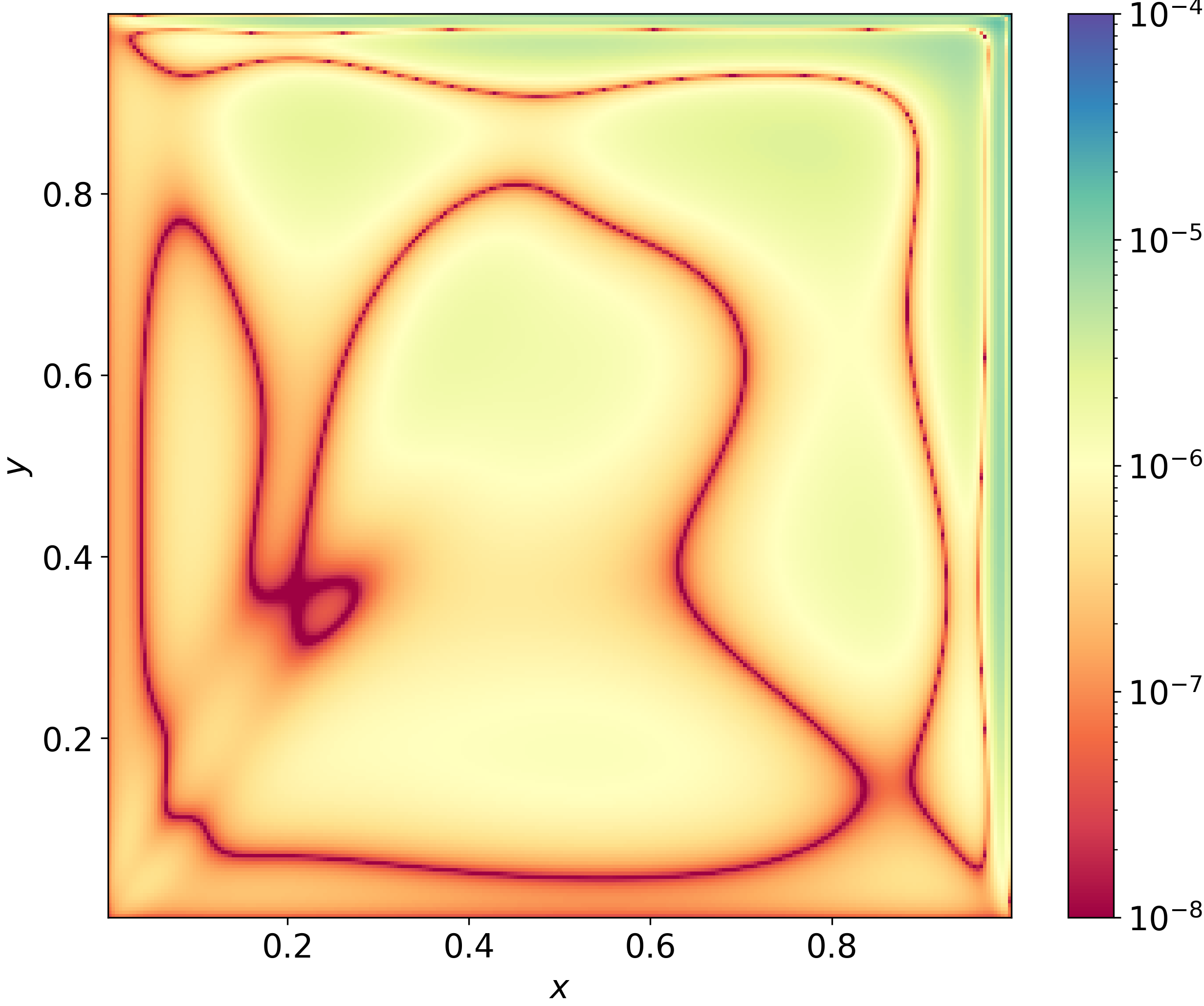}

        \includegraphics[width=0.9\textwidth]{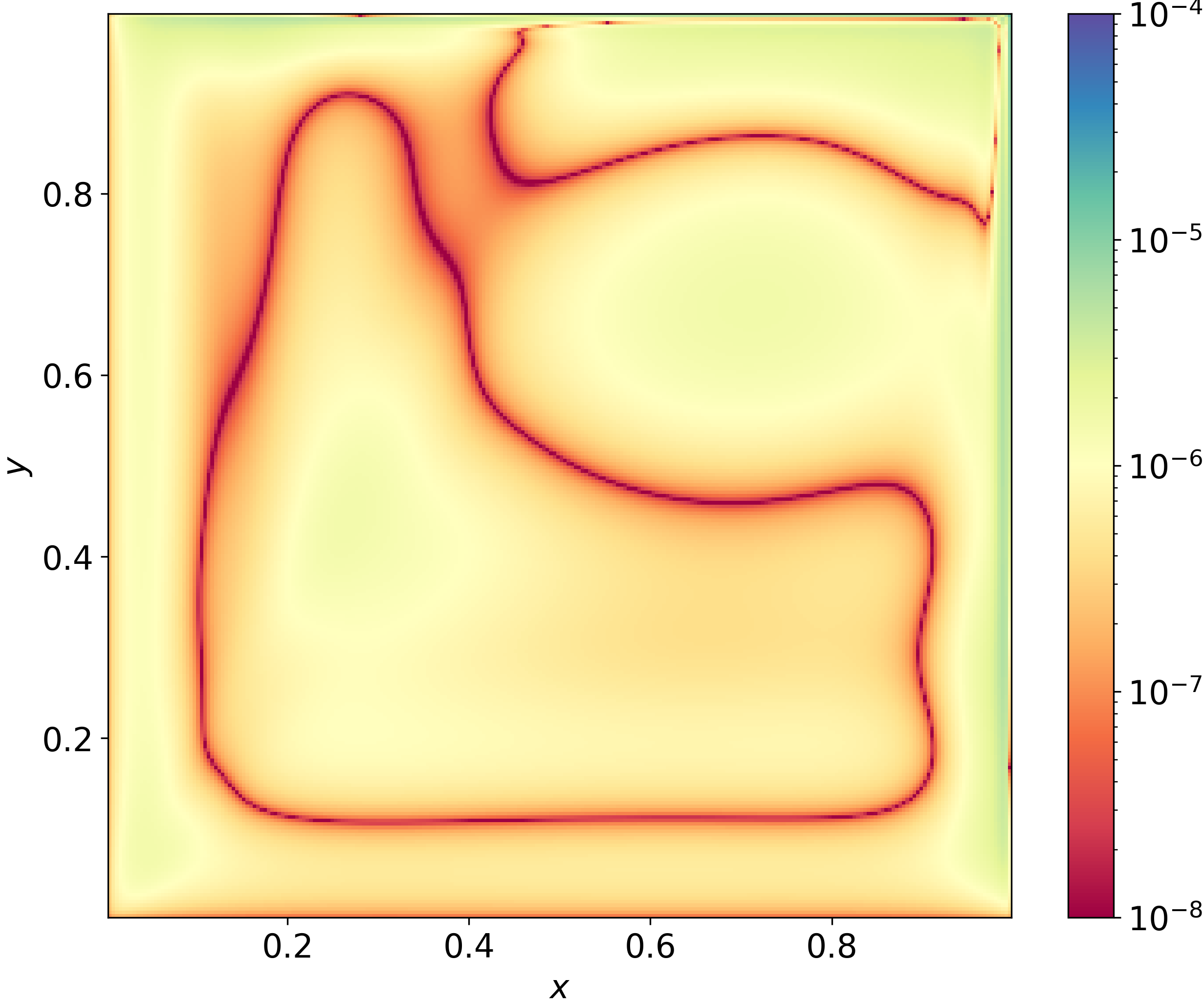}

        \includegraphics[width=0.9\textwidth]{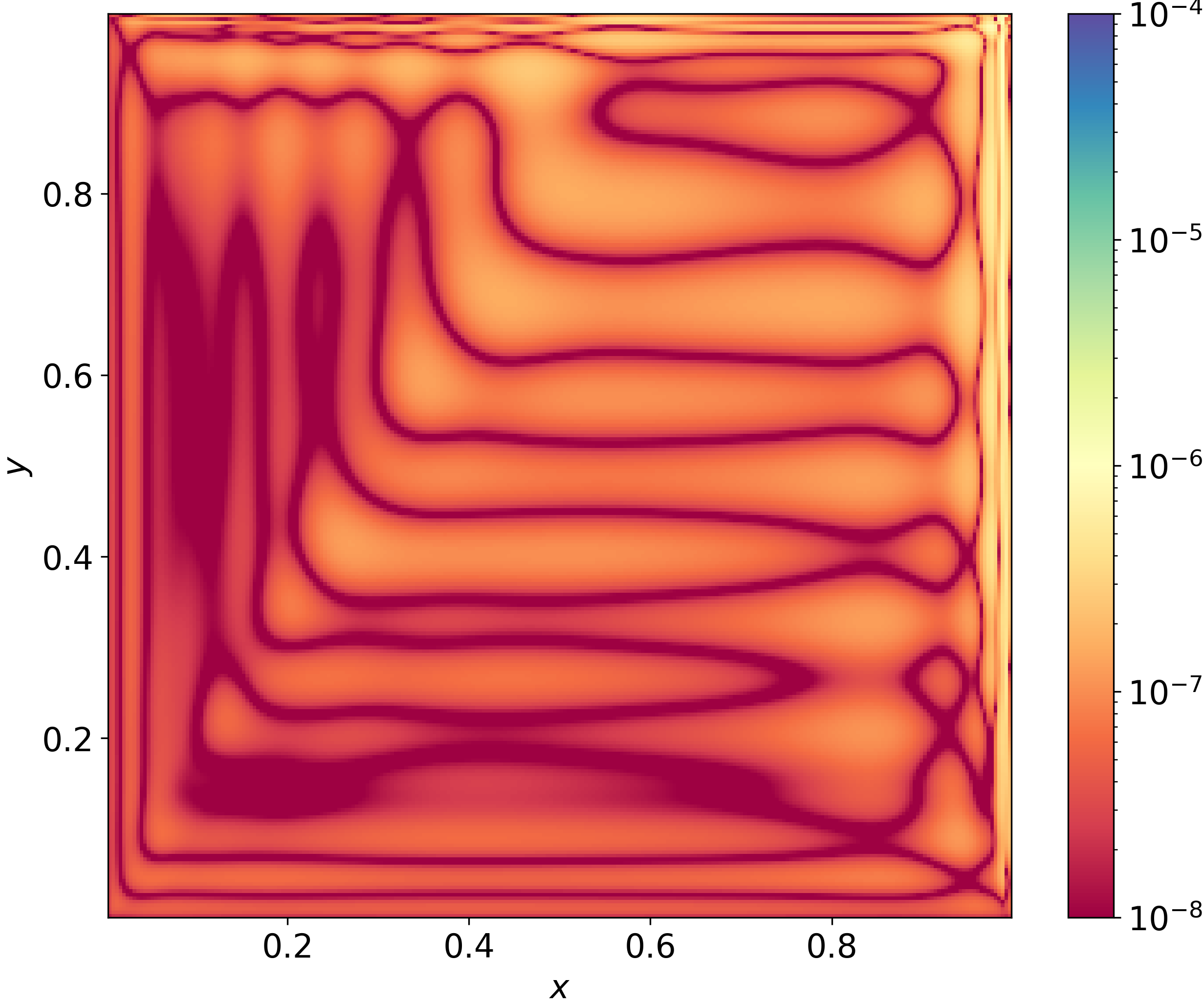}

        \caption{$t=5$}
    \end{subfigure}
    \caption{Plot of state and relative projection error at different time steps for $r=6$ and $m=30$. Top row: full state; second row: POD projection error; third row: Alternating QM projection error; bottom row: Kernel RBF projection error.}
    \label{fig:advdiff_states}
\end{figure}

Next, we compare the different approaches by varying the latent dimension size $r$.
As indicated above, we take the number of augmenting modes to be $m=5r$ for the Kernel QM, Kernel RBF, and Alternating QM approaches.
\Cref{fig:advdiff_error_vs_romsize} shows that the Kernel RBF approach achieves the lowest errors for $r\leq 12$, while each method plateaus with similar error for $r\geq 14$.
Notably, \Cref{fig:advdiff_error_vs_romsize} also shows that the Kernel QM, Kernel RBF, and Alternating QM approaches have significantly smaller training times than Greedy QM, especially for larger values of $r$. 
For this numerical example, we can conclude that the Kernel RBF approach achieves the best performance for small values of $r$, while each method performs similarly for larger values of $r$.
To visualize the error decrease of the Kernel RBF approach over POD on this numerical example, we plot the full state $\bq$ and the relative projection errors $|\bq_j^* -\bg(\bh(\bq_j^*))|/\sqrt{\sum_{j=0}^{N_t}\norm{\bq_j^*}^2}$ using POD, Alternating QM, and Kernel RBF with $r=6$ and $m=5r=30$ in \Cref{fig:advdiff_states}.

%% file: numerics_pseudo_hifire.tex
%!TEX root = main.tex
\subsection{High-speed aerodynamics -- analytic model of surface heating}\label{sec:numerics_pseudo_hifire}

The next problem considered is an analytic function developed to parametrically emulate prominent features of high-speed aerodynamic flowfields.
The physical interpretation for this model problem is
the variation in aerothermal quantities of interest over conical geometries at varying flight conditions.

Specifically, this model emulates the variation in heat flux over a flared-cone geometry, similar to the HIFiRE-1 test vehicle~\cite{BLance_AMKrueger_BAFreno_RMWagnild_2022a}, though with significant simplification of the underlying physical models and geometry.
The test function is defined so that a non-dimensional surface quantity, $s$, corresponding to the heat flux magnitude can be computed on a quasi-three-dimensional domain, defined as a function in cylindrical coordinates $(z, r, \theta)$, where  $z\in [0, 2]$ is the axial coordinate, $r\in [0, R]$, $R>0$, is a radial coordinate, and $\theta\in [0, 2\pi]$ is a circumferential angle measure.
We consider variation of the field in response to $\bmu = [\mu_1 \; \mu_2]\trp \in \cD = [-10^\circ, 10^\circ]\times [0.4, 0.8]$ consisting of two parameters: the angle of attack, $\mu_1\in [-10^\circ, 10^\circ]$, dictating the pitch angle orientation relative to the freestream, and a laminar-to-turbulent transition location parameter, $\mu_2\in [0.4, 0.8]$, controlling the location of turbulent transition (and associated increase in heat flux) on the cone surface, where the dimensionless field quantity $s:[0, 2]\times[0, 2\pi] \times \cD \to \real$ can be expressed as

\begin{align}
\label{eq:hifire}
s(z, \theta; \bmu) &=
\frac{100 (1 - 5 \sin(\mu_1) \sin(\theta))}{(1 + 20 z)^2} + 
50 \sigma(z - \mu_2 (1 + 2 \sin(\mu_1) \sin(\theta))) -
50 \sigma(z - 1.2) 
\\
&\quad
+ \frac{75 \sigma(x - 1.6) (1 - 5 \sin(\mu_1) \sin(\theta))}{\delta + x} + 50. \nonumber
\end{align}

\noindent Here, $\sigma(x)$ is a scaled sigmoid function modified to approximate a sharp discontinuity as

\begin{equation}
    \sigma(x) = \frac{1}{1 + e^{-100x}},
\end{equation}

\noindent and $\delta$ is a small, nonzero value of $10^{-6}$.  

\begin{figure}[H]
    \begin{subfigure}{0.33\columnwidth}
        \centering
        \includegraphics[width=\textwidth]{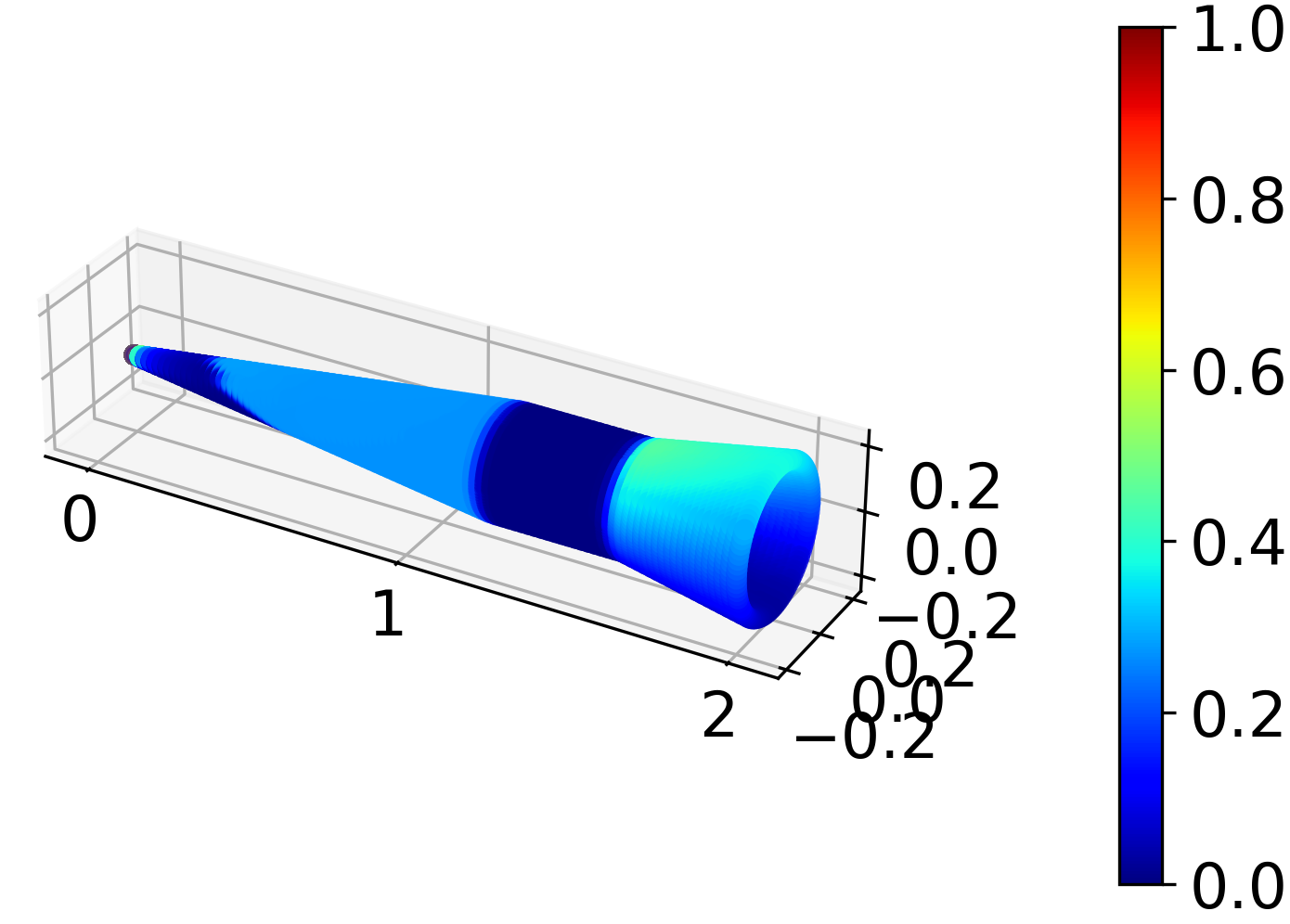}
        \caption{Training, $\bmu=[-10^\circ,\, 0.4]\trp$}
    \end{subfigure}
    \begin{subfigure}{0.33\columnwidth}
        \centering
        \includegraphics[width=\textwidth]{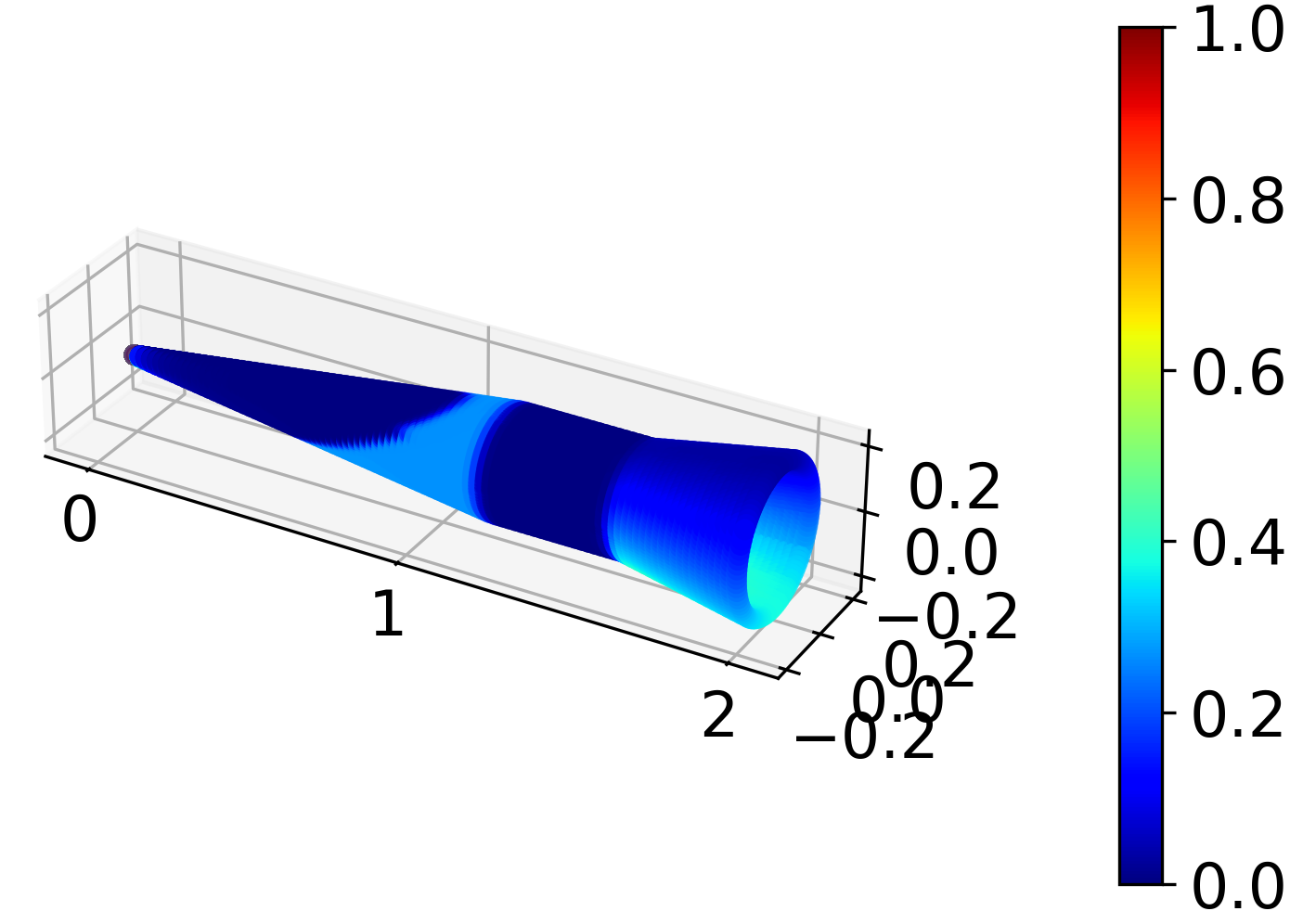}
        \caption{Training, $\bmu=[10^\circ,\, 0.7918]\trp$}
    \end{subfigure}
    \begin{subfigure}{0.33\columnwidth}
        \centering
        \includegraphics[width=\textwidth]{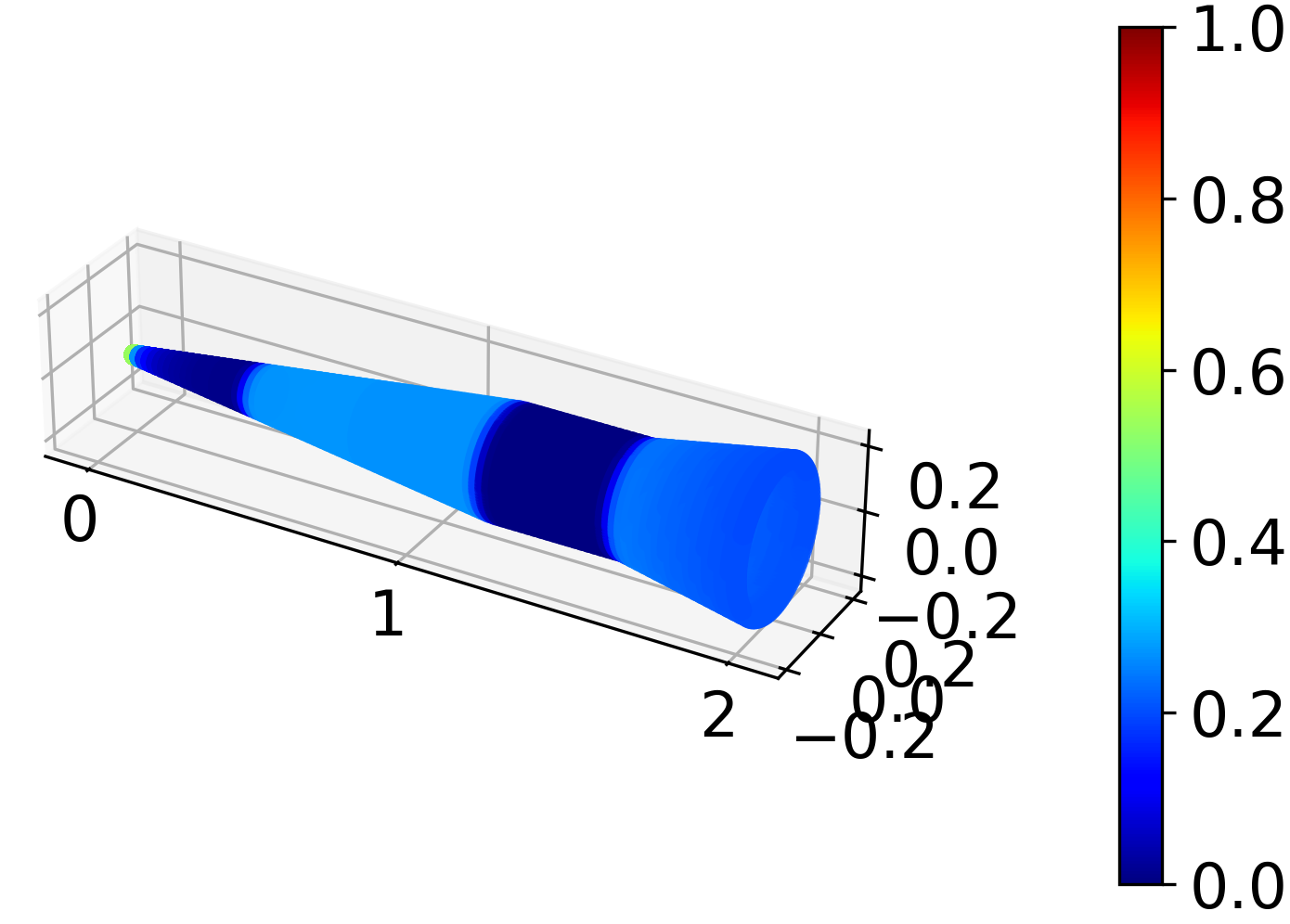}
        \caption{Testing, $\bmu=[0.2041^\circ,\, 0.4082]\trp$}
    \end{subfigure}
    \caption{Field quantity $s$ for several different parameters $\bmu$ on flared-cone geometry.}
    \label{fig:hifire_example_states}
\end{figure}

The first term in \cref{eq:hifire} constitutes a rapid quadratic decay of $s$ over the surface modulated by the angle of attack parameter.        
The second and third terms in the equation represent the onset and end of a laminar-to-turbulent transition front on the conical portion of the vehicle. 
An assumption is made that the transition is pulled forward on the windward surface of the geometry, and in this parameter field the transition front is constrained to the conical region.
The fourth term represents heating augmentation associated with shock-boundary layer at the collar compression corner, resulting in a large jump in $s$ modulated by the angle of attack with a reciprocal decay. 
It is important to note that this model is designed to qualitatively capture features of high-speed heating profiles, but is not based on first-principles physics.
We plot the field quantity $s$ on a flared-cone geometry for several different parameters $\bmu$ in \Cref{fig:hifire_example_states} to demonstrate the effect of the parameters $\bmu$ on $s$. 

\begin{figure}[H]
    \centering
    \includegraphics[width=0.45\textwidth]{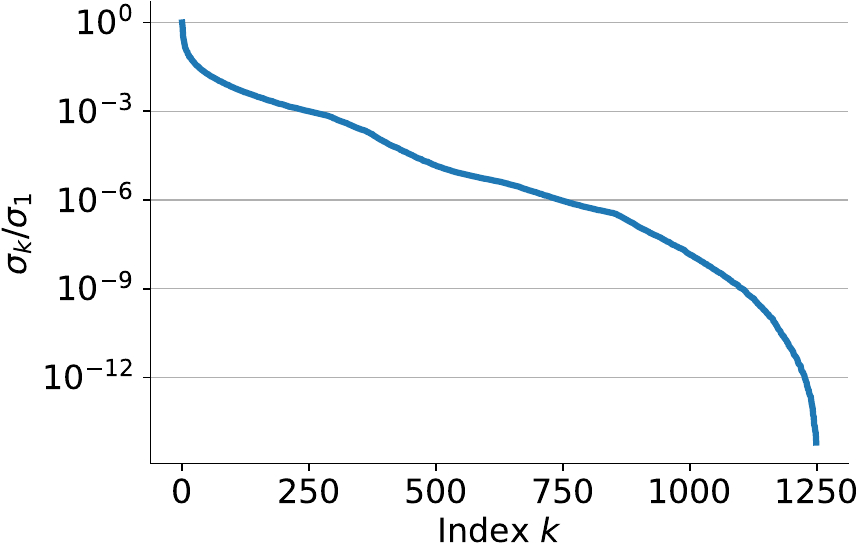}
    \caption{Normalized singular values for the surface heating example.}
    \label{fig:hifire_singvals}
\end{figure}

For a given parameter $\bmu$, we compute $s$ on a uniform grid for 100 equispaced $z$ values in $[0,2]$ and 100 equispaced $\theta$ values in $[0, 2\pi]$, resulting in snapshots $\bq(\bmu)\in \real^N$, $N=10,000$. 
To collect training and testing data for this problem, 
we compute $\bq(\bmu)$ over a uniform grid of 50 equispaced $\mu_1$ values in $[-10^\circ,10^\circ]$ and 50 equispaced $\mu_2$ values in $[0.4, 0.8]$, resulting in $M=2500$ total snapshots, ordered as

$$
\bQ^0 :=
\begin{bmatrix}
    \bq\left(\begin{bmatrix} \mu_1^{(1)}\\ \mu_2^{(1)}\end{bmatrix}\right) & \dots &
    \bq\left(\begin{bmatrix} \mu_1^{(1)}\\ \mu_2^{(50)}\end{bmatrix}\right) & 
    \bq\left(\begin{bmatrix} \mu_1^{(2)}\\ \mu_2^{(1)}\end{bmatrix}\right) & \dots &
    \bq\left(\begin{bmatrix} \mu_1^{(50)}\\ \mu_2^{(50)}\end{bmatrix}\right) 
\end{bmatrix} \in \real^{N\times M}.
$$
We then scale $\bQ^0$ as 
$$\bQ = \frac{1}{\max(\bQ^0_{ij})-\min(\bQ^0_{ij})}\bQ^0,$$
and separate every other column of $\bQ$ into training and testing sets, $\bQ_{\rm train}$ and $\bQ_{\rm test}$, respectively.
The offset $\obq$ is taken to be the mean of the training data $\bQ_{\rm train}$, and
the error metric used is the mean relative $\ell_2$ projection error
\begin{equation}\label{eq:mean_relative_projection_error}
    E=\frac{1}{M_{\rm test}}\sum_{j=1}^{M_{\rm test}}\frac{\norm{\bq_j^{\rm test}-\bg(\bh(\bq_j^{\rm test}))}_2}{\norm{\bq_j^{\rm test}}_2},
\end{equation}
where $M_{\rm test}=1250$. 
\Cref{fig:hifire_singvals} plots the decay of the normalized singular values of the training data.
Notice that the singular values for this example decay more slowly than for the boundary layer example, indicating that the current example is a more demanding test case for our KM approach. 

For this example, the Kernel RBF approach uses the Gaussian kernel with shape parameter $\epsilon = 10^{-1}.$
The Kernel QM and Kernel RBF approaches both use the input normalization discussed in \Cref{rmk:input_normalization}, and
the Kernel QM, Kernel RBF, Alternating QM, and Greedy QM approaches use the regularization values in \Cref{tbl:hifire_regs}. 
For further details on the selection of the RBF kernel, shape parameter, and regularization values for this example, see \Cref{app:hifire}. 
\begin{table}[H]
    \centering
    \begin{tabular}{c|c}
        Manifold type & regularization $\lambda$ \\ \hline
        Kernel QM & $10^{-9}$ \\ 
        Kernel RBF & $10^{-9}$ \\
        Alternating QM & $10^{-5}$ \\
        Greedy QM & $10^{-4}$ 
    \end{tabular}
    \caption{Fixed regularization values for different dimensionality reduction approaches for the surface heating example.}
    \label{tbl:hifire_regs}
\end{table}

\begin{figure}[H]
    \centering
    \includegraphics[width=0.6\textwidth]{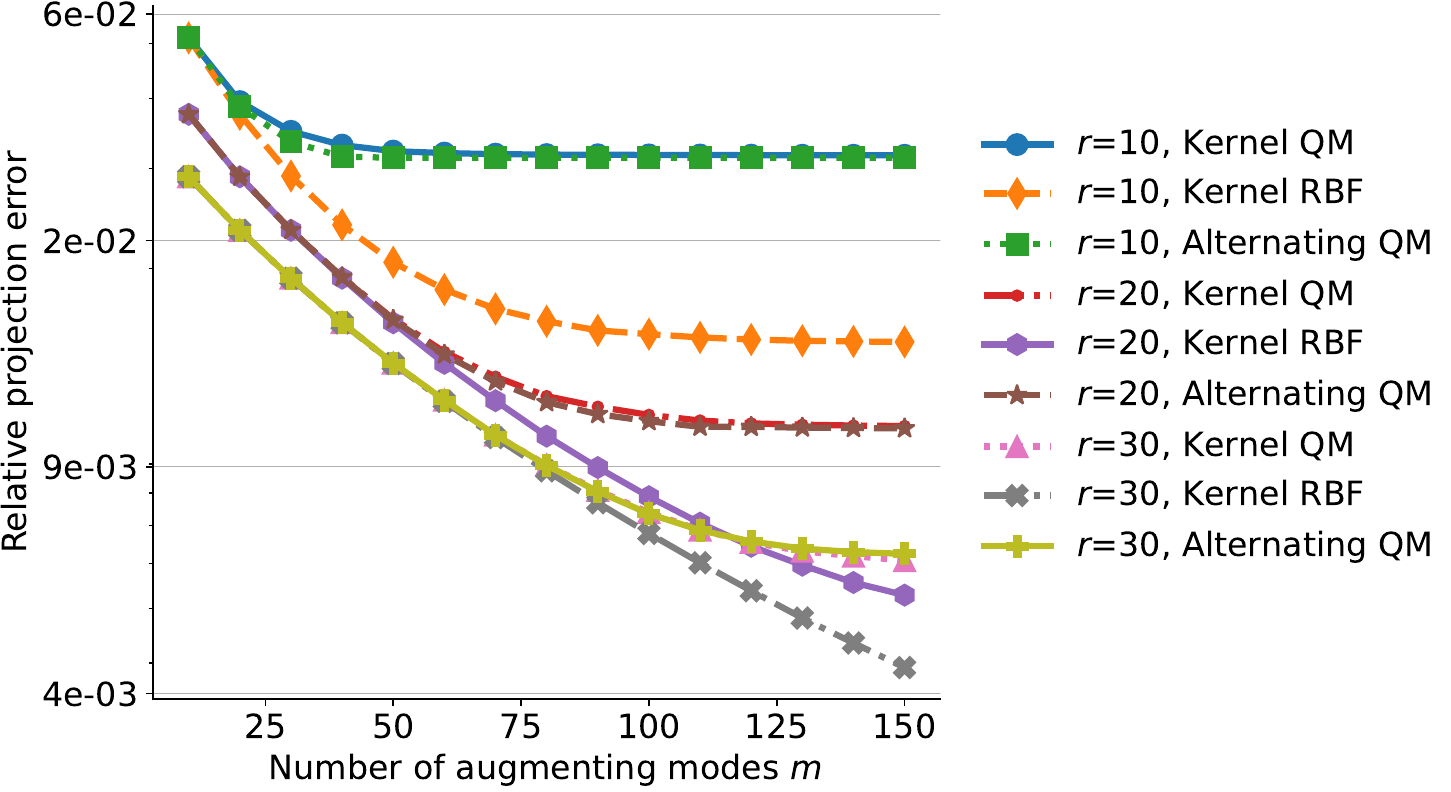}
    \caption{Relative projection error as a function of number of augmenting modes $m$ for several different latent dimension sizes $r$ for the surface heating example.}
    \label{fig:hifire_error_vs_augmenting_modes}
\end{figure}

First, we vary the number of augmenting modes $m \in [10, 150]$ for the Kernel QM, Kernel RBF, and Alternating QM approaches for $r\in \set{10, 20, 30}$ to examine the effect of the number of augmenting modes on the mean relative projection error.
As in the advection-diffusion-reaction case, 
the Greedy QM approach is excluded from this test because it does not include an augmenting basis.
Like the previous example,
\Cref{fig:hifire_error_vs_augmenting_modes} shows that for a fixed $r$, the error plateaus as $m$ exceeds a small integer multiple of $r$ for each method.
Interestingly, the errors begin to plateau for larger values of $m$ for this problem than for the advection-diffusion-reaction problem. 
For $r=10$, the Alternating QM and Kernel QM errors plateau for $m\geq40$ augmenting modes, while for $r=20$, the errors plateau for $m\geq 100$ modes, and for $r=30$, they plateau for $m\geq 130$ modes. 
On the other hand, the Kernel RBF approach plateaus for $m\geq100$ for $r=10$, but does not plateau in the tested range of $m$ for $r=20$ and $r=30$.
Thus, in contrast with the advection-diffusion-reaction problem, the additional accuracy benefit of using an RBF kernel over a QM is not only restricted to smaller values of $r$.
For the remaining comparisons on this example, we fix $m=10r$ for each of the Kernel RBF, Kernel QM, and Alternating QM approaches.

\begin{figure}[H]
    \centering
    \includegraphics[width=\textwidth]{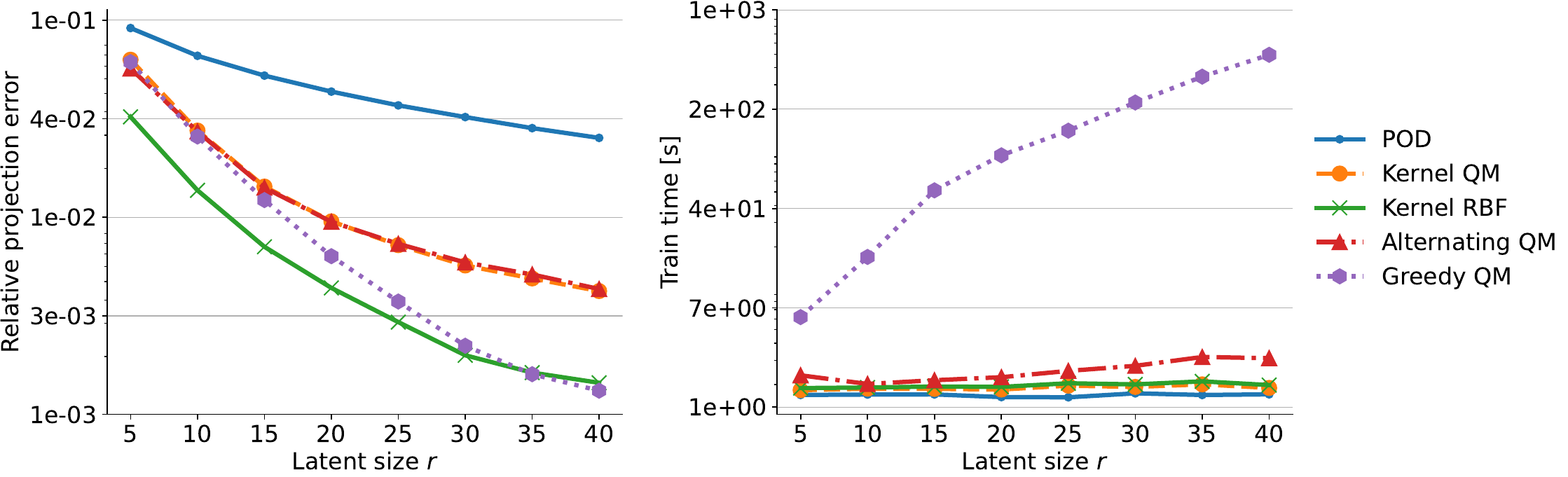}
    \caption{Relative projection error and train time as a function of the latent dimension size $r$ for different dimensionality reduction approaches for the surface heating example.}
    \label{fig:hifire_error_vs_romsize}
\end{figure}

We now compare the different approaches by computing the error as a function of the latent space dimension $r$.
From \Cref{fig:hifire_error_vs_romsize}, we see that each nonlinear augmentation approach outperforms POD.
The Kernel QM and Alternating QM have nearly identical errors for each $r$. 
The Kernel RBF approach attains the lowest errors for $r\leq 30$, 
while Greedy QM has roughly the same error as Kernel RBF for $r=35$ and slightly lower error for $r=40$.
Greedy QM yields similar errors to the other QM approaches for $r\leq 15$, but decreases more rapidly for $r>15$. 
The KM approaches have smaller training times than both the Alternating QM and Greedy QM approaches,
and the training time for the Greedy QM approach increases substantially with increasing $r$.
Since the Kernel RBF approach has the smallest errors for $r\leq 30$ and has nearly the same error as Greedy QM for $r=35, 40$ with substantially lower training times, we can conclude that Kernel RBF has the best performance for this example.
Furthermore, as noted above, unlike the advection-diffusion-reaction example, the Kernel RBF has accuracy benefits over QMs for each value of $r$ tested rather than for only small values of $r$. 

\begin{figure}[H]
    \begin{subfigure}{0.49\columnwidth}
        \centering
        \includegraphics[width=0.8\textwidth]{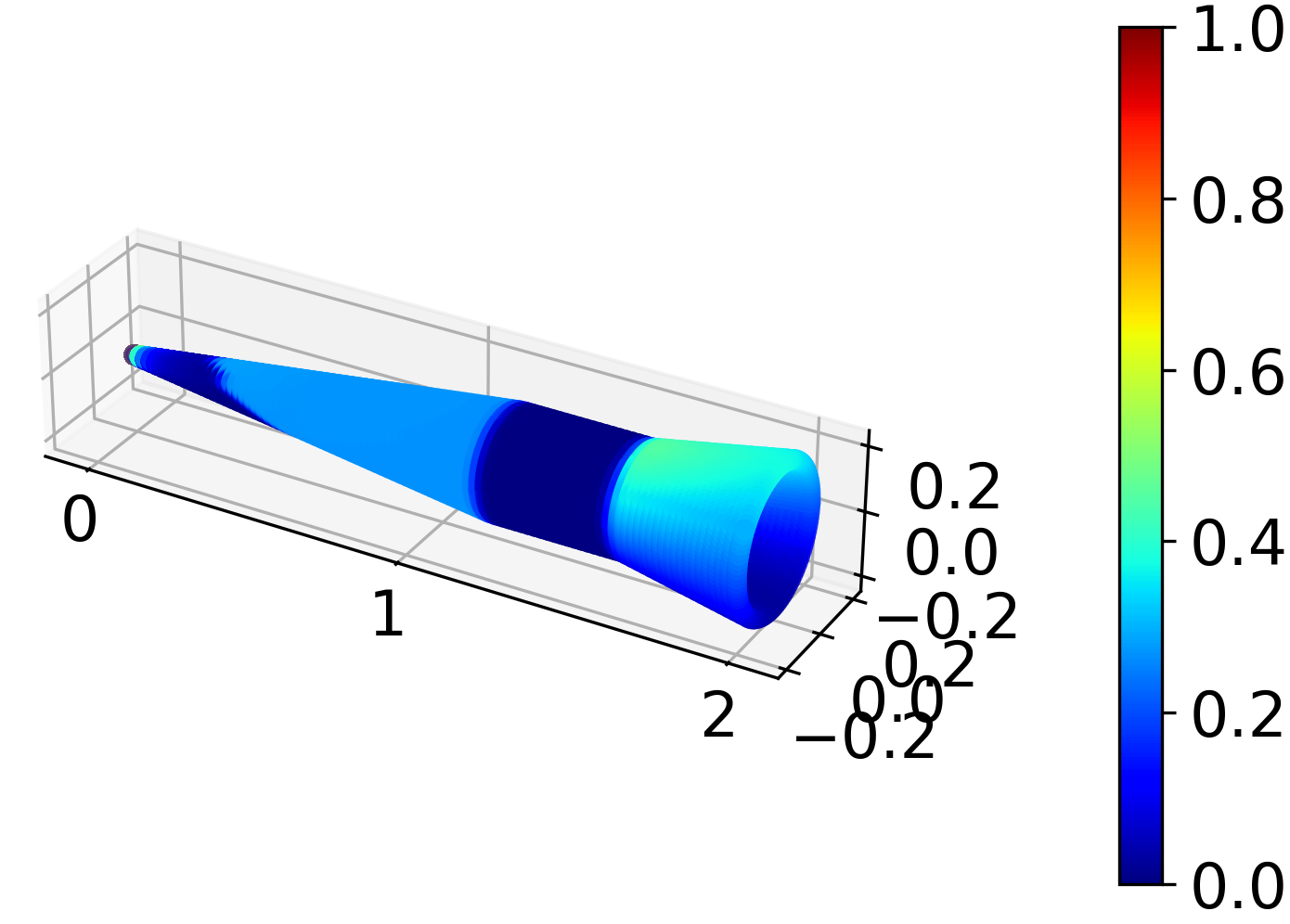}
        \caption{Full state.}
    \end{subfigure}
    \begin{subfigure}{0.49\columnwidth}
        \centering
        \includegraphics[width=0.8\textwidth]{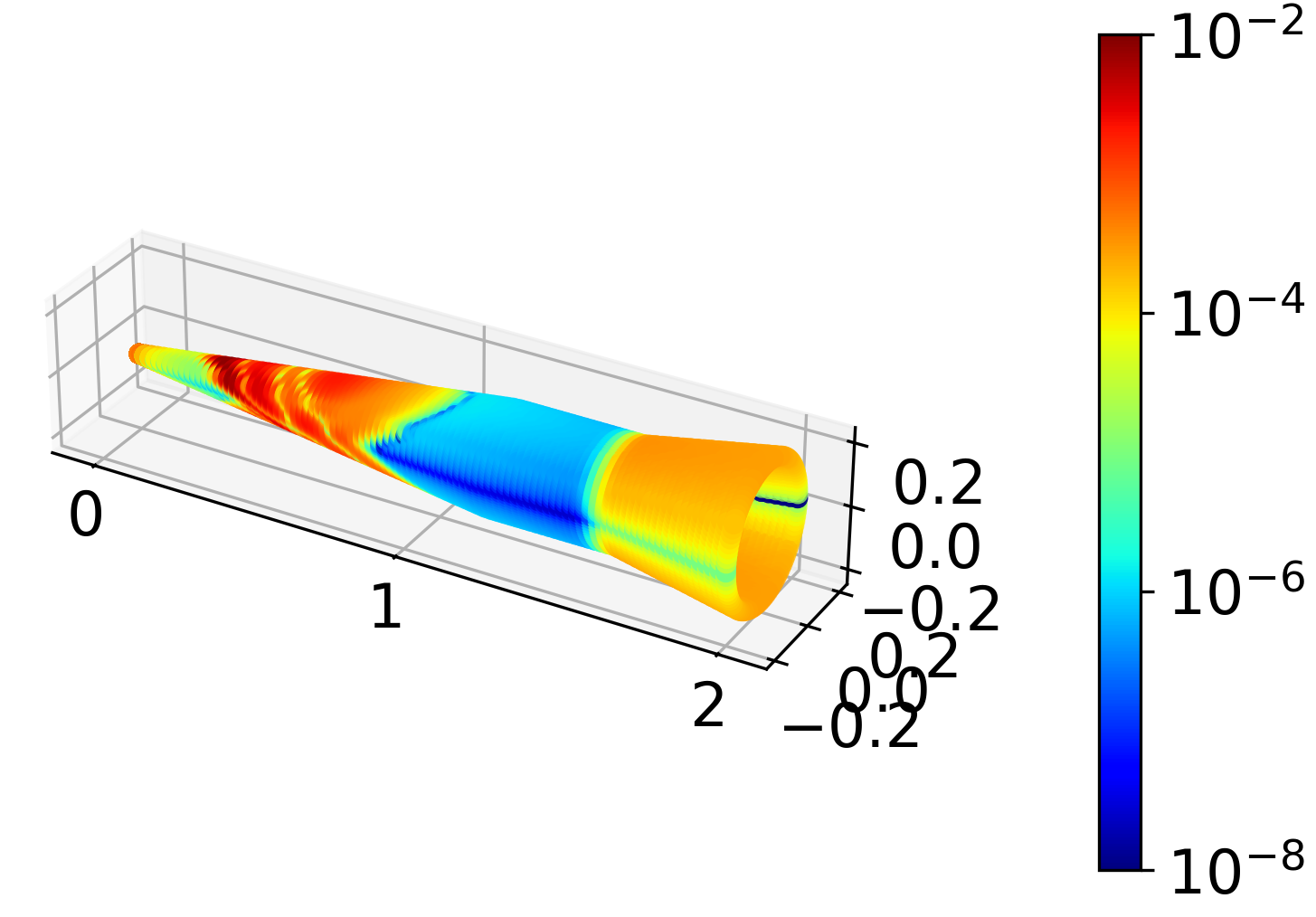}
        \caption{POD projection error.}
    \end{subfigure}
    \begin{subfigure}{0.49\columnwidth}
        \centering
        \includegraphics[width=0.8\textwidth]{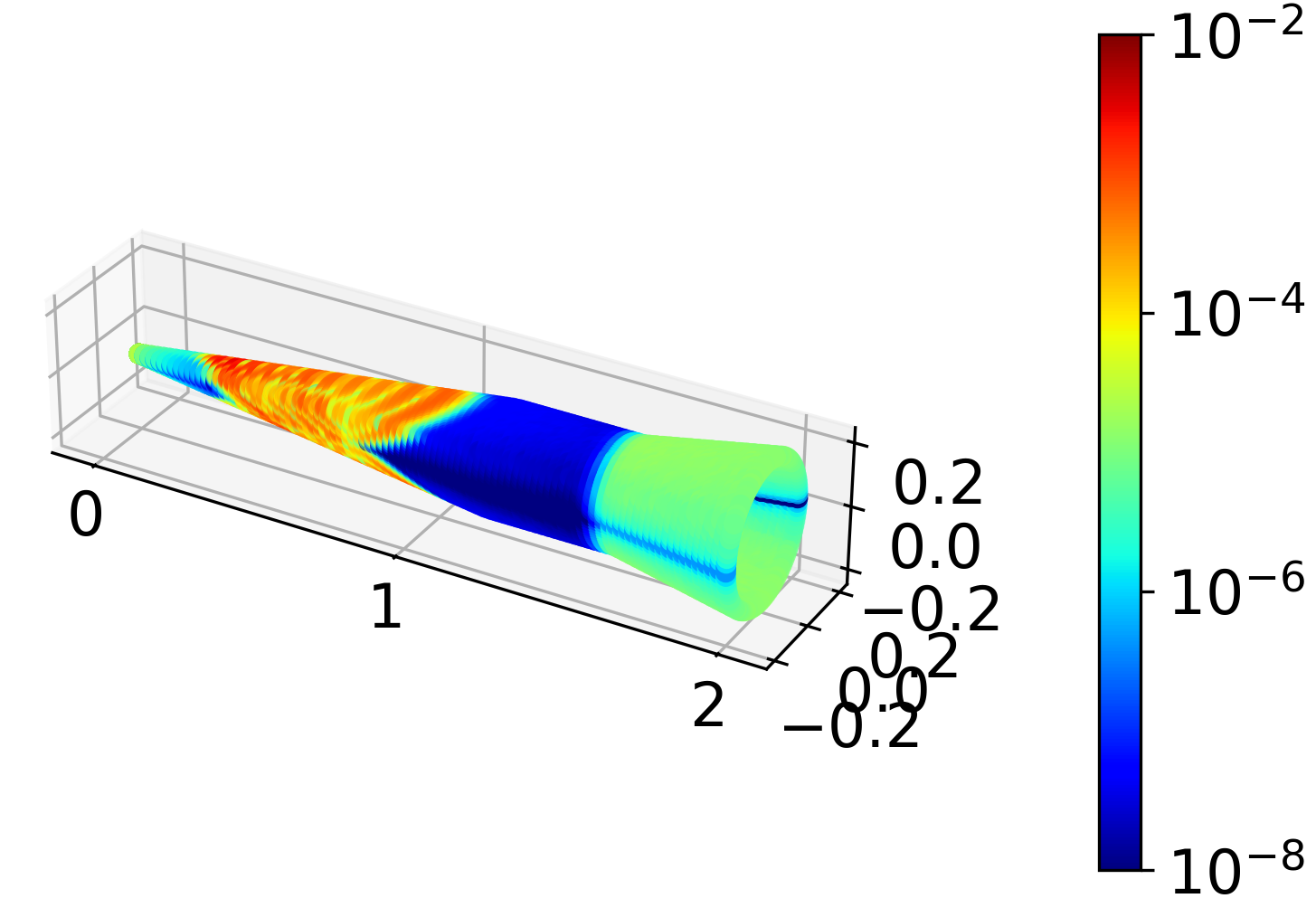}
        \caption{Alternating QM projection error.}
    \end{subfigure}
    \begin{subfigure}{0.49\columnwidth}
        \centering
        \includegraphics[width=0.8\textwidth]{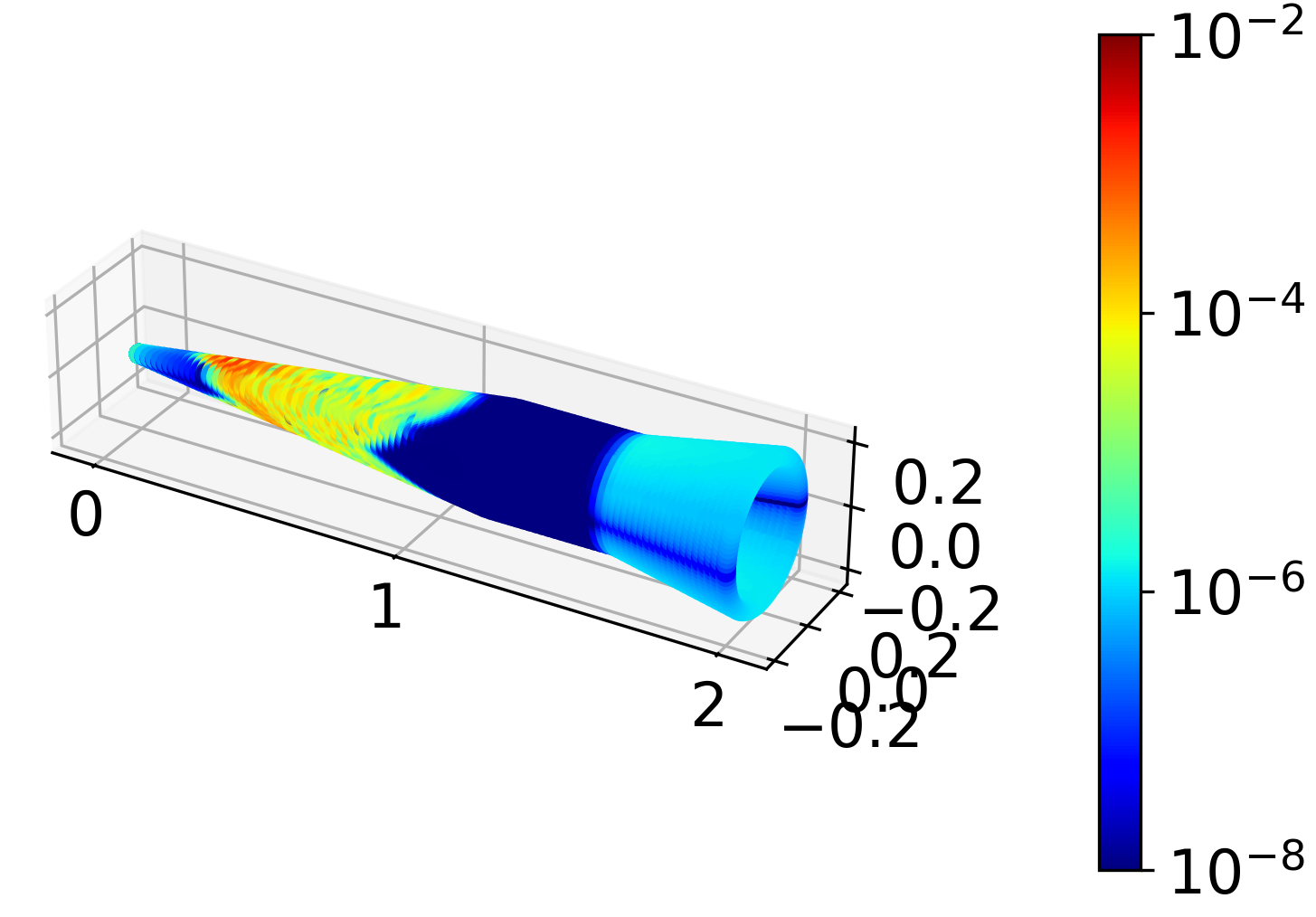}
        \caption{Kernel RBF projection error.}
    \end{subfigure}
    \caption{Plot of full state and relative projection errors for $r=10$, $m=100$ for the surface heating example.}
    \label{fig:hifire_states}
\end{figure}

We plot the full state $\bq(\bmu)$ for $\bmu=[-10^\circ,\, 0.4082]\trp$ and the relative projection errors $|\bq(\bmu) -\bg(\bh(\bq(\bmu)))|/\norm{\bq(\bmu)}_2$ for POD, Alternating QM, and Kernel RBF in \Cref{fig:hifire_states} for $r=10$ and $m=100$. 
Notice that the POD projection errors are largest towards the tip of the conical geometry. These errors are decreased in the Alternating QM projection error, and decreased further in the Kernel RBF projection error.

%% file: numerics_bracket.tex
%!TEX root = main.tex
\subsection{Solid mechanics -- 3D flexible bracket}\label{sec:numerics_bracket}
The third example is a moderately sized 3D hyper-elastic solid dynamics problem, referred to herein as the flexible bracket.  
The governing PDEs are the following equations of motion, written in terms of displacement vector $\mathbf{d} \in \mathbb{R}^3$,  material density $\rho>0$, and the first Piola-Kirchhoff stress $\mathbf{P}\in\mathbb{R}^{3\times 3}$:
\begin{equation} \label{eq:3delasticity}
    \rho \ddot{\mathbf{d}} = \nabla \cdot  \mathbf{P}, \hspace{0.5cm} \text{on } \Omega \subset \mathbb{R}^3.
\end{equation}
In \eqref{eq:3delasticity}, $\ddot{\mathbf{d}}:=\frac{\partial^2 \mathbf{d}}{\partial t^2}$ represents the acceleration. 
Encoded into $\mathbf{P}$ is a constitutive model for the underlying material.  Herein, we specify a Neohookean-type material model extended to the compressible regime, with Young's modulus $E = 200\times 10^9$ GPa, Poisson’s ratio $\nu= 0.25$ and density $\rho= 7800$ kg/m$^3$.  These properties correspond to a real material, namely steel.  For more details about the Neohookean-type constitutive model used to generate the results herein, the reader is referred to \cite{MOta:2011}.  We emphasize that the governing PDE \eqref{eq:3delasticity} is nonlinear and contains generic (non-polynomial) nonlinearities, as the Helmholtz free-energy density corresponding to the Neohookean constitutive model is nonlinear.

\begin{figure}
\begin{subfigure}[t]{.5\textwidth}\centering
  \includegraphics[width=\textwidth]{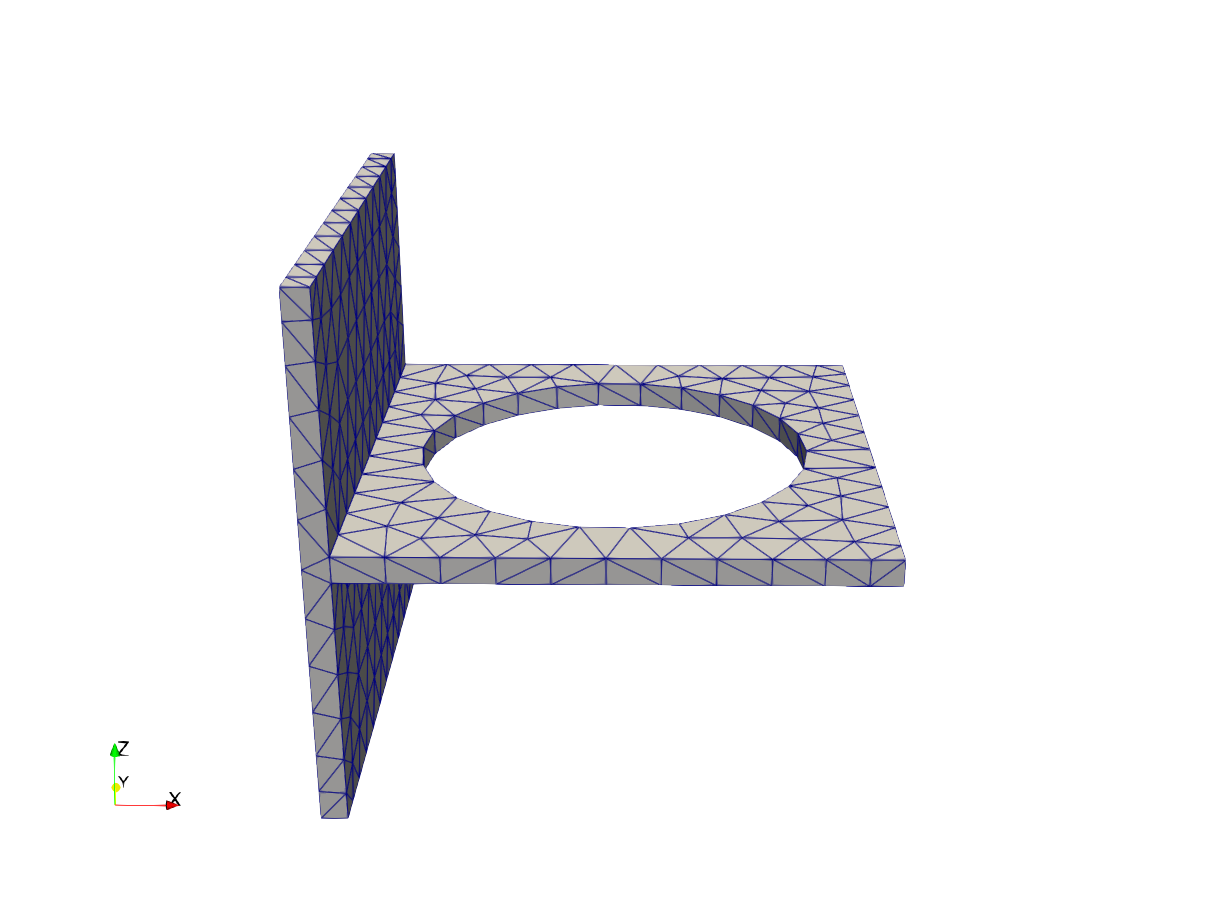}
  \caption{Flexible bracket geometry and mesh}
\end{subfigure}
\begin{subfigure}[t]{.5\textwidth}\centering
\raisebox{8ex}{\includegraphics[width=0.65\textwidth]{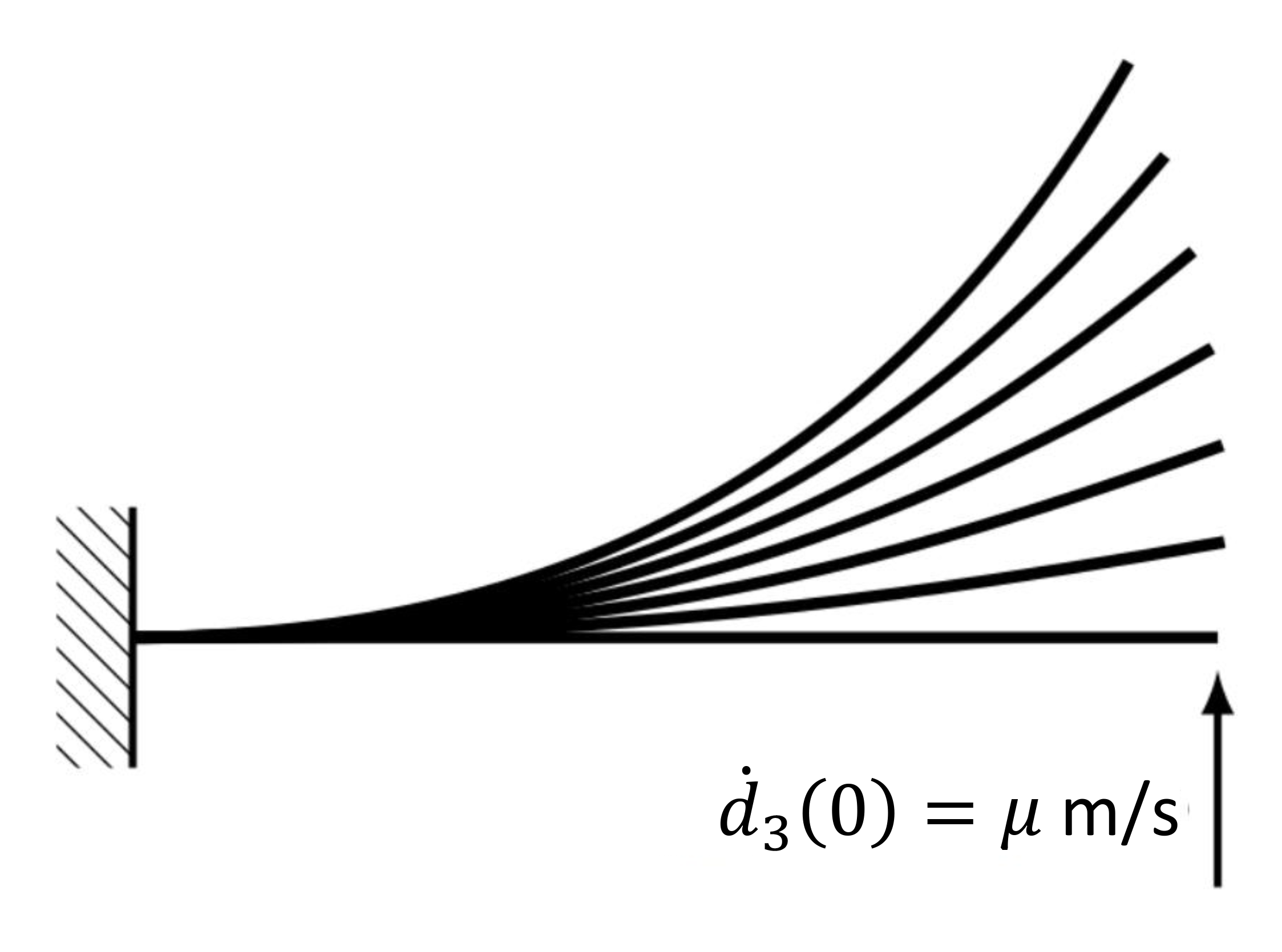}}
  \caption{1D cartoon of $z$--velocity initial condition}
\end{subfigure}
\caption{Geometry and mesh for the 3D flexible bracket example, along with a one-dimensional (1D) cartoon illustrating the problem setup.  } 
\label{fig:bracket}
\end{figure}

The flexible bracket geometry, mesh and initial conditions for our test case are shown in \Cref{fig:bracket}.  In the $xy-$plane is a square plate that is  0.205 m wide and features a circular whole with diameter 0.075 m.  This plate is connected to a vertical plate of the same dimensions in the $yz-$plane.  On the left side, at $x = -0.105$ m, the bracket is clamped, meaning we set a boundary condition of $\mathbf{d} = \mathbf{0}$.  The problem is initialized by setting a constant initial condition $\dot{\mathbf{d}}(0; \mu) := \frac{\partial \mathbf{d}}{\partial t}(0; \mu) = \left( \begin{array}{ccc} 0, & 0, & \mu \end{array}\right)^\intercal \in \mathbb{R}^3 $ m/s for a specified value of the $z$--velocity $\mu\in \mathbb{R}$ on the right-most side of the bracket, corresponding to $x = 0.1$ m (see \Cref{fig:bracket}(b)).  
As a result of the velocity initial condition, the flexible bracket begins to vibrate and flap.  The presence of the circular hole weakens the bracket as well as complicates the propagation and interaction of 3D waves moving throughout the bracket, leading to truly nonlinear 3D dynamics.  A linear version of this problem was considered in \cite{Gruber:2025}.

\begin{figure}
\begin{subfigure}[t]{.33\textwidth}\centering
  \includegraphics[width=\textwidth]{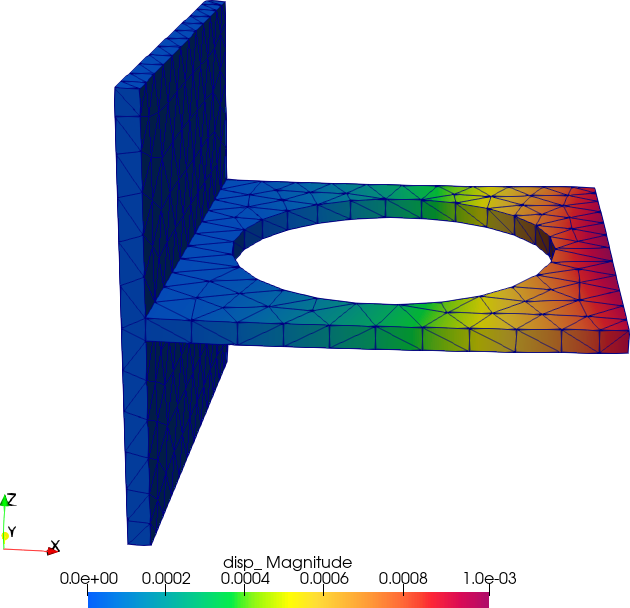}
  \caption{$\mu = 10$ m/s.}
\end{subfigure}
\begin{subfigure}[t]{.33\textwidth}\centering
\includegraphics[width=\textwidth]{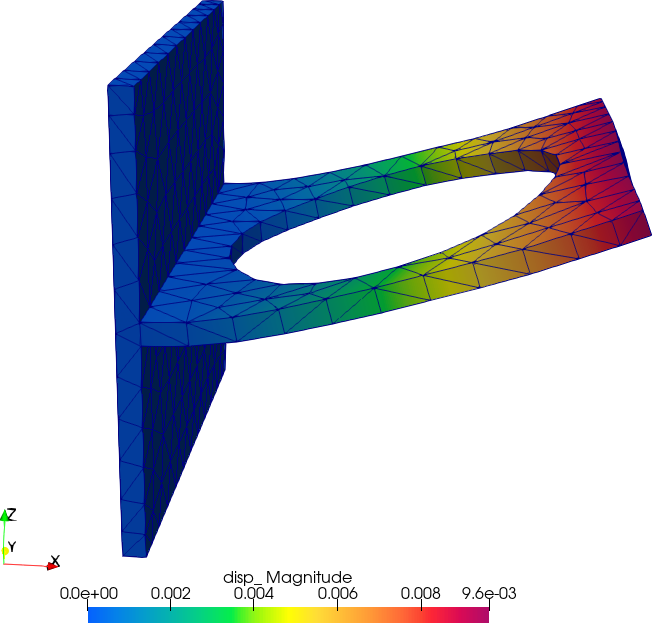}
  \caption{$\mu = 100$ m/s}
\end{subfigure}
\begin{subfigure}[t]{0.33\textwidth}\centering
\includegraphics[width=\textwidth]{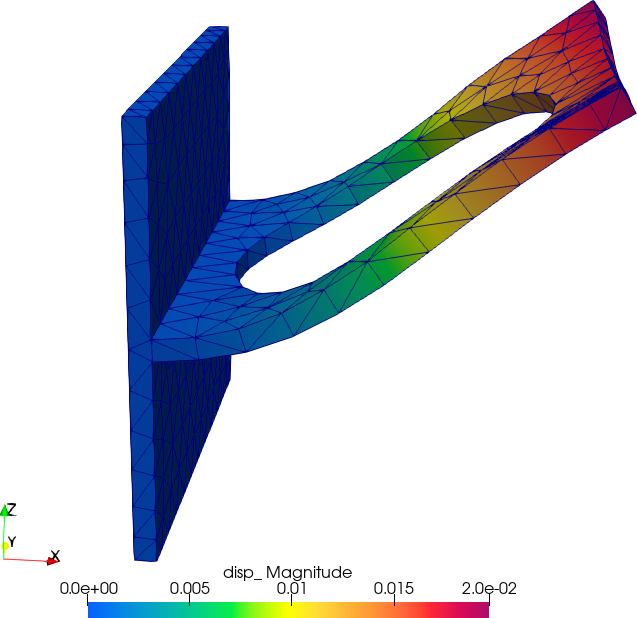}
  \caption{$\mu = 200$ m/s}
\end{subfigure}
\caption{Displacement magnitude solutions to the flexible bracket problem, scaled by a factor of
five for visualization purposes, at the final time t = $20.0 \times 10^{-3}$ seconds for different initial $z$--velocities $\mu$. }
\label{fig:bracket_ics}
\end{figure}

\begin{figure}[H]
    \centering
    \includegraphics[width=0.45\textwidth]{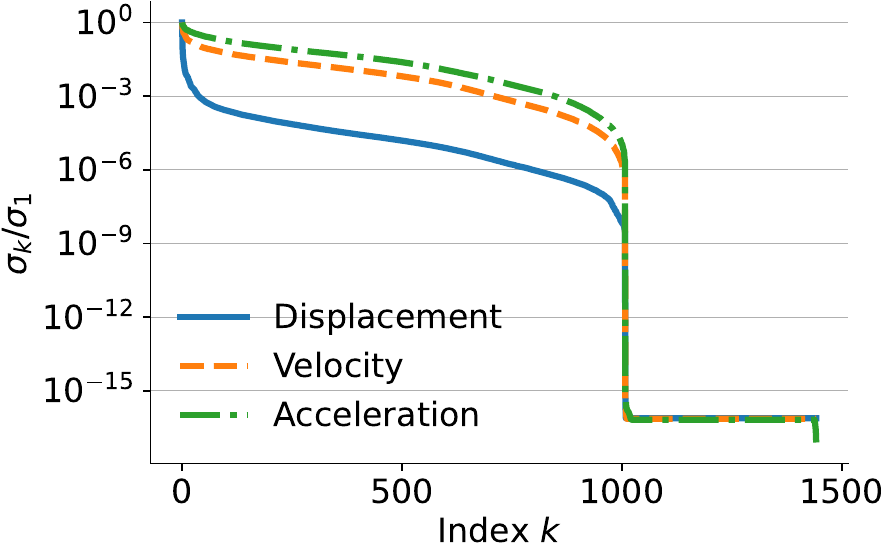}
    \caption{Singular values for bracket displacement, velocity, and acceleration fields.}
    \label{fig:bracket_all_singvals}
\end{figure}

To generate training data for our KM methods, we first discretize and solve \cref{eq:3delasticity} using a {\tt Julia} finite element code known as {\tt Norma.jl} \cite{norma}.  The flexible bracket geometry is discretized using a unstructured four-node tetrahedral mesh consisting of 1174 elements having a total of 481 nodes, as shown in \Cref{fig:bracket}(a).  The resulting problem has 1443 degrees of freedom.  The semi-discrete version of \cref{eq:3delasticity} is then advanced forward in time from time 0 to $20.0 \times 10^{-3}$ seconds using an implicit Newmark time-integration scheme with parameters $\beta = 0.25$ and $\gamma = 0.5$.  For the cases with initial condition $\mu \leq 150$ m/s, a time-step of $\Delta t = 1.0\times 10^{-5}$ seconds is utilized.  For the harder $\mu = 200$ m/s initial condition case, we employ a smaller time-step of $\Delta t = 1.0\times 10^{-6}$ seconds.  
Each of the solutions are then uniformly downsampled to retain $1001$ snapshots per sample.
To create the training data for our methods, we simulate the flexible bracket for four different initial conditions on the $z$--velocity $\mu$: 10 m/s, 50 m/s, 150 m/s and 200 m/s.  
We then test the different dimensionality reduction approaches discussed herein on the displacement, velocity, and acceleration fields for the $\mu = 100$ m/s case, which is outside our training set.  
\Cref{fig:bracket_ics} shows the norm of the displacement vector, $|| \mathbf{d}||_2$, amplified by a factor of five, at the final simulation time for various initial conditions $\mu$.  The reader can observe that the set of $\mu$ considered in our study gives rise to nontrivial differences in the solution, making this an excellent test case for evaluating the predictive power of the various KM approaches.  
\Cref{fig:bracket_all_singvals} shows the singular value decay for the first $1500$ normalized singular values for the flexible bracket test case.  
The fact that there is an incredibly slow singular value decay up to index 1000 makes the flexible bracket problem a good candidate for the KM approaches described and proposed herein.

\begin{figure}[H]
    \centering
    \includegraphics[width=\textwidth]{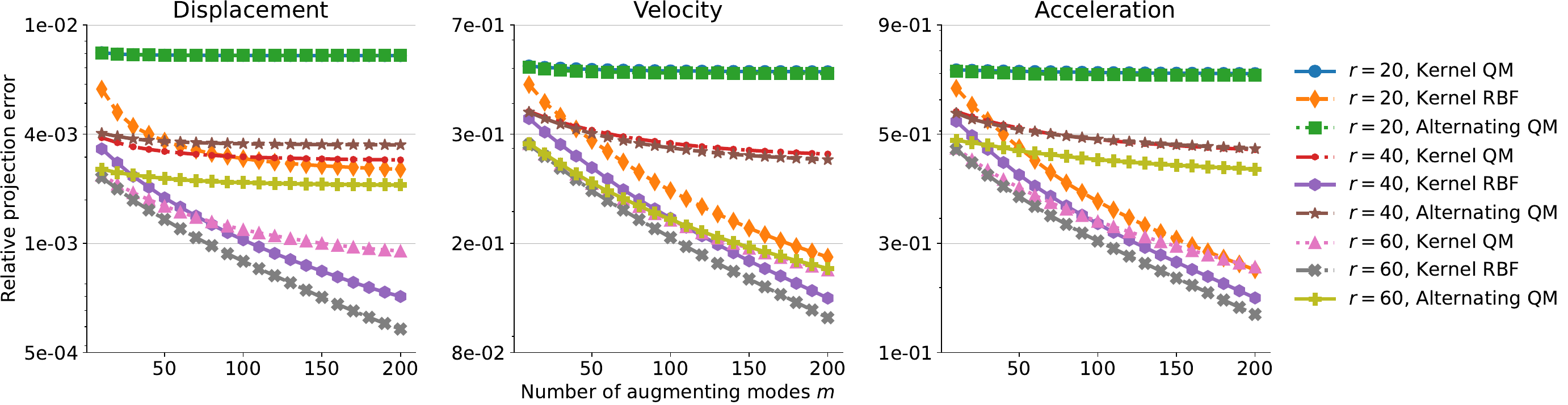}
    \caption{Relative projection error as a function of number of augmenting modes $m$ for several different latent dimension sizes $r$ for the 3D bracket example.}
    \label{fig:bracket_error_vs_augmenting_modes}
\end{figure}

\begin{figure}[H]
    \centering
    \includegraphics[width=\textwidth]{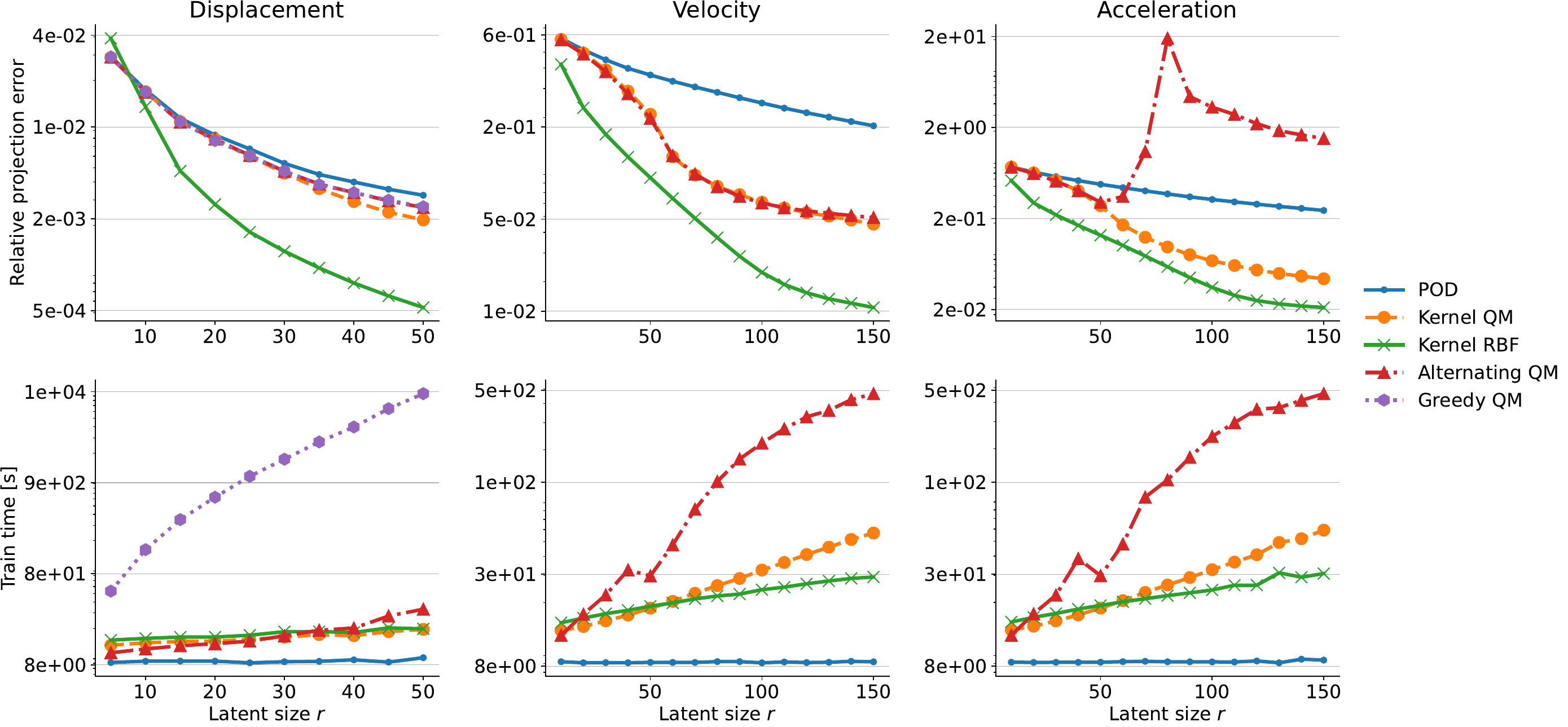}
    \caption{Relative projection error and train time as a function of the latent dimension size $r$ for different dimensionality reduction approaches for the 3D bracket example.
    }
    \label{fig:bracket_error_vs_romsize}
\end{figure}

Results comparing the performance of the various dimension reduction approaches are summarized in \Cref{fig:bracket_error_vs_augmenting_modes}--\Cref{fig:bracket_solns_and_figs}.  
We employ input normalization as described in \Cref{rmk:input_normalization} for each KM approach tested.
The error metric that we use for this example is the relative $\ell^2$ projection error \cref{eq:relative_l2_projection_error}, 
where $\bq_j^*$ is test data for the displacement solution corresponding to a given value of the parameter $\mu$ (the $z$--velocity initial condition)
at time step $j$.

For this example, we compute separate nonlinear-augmentation manifolds for the displacement, velocity, and acceleration fields. 
For each field, the Kernel RBF approach uses the linear Mat\'{e}rn kernel with shape parameter $\epsilon = 10^{-1}.$
Each Kernel QM and Kernel RBF computed uses the input normalization as discussed in \Cref{rmk:input_normalization}.
\Cref{tbl:hifire_regs} lists the regularization values used for Kernel QM, Kernel RBF, Alternating QM, and Greedy QM for the displacement, velocity, and acceleration fields. 
Note that the regularization values needed for the velocity and acceleration Alternating QMs are quite large. 
This is likely due to the complexity of the underlying structures in the data, which in turn increases the values of $r$ and $m$ necessary to obtain low errors, thus resulting in increasingly poor conditioning of the system \cref{eq:feature_map_coef_minimizer_equation} as $r$ and $m$ increase. 
This poor conditioning is mitigated in the KM approaches because the size of \cref{eq:coefficient_equation} is independent of $r$ and because normalizing the inputs also improves the conditioning of \cref{eq:coefficient_equation}.
For further details on the selection of the RBF kernel, shape parameter, and regularization values for this example, see \Cref{app:bracket}. 

\begin{table}[H]
    \centering
    \begin{tabular}{c|c|c|c}
        Manifold type & Displacement reg.\ & Velocity reg. & Acceleration reg. \\ \hline
        Kernel QM & $10^{-7}$ & $10^{-7}$ & $10^{-7}$  \\ 
        Kernel RBF & $10^{-7}$ & $10^{-7}$ & $10^{-7}$  \\
        Alternating QM & $10^{-11}$ & $10^{6}$ & $10^{8}$ \\ 
        Greedy QM & $10^{-11}$ & NA & NA
    \end{tabular}
	\caption{Fixed regularization $\lambda$ values for different dimensionality reduction approaches in 3D solid mechanics flexible bracket example for the displacement, velocity, and acceleration fields.}
    \label{tbl:bracket_regs}
\end{table}

We examine the effect of the number of augmenting modes $m$ on the displacement, velocity, and acceleration projection errors for the Kernel QM, Kernel RBF, and Alternating QM approaches for $r=20, 40, 60$. 
Unlike the previous examples, \Cref{fig:bracket_error_vs_augmenting_modes} shows that Kernel QM and Alternating QM show little improvement in error as $m$ increases for $r=20, 40$ across each of the displacement, velocity, and acceleration fields. 
For $r=60$, the Kernel QM error improves compared to Alternating QM and decreases as $m$ increases, whereas $r=60$ Alternating QM only decreases noticeably as $m$ increases for the velocity field.
In contrast, the Kernel RBF substantially improves in error as $m$ increases for each of the fields and for each value of $r$ tested. 
Only for the $r=10$ displacement case does Kernel RBF plateau; in the other cases, the Kernel RBF error decreases monotonically.
These results indicate that the quadratic nonlinear augmentations of the QM approaches poorly model the corrections needed to improve the accuracy for this problem, while RBF-based augmentations are a better model for the nonlinear correction terms.

Since Kernel RBF has monotonically decreasing error as $m$ increases for this example, we set $m=5r$ for each of the Kernel QM, Kernel RBF, and Alternating QM methods in the following comparisons. 
Lastly, we note that, although the trends among the different nonlinear augmentation methods are similar for displacement, velocity, and acceleration, the errors obtained for velocity and acceleration are significantly larger than for displacement.  
This result is expected for solid dynamic problems: generally, the velocity error is one order of magnitude higher than the displacement error, and the acceleration error is one order of magnitude higher than the velocity error. 

Next, in \Cref{fig:bracket_error_vs_romsize}, we compare the relative errors and train times of the different approaches as a function of latent dimension size for the displacement, velocity, and acceleration fields.
As noted above, the velocity and acceleration fields yield much larger errors than displacement, thus requiring much larger values of $r$ to yield errors similar to displacement. 
Consequently, we omit Greedy QM from the velocity and acceleration comparisons due to the excessive training times required for larger $r$ values.
\Cref{fig:bracket_error_vs_romsize} shows that for the displacement field, each of the QMs yield nearly identical errors for $r\leq 30$, while Kernel QM yields slightly smaller errors than Alternating QM and Greedy QM for $r\geq 35$.
Each of the QMs only yield slight error improvements over POD.
Kernel RBF yields slighly larger error than POD at $r=5$, but rapidly decreases in error as $r$ increases, with nearly an order of magnitude improvement in error over the other approaches. 
For the velocity field, Kernel QM and Alternating QM have nearly identical errors, and yield nearly an order of magnitude improvement over POD for $r>50$. 
Kernel QM performs similarly for the acceleration field, while Alternating QM sharply increases in error for $r>50$.
We note that since we use the fixed regularization parameters in \Cref{tbl:bracket_regs} for each value of $r$, the Alternating QM error for $r>50$ may be improved by fine-tuning the regularization for different values of $r$. 
Kernel RBF still yields the smallest errors among the different methods for the velocity and acceleration fields. 
Lastly, we note that Greedy QM still requires the largest train times for the displacement field, while Alternating QM requires the largest train times for the velocity and acceleration fields among the compared methods. 

Finally, in \Cref{fig:bracket_solns_and_figs}, we plot the full solution states and the relative projection errors for the displacement, velocity, 
and acceleration fields at the final time $t=20.0\times 10^{-3}$ seconds.  Projection errors are shown for the POD, Alternating QM and Kernel RBF KMs.  
We utilize a value of $r=25$ and $m=125$ when predicting the displacement, a value of $r=50$ and $r=250$ when predicting the velocity and a value of $r=100$ and $m=500$ when predicting the acceleration. 
The reader can observe that the results in \Cref{fig:bracket_solns_and_figs} are consistent with the previously-presented results in \Cref{fig:bracket_error_vs_augmenting_modes} and \Cref{fig:bracket_error_vs_romsize}.  
The Kernel RBF KM is the most accurate by at least an order of magnitude.  
Generally, POD is the least accurate, with the exception of the acceleration field.  While it may initially seem surprising that the Alternating QM is the least accurate when it comes to the acceleration field, this result is consistent with the spike in the acceleration projection error that occurs around $r=100$, as shown in \Cref{fig:bracket_error_vs_romsize}.  The noisiness in the velocity and acceleration solutions in \Cref{fig:bracket_solns_and_figs} is expected, and can be attributed to the complex nonlinear interaction of propagating waves within the bracket geometry.  

\newcommand{\brwidth}{0.92}
\begin{figure}[H]
\begin{subfigure}[t]{.33\textwidth}\centering
  \includegraphics[width=\brwidth\textwidth]{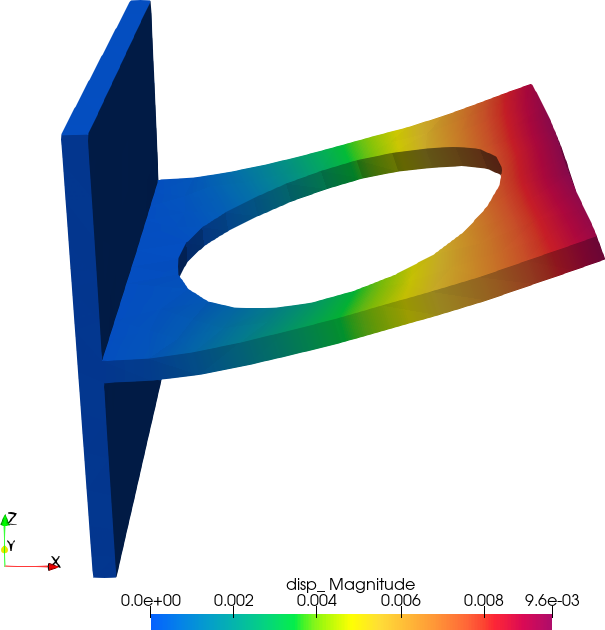}
\end{subfigure}%
\begin{subfigure}[t]{.33\textwidth}\centering
\includegraphics[width=\brwidth\textwidth]{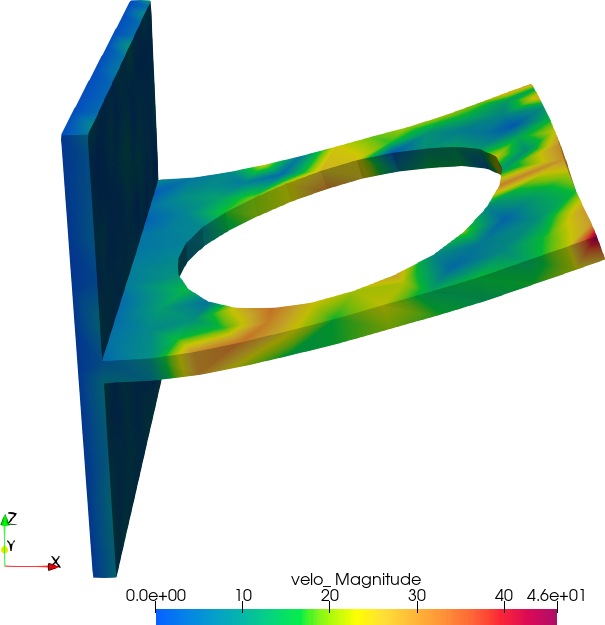}
\end{subfigure}
\begin{subfigure}[t]{0.33\textwidth}\centering
\includegraphics[width=\brwidth\textwidth]{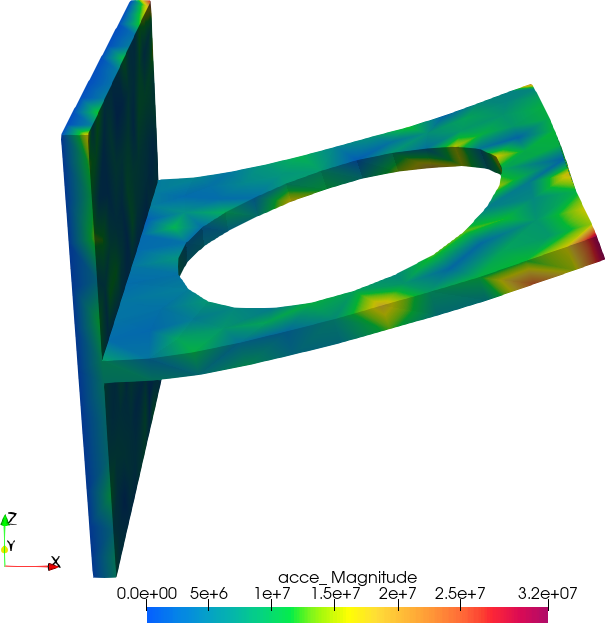}
\end{subfigure}
\begin{subfigure}[t]{.33\textwidth}\centering
  \includegraphics[width=\brwidth\textwidth]{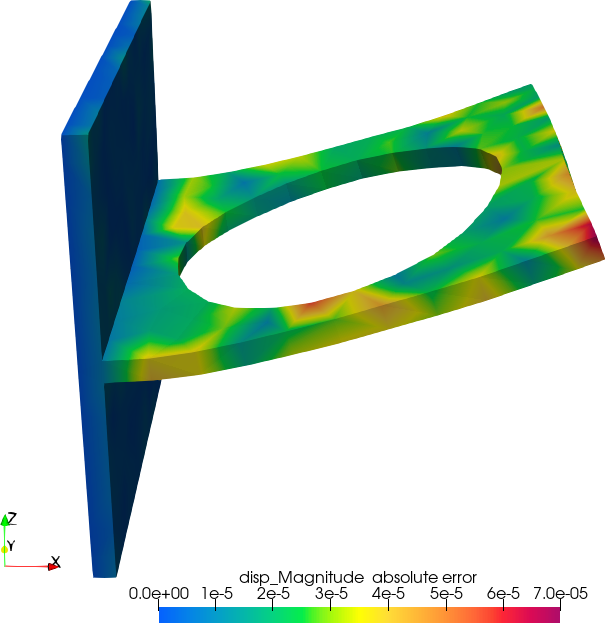}
\end{subfigure}%
\begin{subfigure}[t]{.33\textwidth}\centering
  \includegraphics[width=\brwidth\textwidth]{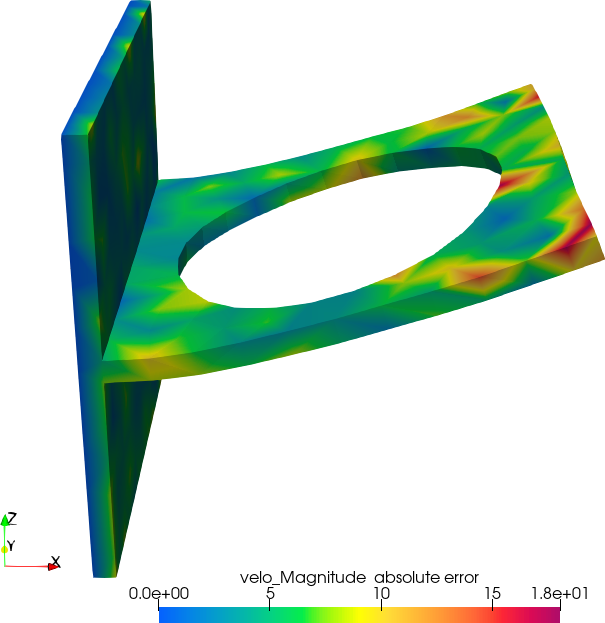}
\end{subfigure}%
\begin{subfigure}[t]{.33\textwidth}\centering
  \includegraphics[width=\brwidth\textwidth]{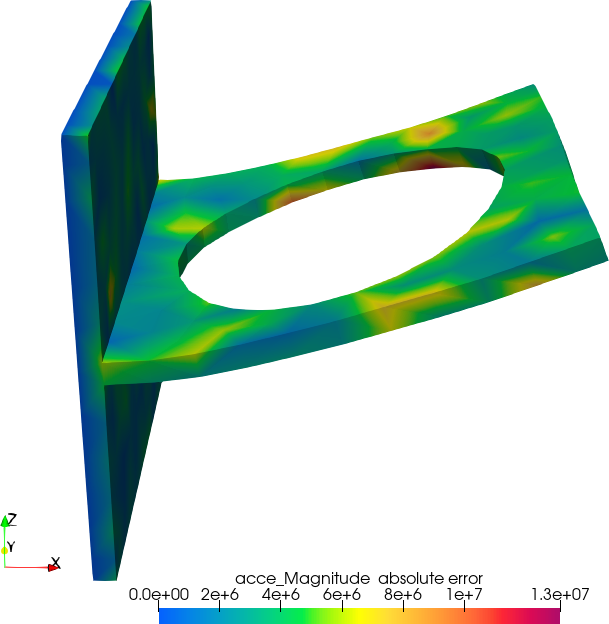}
\end{subfigure}%

\begin{subfigure}[t]{.33\textwidth}\centering
  \includegraphics[width=\brwidth\textwidth]{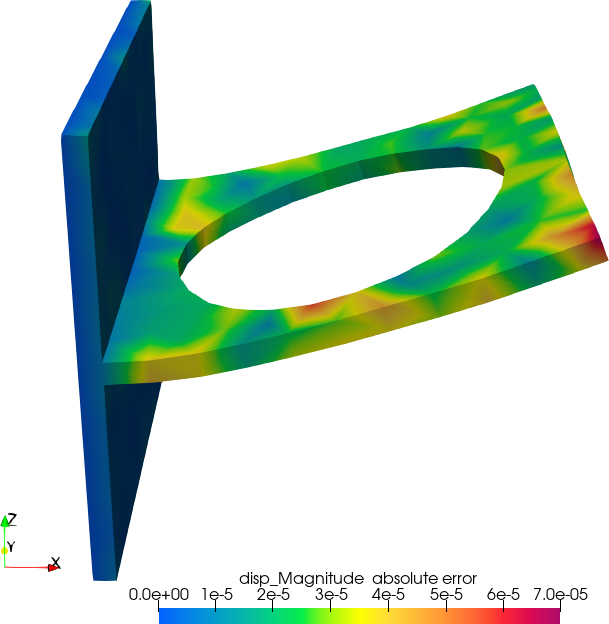}
\end{subfigure}%
\begin{subfigure}[t]{.33\textwidth}\centering
  \includegraphics[width=\brwidth\textwidth]{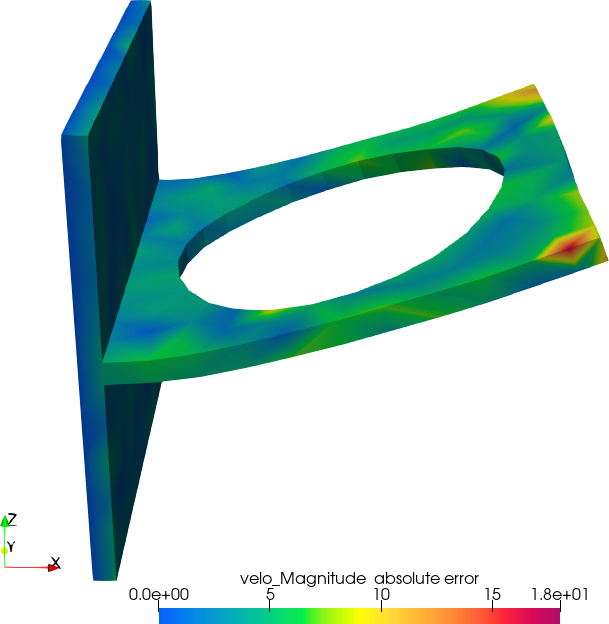}
\end{subfigure}%
\begin{subfigure}[t]{.33\textwidth}\centering
  \includegraphics[width=\brwidth\textwidth]{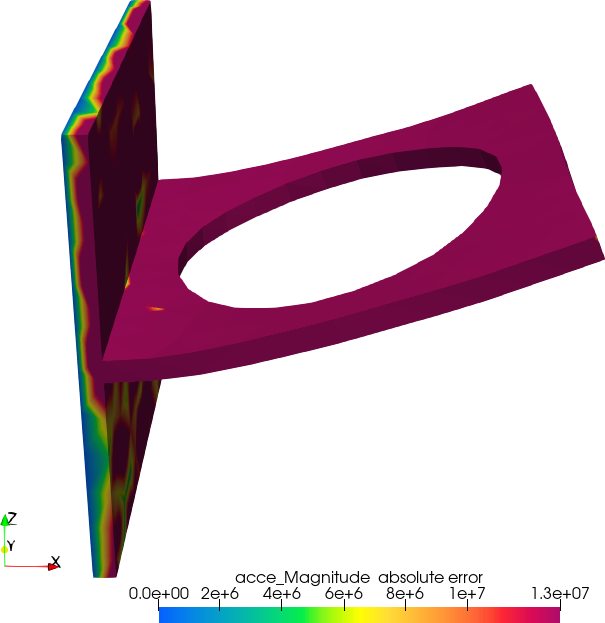}
\end{subfigure}%

\begin{subfigure}[t]{.33\textwidth}\centering
  \includegraphics[width=\brwidth\textwidth]{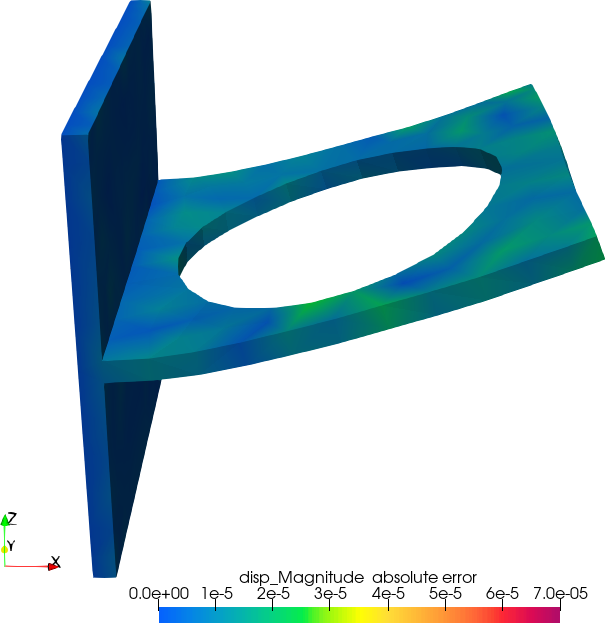}
  \caption{Displacement, $r=25$, $m=125$}
\end{subfigure}%
\begin{subfigure}[t]{.33\textwidth}\centering
  \includegraphics[width=\brwidth\textwidth]{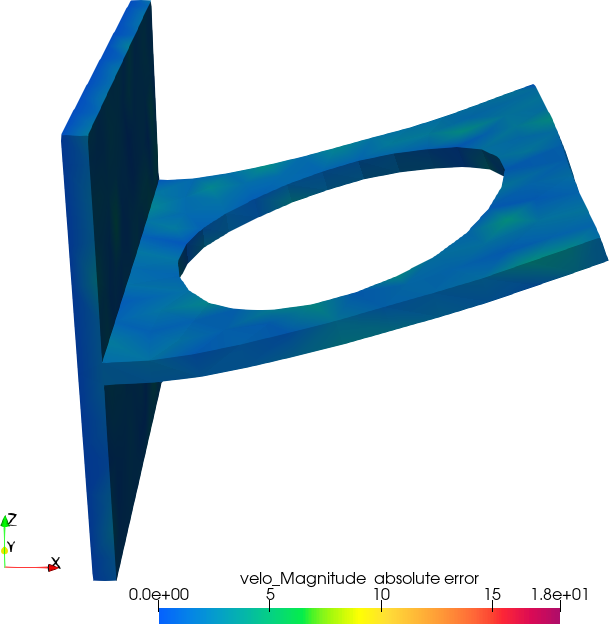}
  \caption{Velocity, $r=50$, $m=250$}
\end{subfigure}%
\begin{subfigure}[t]{.33\textwidth}\centering
  \includegraphics[width=\brwidth\textwidth]{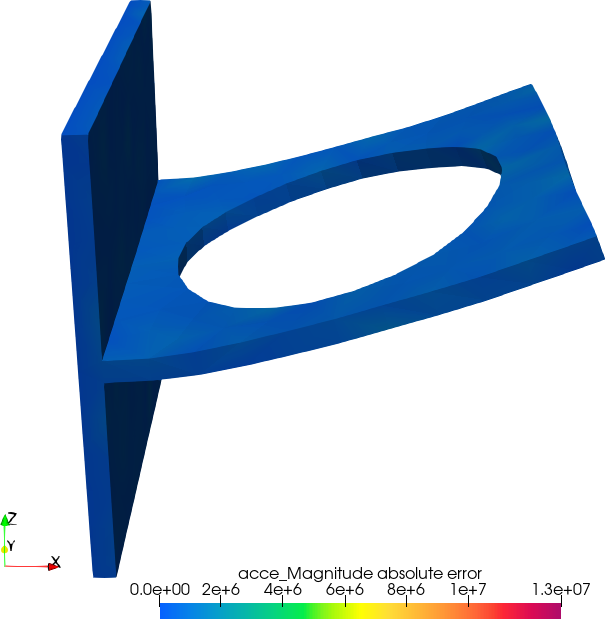}
  \caption{Acceleration, $r=100$, $m=500$}
\end{subfigure}%
\caption{Plots of full states and relative projection errors for the displacement, velocity and acceleration fields at the final time $t = 20.0 \times 10^{-3}$ seconds.  Top row: full state; second row: POD projection error; third row: Alternating QM projection error; bottom row: Kernel RBF projection error.  All figures are scaled by a factor of five times the displacement. }
\label{fig:bracket_solns_and_figs}
\end{figure}

%% file: numerics_doublemach.tex
%!TEX root = main.tex
\subsection{2D Euler -- double Mach reflection}\label{sec:numerics_doublemach}

\begin{figure}[H]
    \centering
    \includegraphics[width=0.3\textwidth]{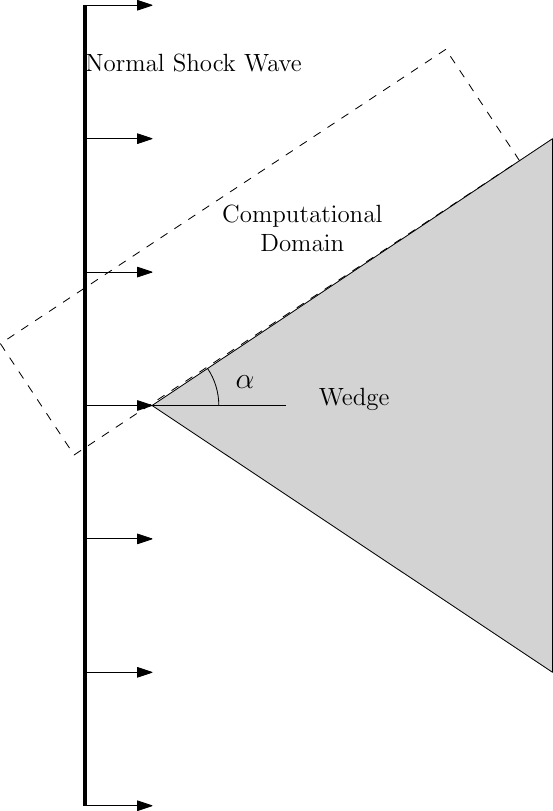}
    \caption{Cartoon of double Mach reflection computational domain. }
    \label{fig:doublemach_cartoon}
\end{figure}

Our last numerical example is the double Mach reflection, a commonly used verification and validation test case for a compressible Euler equation solver originally proposed by Woodward and Colella \cite{woodward1984numerical}. The double Mach reflection case models a shocktube experiment in which a planar shock is driven down a tube containing a wedge, as shown in \Cref{fig:doublemach_cartoon}. 

\begin{figure}[H]
    \centering
    \includegraphics[width=0.8\textwidth]{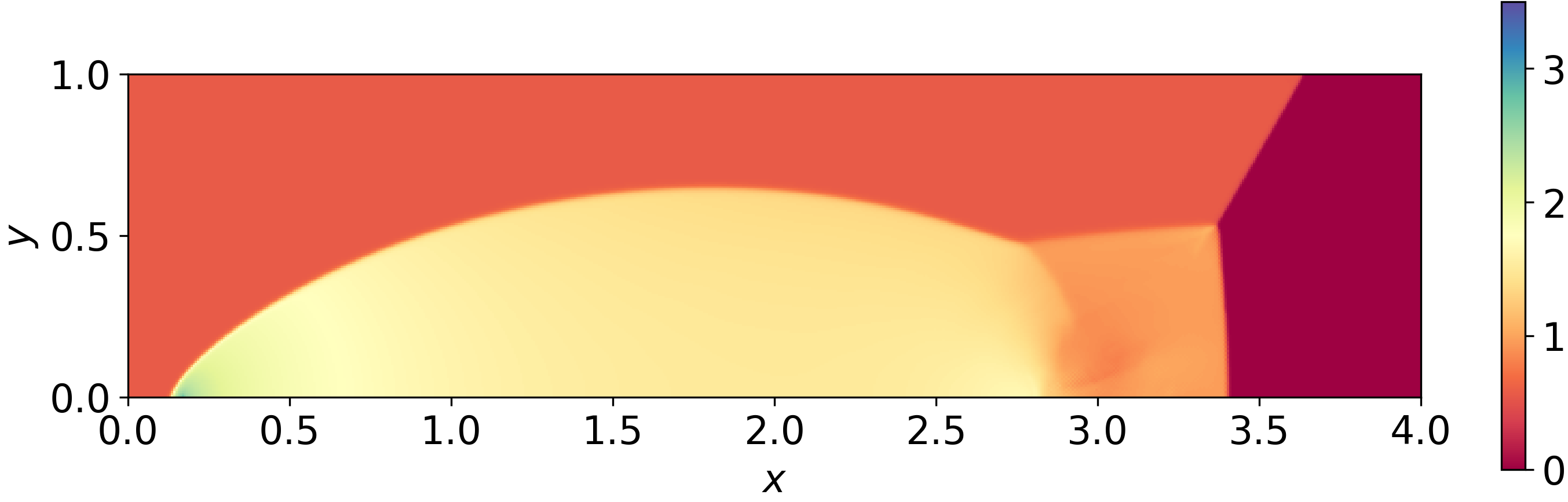}
    \caption{Pressure at $t=0.25$ for wedge half-angle $\alpha=30^\circ$. }
    \label{fig:doublemach_baseline}
\end{figure}

When the shock interacts with the wedge, a complicated shock interaction occurs, as seen in \Cref{fig:doublemach_baseline}. This shock structure is a challenge to the shock capturing schemes utilized by Euler solvers. As in the original case proposed by Woodward and Colella, we consider a Mach 10 shock, a specific heat ratio $\gamma=1.4$ and an upstream non-dimensionalized pressure of $p=1.0$. 

To parametrize the problem, we consider the pressure fields $p(x,y)$ computed for different wedge half-angles $\alpha$. This results in a range of different shock structures that are difficult to capture accurately with a low-dimensional linear subspace. 

The double Mach reflection is modeled by the 2D conservative Euler equations
\begin{align}\label{eq:euler_pde}
    \pdt
    \begin{bmatrix}
        \rho \\ \rho u \\ \rho v \\ \rho E
    \end{bmatrix}
    + \pdx
    \begin{bmatrix}
        \rho u \\ \rho u^2+p \\ \rho u v \\ (E+p) u
    \end{bmatrix}
    + \pdx
    \begin{bmatrix}
        \rho v  \\ \rho u v \\ \rho v^2+p \\ (E+p) v
    \end{bmatrix}
    = 0,
\end{align}
where $u$ is the $x$-velocity, $v$ is the $y$-velocity, $\rho$ is the fluid density, $p$ is the pressure, and $E$ is the energy.
The system is closed by the state equation
\begin{align}\label{eq:state_equation}
    p = (\gamma - 1)\left(\rho E - \frac{1}{2} \rho (u^2 + v^2)\right),
\end{align}
where $\gamma=1.4$ is the specific heat ratio.
The spatial domain is the unit square $\Omega = (0,4) \times (0,1)$ and the time domain is $(0, 0.25)$.

\begin{figure}[H]
    \centering
    \includegraphics[width=0.45\textwidth]{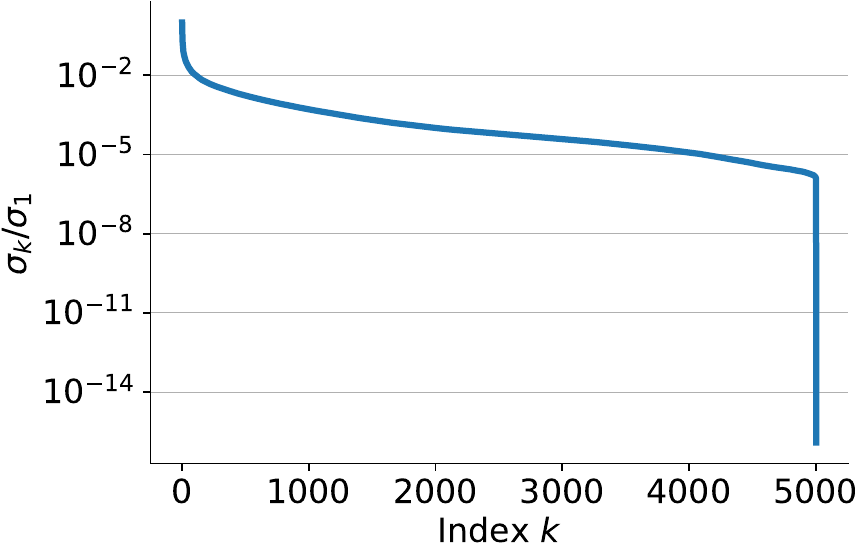}
    \caption{Normalized singular values for double Mach reflection example.}
    \label{fig:doublemach_singvals}
\end{figure}

We collect snapshots using the open-source Python library \texttt{pressio-demoapps}\footnote{\href{https://pressio.github.io/pressio-demoapps/index.html}{pressio.github.io/pressio-demoapps}}
to simulate \cref{eq:euler_pde}, which uses a cell-centered finite volume scheme.
For this example, we use a $600\times150$ uniform Cartesian mesh, resulting in a discretization with state dimension $n_q = 600\times150\times 4 = 360,000$, and a WENO5 scheme for inviscid flux reconstruction.
The time stepping is done using \texttt{pressio-demoapps}' SSP3 scheme for times $t\in (0, 0.25)$ with time step $\Delta t=2.5\times10^{-4}$.
We plot decay of the normalized singular values in \Cref{fig:doublemach_singvals}.  
It can be observed that the singular value decay for this example is exceptionally slow, indicating that this is a particularly difficult test case. 
The error metric that we use for this example is the relative $\ell^1$ projection error 
\begin{equation}\label{eq:relative_l1_projection_error}
    E=\frac{\max_{j\in \set{0, \dots, N_t}}\norm{\bq_j^*-\bg(\bh(\bq_j^*))}_1}{\max_{j\in \set{0, \dots, N_t}}\norm{\bq_j^*}_1},
\end{equation}
where $\bq_j^*$ is test data at time step $j$.

\begin{figure}[H]
    \begin{subfigure}{0.33\columnwidth}
        \centering
        \includegraphics[width=\textwidth]{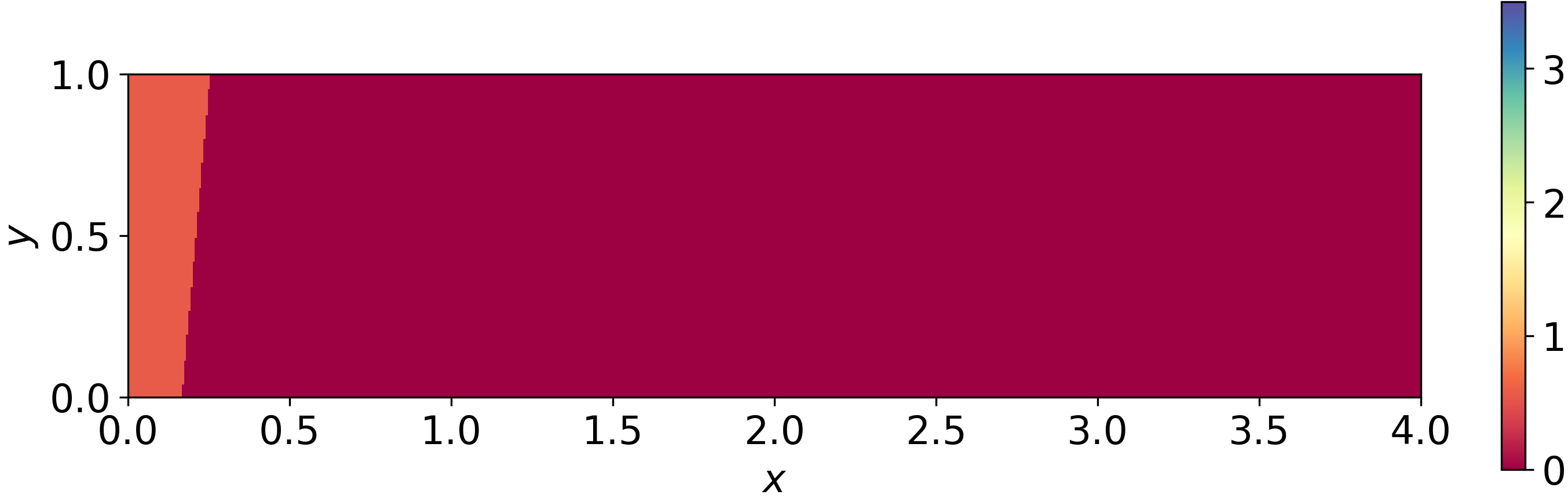}
        \caption{Training, $\alpha=5^\circ$}
    \end{subfigure}
    \begin{subfigure}{0.33\columnwidth}
        \centering
        \includegraphics[width=\textwidth]{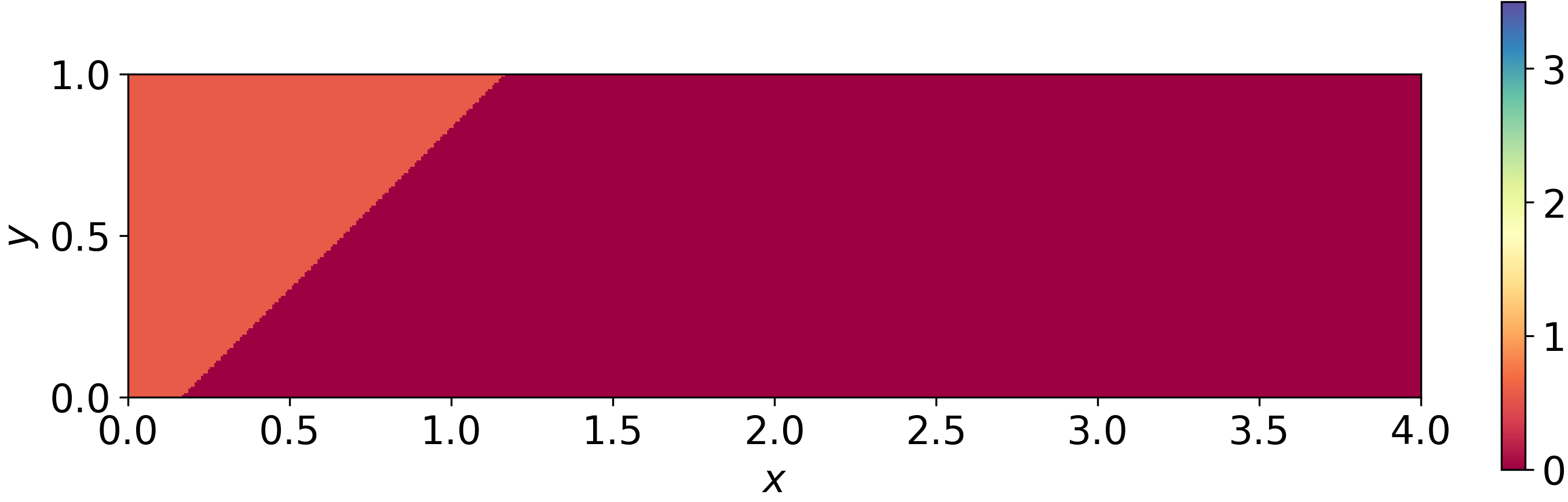}
        \caption{Training, $\alpha=45^\circ$}
    \end{subfigure}
    \begin{subfigure}{0.33\columnwidth}
        \centering
        \includegraphics[width=\textwidth]{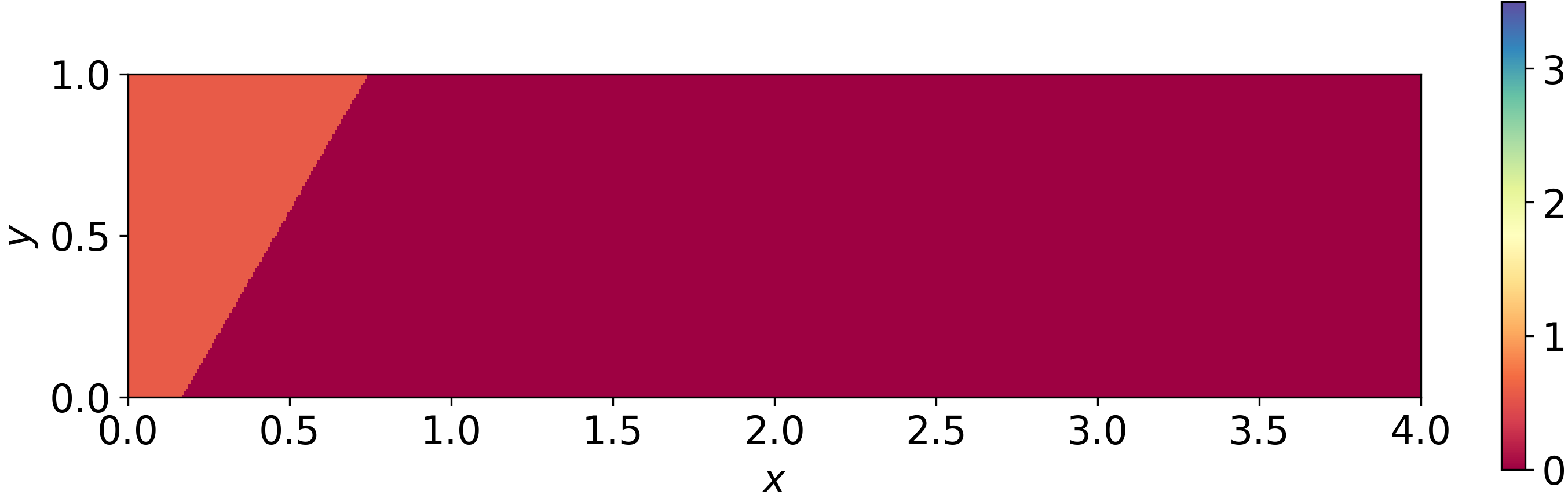}
        \caption{Testing, $\alpha=30^\circ$}
    \end{subfigure}
    \caption{Parameterized initial pressure.}
    \label{fig:doublemach_initial_pressures}
\end{figure}

\Cref{fig:doublemach_initial_pressures} plots the initial pressure for varying wedge angles. 
Notice that each initial pressure contains a sharp front. The training set is comprised of $\alpha=5^\circ, 15^\circ, 25^\circ, 35^\circ, 45^\circ$, and all testing is done with $\alpha=30^\circ$. 


For this example, the Kernel RBF approach uses the inverse quadratic RBF kernel with shape parameter $\epsilon = 10^{-3}.$
The Kernel QM and Kernel RBF approaches do use the input normalization discussed in \Cref{rmk:input_normalization}, and
the Kernel QM, Kernel RBF, Alternating QM, and Greedy QM approaches use the regularization values in \Cref{tbl:doublemach_regs}. 
Note that similar to the 3D bracket example, the optimal regularization values for this problem are especially large. 
This is likely due to the sharp shocks that appear in the training data, which require large POD basis sizes to resolve, thus worsening the conditioning of the systems \cref{eq:feature_map_coef_minimizer_equation} and \cref{eq:coefficient_equation} for training the Alternating QM and KMs, respectively. 
For further details on the selection of the RBF kernel, shape parameter, and regularization values for this example, see \Cref{app:euler}. 

\begin{table}[H]
    \centering
    \begin{tabular}{c|c}
        Manifold type & regularization $\lambda$ \\ \hline
        Kernel QM & $10^{6}$ \\ 
        Kernel RBF & $10^{-3}$ \\
        Greedy QM & $10^{10}$ \\ 
        Alternating QM & $10^{8}$ 
    \end{tabular}
	\caption{Fixed regularization values for different dimensionality reduction approaches for the 2D Euler double Mach reflection example.}
    \label{tbl:doublemach_regs}
\end{table}

\begin{figure}[H]
    \centering
    \includegraphics[width=0.6\textwidth]{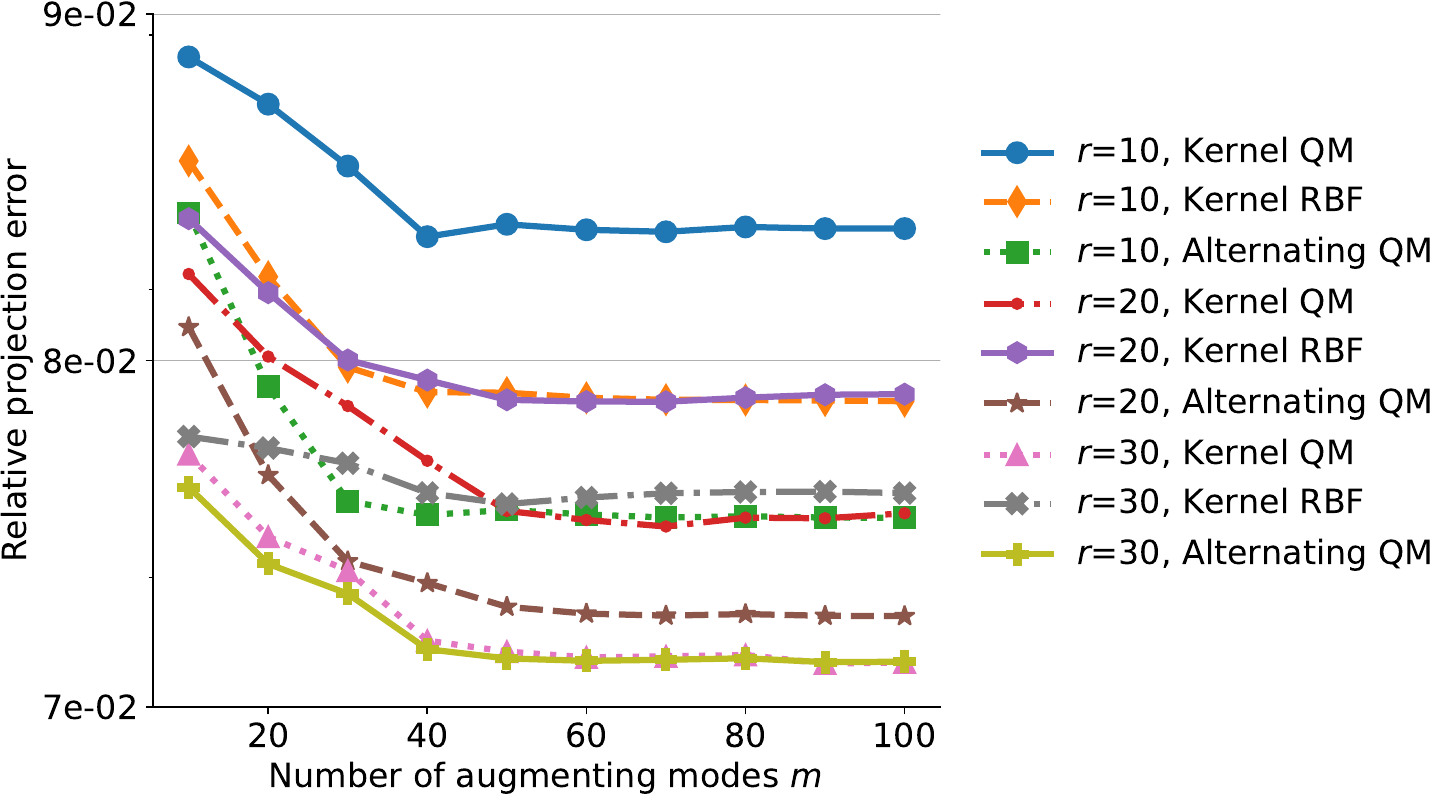}
    \caption{Relative projection error as a function of number of augmenting modes $m$ for several different latent dimension sizes $r$ for the 2D Euler double Mach reflection example.}
    \label{fig:doublemach_error_vs_augmenting_modes}
\end{figure}

\Cref{fig:doublemach_error_vs_augmenting_modes} plots the relative projection error as a function of the number of augmenting modes for $r=10, 20, 30$ for Kernel QM, Kernel RBF, and Alternating QM. 
For $r=10$, Alternating QM outperforms both Kernel QM and Kernel RBF, yielding smaller relative errors and plateaus in error for $r\geq30$. 
Kernel RBF yields smaller errors than Kernel QM and plateaus in error for $r\geq 40$, while Kernel QM with $r=10$ plateaus in error for $r\geq 40$.
For $r=20$, Alternating QM again outperforms both Kernel QM and Kernel RBF and plateaus in error for $r\geq 60$. 
In contrast with the $r=10$ case, Kernel QM outperforms Kernel RBF and plateaus in error for $r\geq 60$. 
Interestingly, $r=20$ Kernel RBF has nearly identical performance as $r=10$. 
For $r=30$, Alternating QM and Kernel QM have nearly identical performance as $m$ increases, and plateau in error for $r\geq 50$, 
while Kernel RBF has the largest errors for $r=30$. 
However, note that all of the relative errors computed for this comparison are in the range $[0.07, 0.09]$, so each dimensionality reduction approach has overall similar performance. 
For the following comparison, we fix $m=4r$ for Kernel QM, Kernel RBF, and Alternating QM.

\begin{figure}[H]
    \centering
    \includegraphics[width=\textwidth]{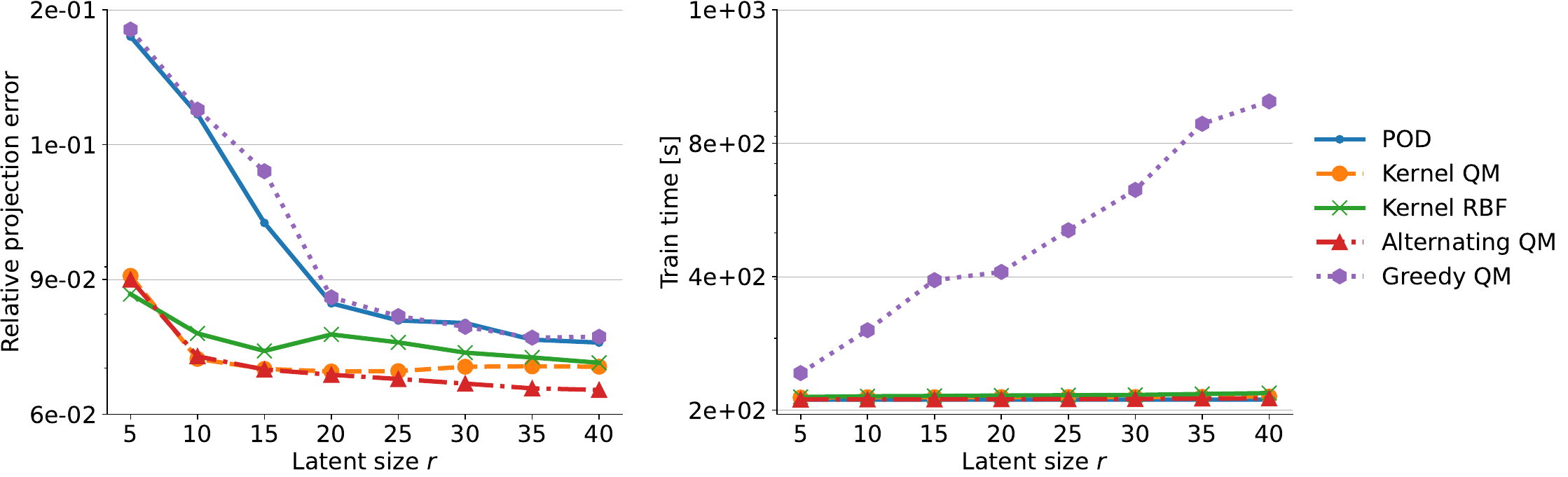}
    \caption{Relative projection error and train time as a function of the latent dimension size $r$ for different dimensionality reduction approaches for the 2D Euler double Mach reflection example. }
    \label{fig:doublemach_error_vs_romsize}
\end{figure}

\Cref{fig:doublemach_error_vs_romsize} compares the different approaches by plotting the relative projection error as a function of the latent dimension size $r$ for $r\in [5, 40].$
Notice that Kernel QM and Alternating QM have nearly identical errors for $r\leq 20$, and Kernel QM slightly increases in error for $r\geq 25$ while Alternating QM continues to decrease in error. 
Kernel RBF has slightly smaller errors than Kernel QM and Alternating QM for $r=5$, but has larger errors than Kernel QM and Alternating QM for $r\geq 10$. 
Greedy QM has errors nearly identical to POD for each value of $r$, with $r=10$ yielding slightly higher error than POD. 
This is because Greedy QM requires an especially large regularization value for this particular problem, thus effectively negating the benefits of the additional quadratic term; see \Cref{fig:doublemach_error_vs_reg}. 
Unlike the previous examples, Alternating QM overall outperforms the KM and Greedy QM approaches for this particular example.
Thus we cannot conclude that the RBF KM approach will always outperform other nonlinear-augmentation manifolds. 
However, as noted above, the difference in performance among Kernel QM, Kernel RBF, and Alternating QM is slight; the error obtained for each method and for each value of $r$ lies in the range $[0.06, 0.09]$. 
Furthermore, the relative error for POD with $r\geq 20$ also lies in this range, indicating that the nonlinear-augmentation approaches provide limited accuracy gains over POD for this particular example.

\begin{figure}[H]
    \begin{subfigure}{0.33\columnwidth}
        \centering
        \includegraphics[width=\textwidth]{figures/doublemach/fom_pressure_angle30_timestep0.png}

        \includegraphics[width=\textwidth]{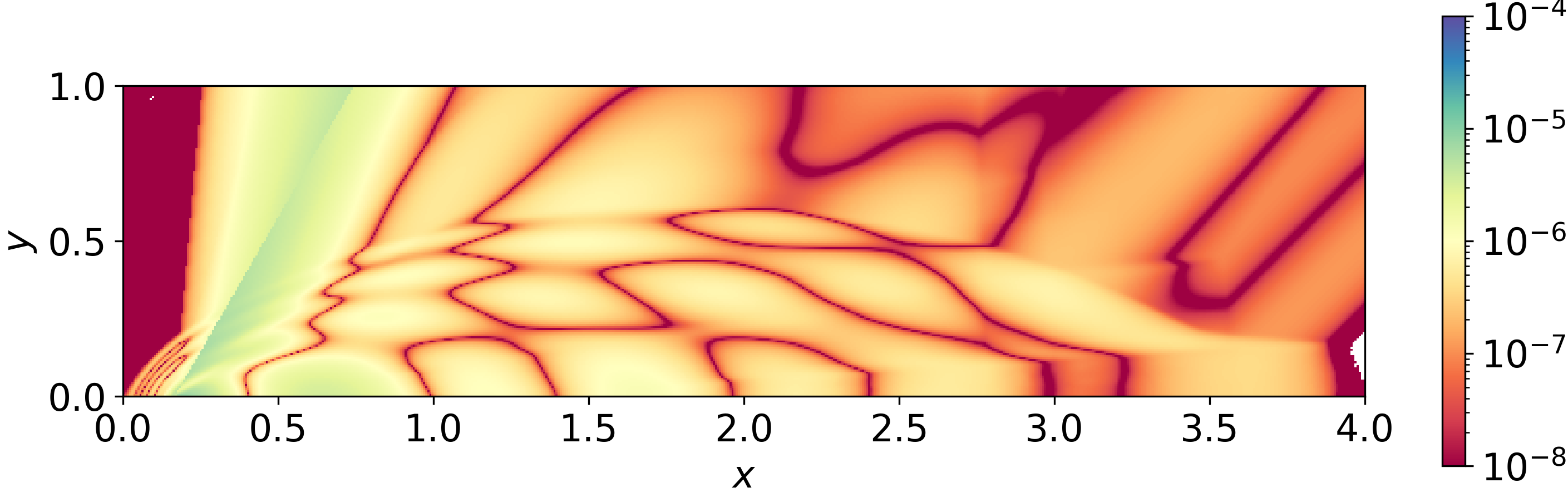}
                
        \includegraphics[width=\textwidth]{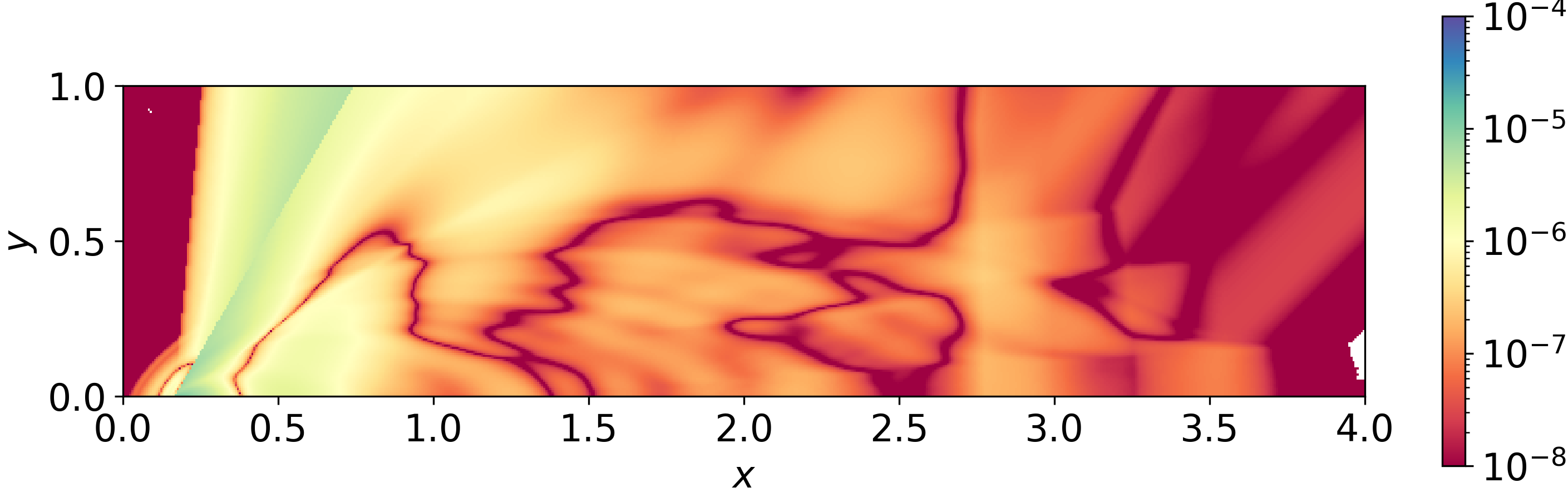} 

        \includegraphics[width=\textwidth]{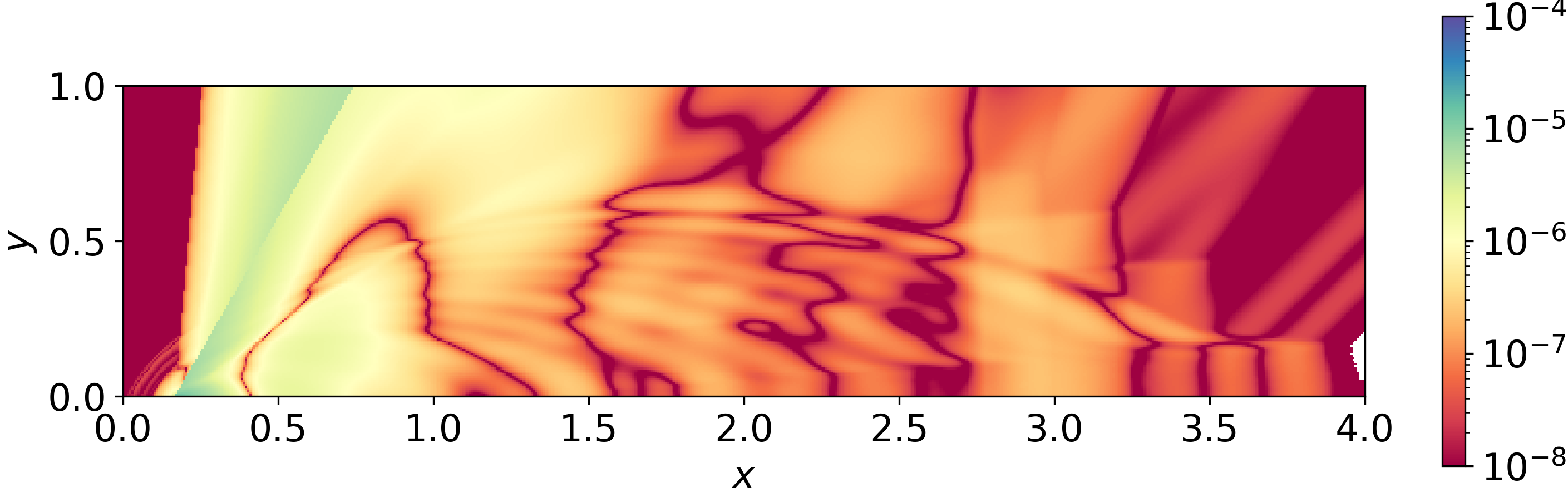}

        \caption{$t=0$}
    \end{subfigure}
    \begin{subfigure}{0.33\columnwidth}
        \centering
        \includegraphics[width=\textwidth]{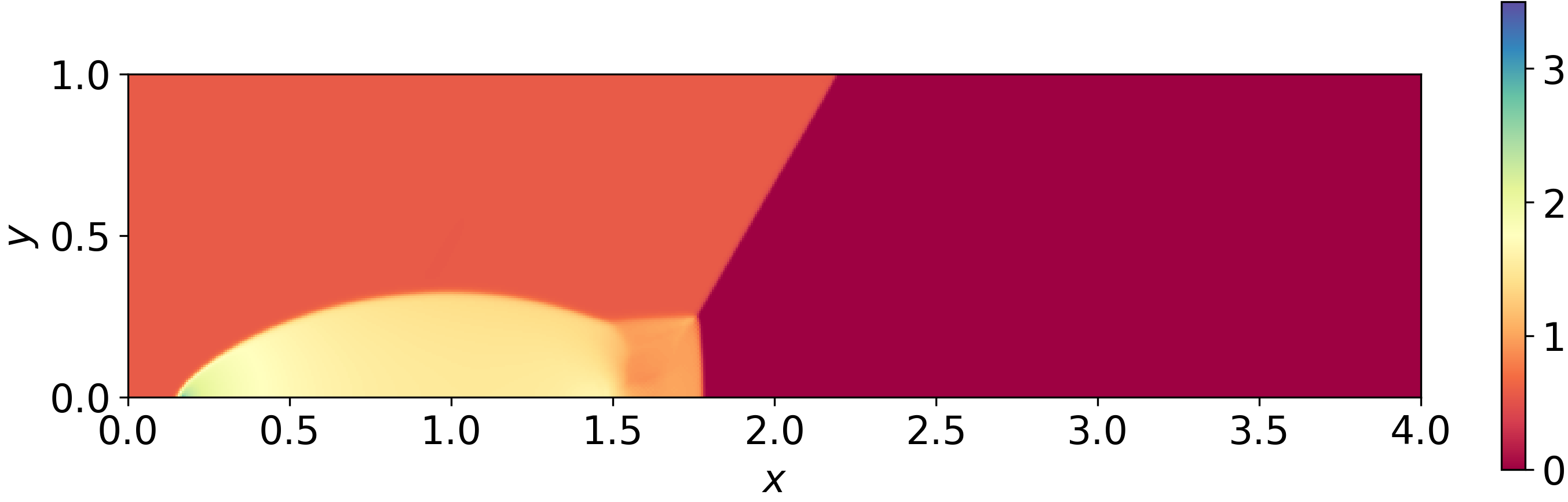}

        \includegraphics[width=\textwidth]{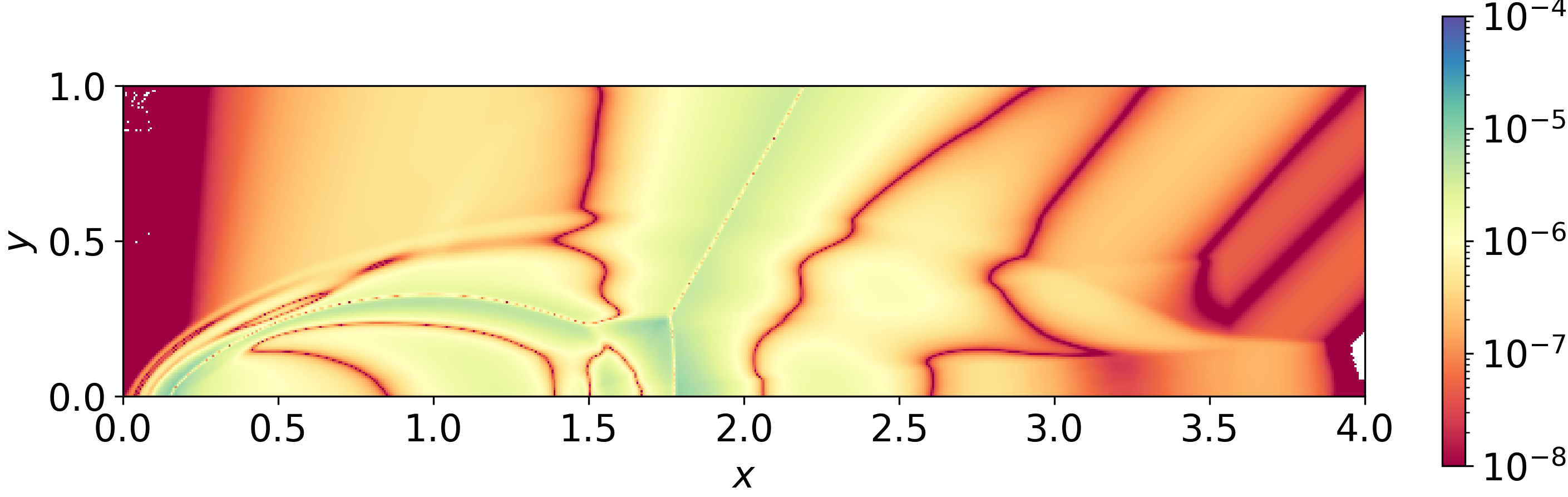}
                
        \includegraphics[width=\textwidth]{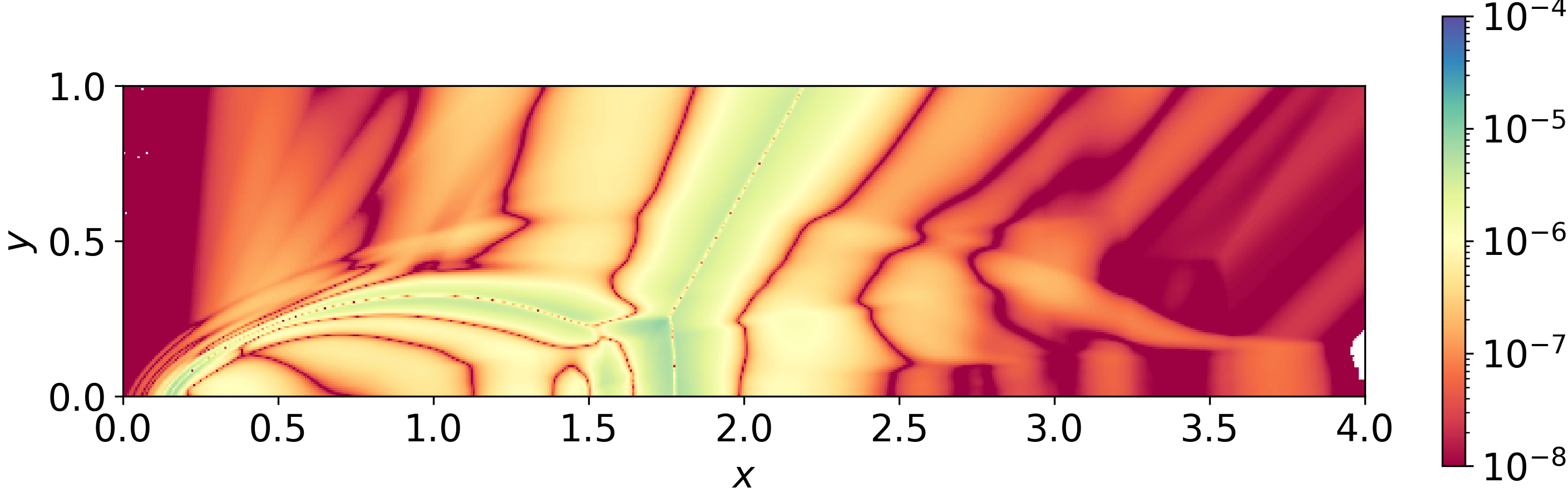}  

        \includegraphics[width=\textwidth]{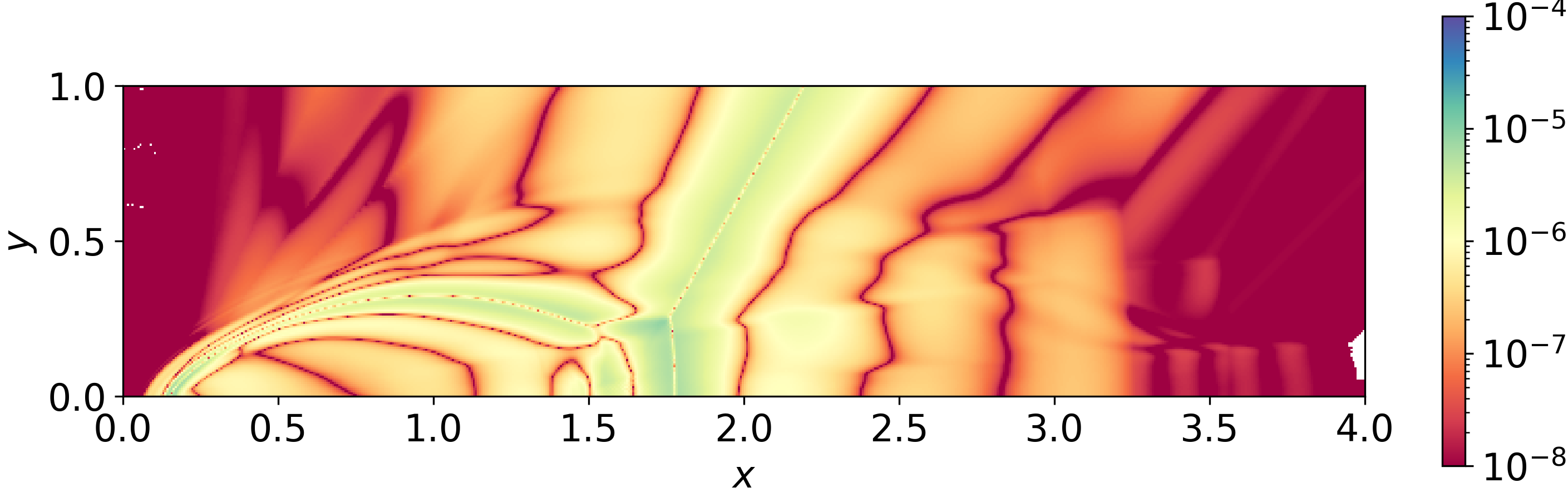}

        \caption{$t=0.125$}
    \end{subfigure}
    \begin{subfigure}{0.33\columnwidth}
        \centering
        \includegraphics[width=\textwidth]{figures/doublemach/fom_pressure_angle30_timestep1000.png}

        \includegraphics[width=\textwidth]{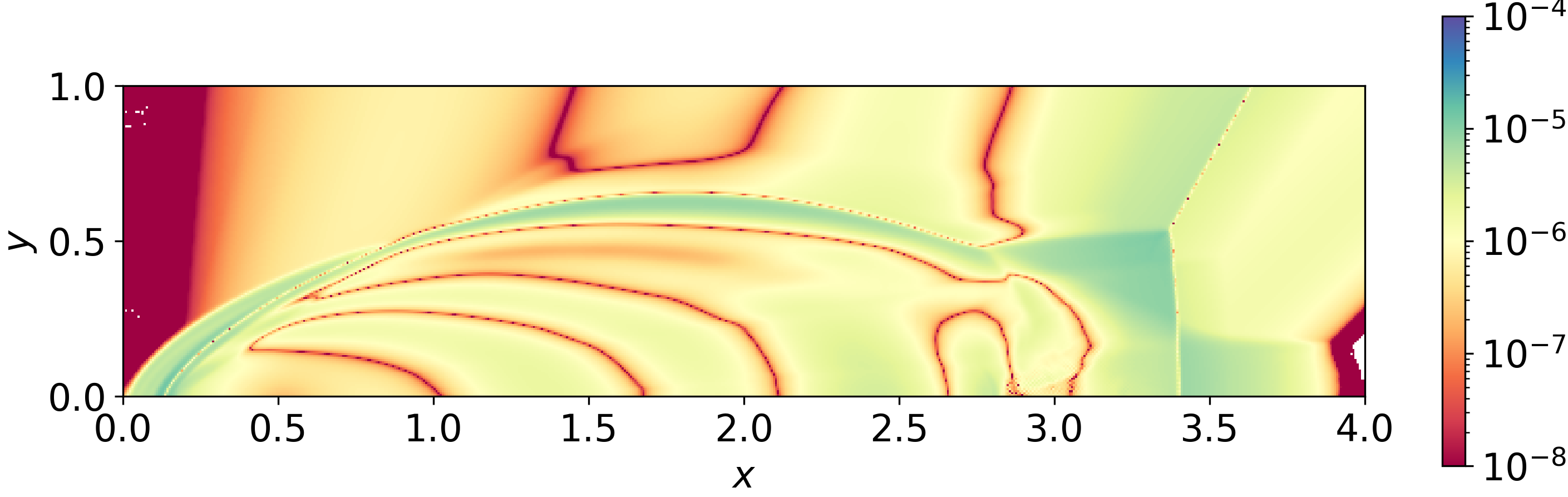}

        \includegraphics[width=\textwidth]{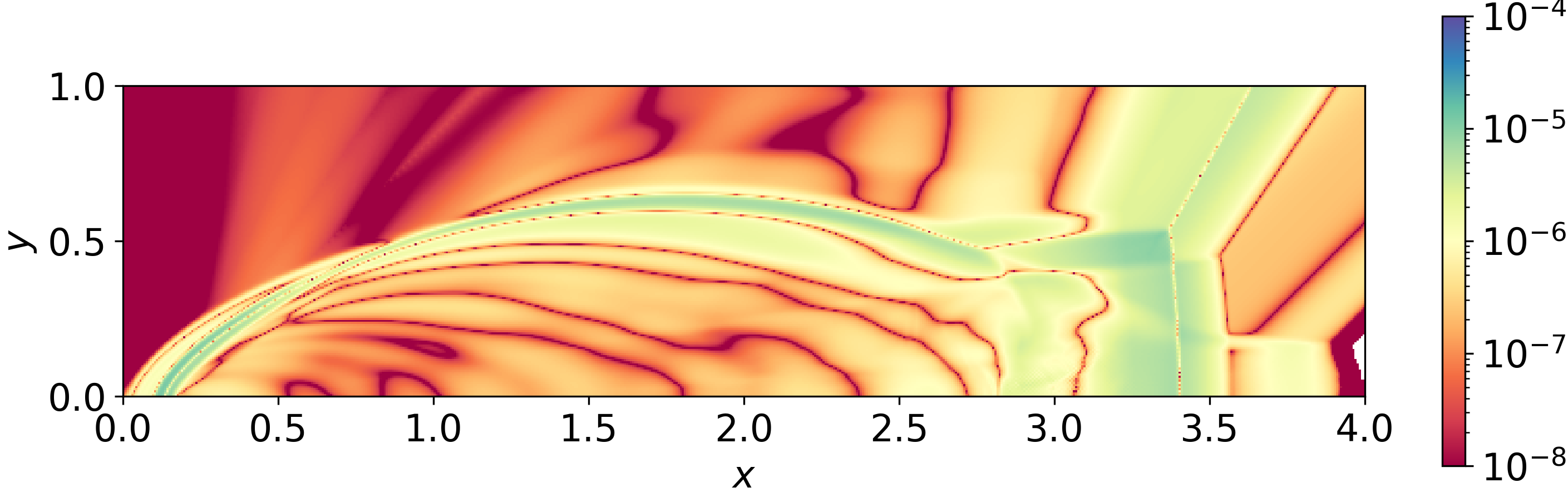} 
        
        \includegraphics[width=\textwidth]{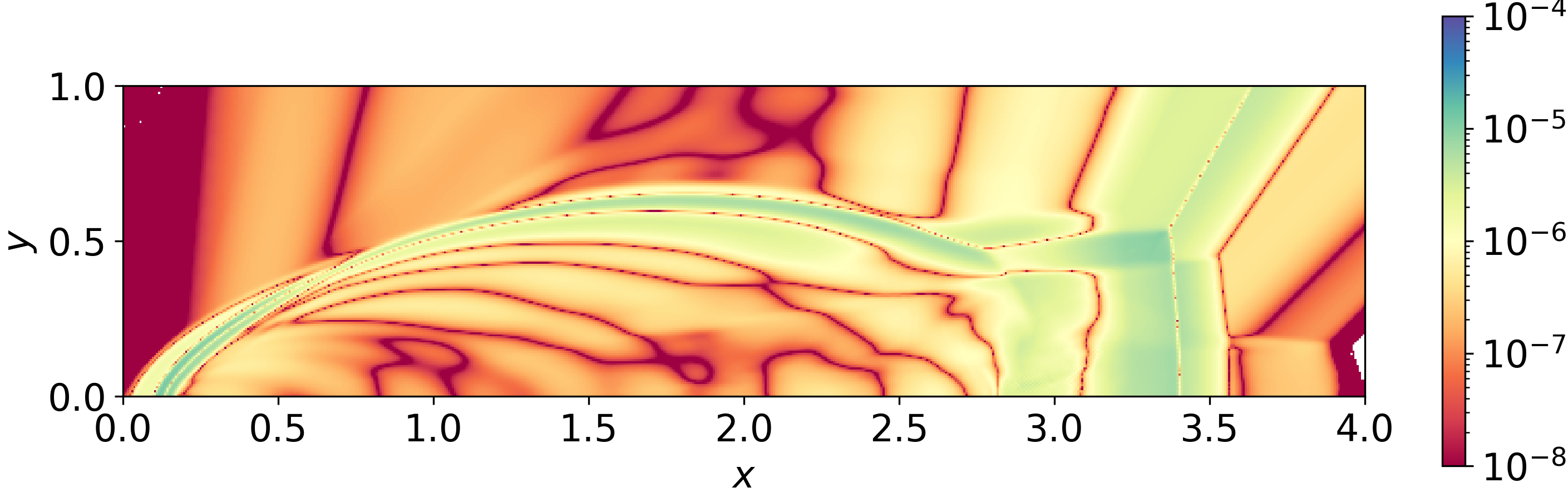}
        \caption{$t=0.25$}
    \end{subfigure}
	\caption{Plot of state and relative projection error at different time steps for $r=10$, $m=40$ for the 2D Euler double Mach reflection example. 
        Top row: full state; 
        second row: POD projection error; 
        third row: Alternating QM projection error; 
        bottom row: Kernel RBF projection error.}
    \label{fig:doublemach_pressures}
\end{figure}

Lastly, \Cref{fig:doublemach_pressures} plots the full state pressure for $\alpha=30^\circ$ and times $t=0, 0.125, 0.25$ and the corresponding projection errors for POD, Kernel RBF, and Alternating QM for $r=10$ and $m=40$. 
For each approach, the error is largest near the shock fronts. 
We also see that Kernel RBF and Alternating QM have smaller errors than POD in regions further away from the shock fronts.

%% file: conclusion.tex
%!TEX root = main.tex
\section{Discussion and conclusion}\label{sec:conclusion}
This paper generalizes recent work on QMs and the GP/RBF-based nonlinear closure approach of \cite{deParga:2025} by developing nonlinear-augmentation dimensionality reduction relying on regularized kernel interpolation. 
Like previous approaches, our method augments POD with a nonlinear correction term in the decoder map, which we model as an optimal nonlinear function lying in a given RKHS.
We leverage the Representer Theorem to fit a nonlinear mapping from low-order POD coefficients to high-order POD coefficients,  
which enables one to fit nonlinear augmentation terms with arbitrary nonlinearities through the choice of kernel.
We show that feature-map based nonlinear augmentations, specifically QMs, as well as the GP/RBF approach, can be obtained as special cases of our method. 

Our numerical examples compare the KM approach to POD and several recent state-of-the-art QM approaches on data from several example problems and show that KMs using RBF kernels can yield substantial accuracy gains in the small latent dimension size regime. 
These results suggest that mappings from low-order POD modal coefficients to an optimal nonlinear correction often cannot be sufficiently approximated using quadratic nonlinearities, making RBF kernels an attractive alternative.
Additionally, the results on the double Mach reflection test case show that nonlinear-augmentation manifolds may only provide slight improvements over POD for especially difficult test cases.
This is because nonlinear-augmentation approaches still rely on an underlying POD, and therefore will at best match the projection error of linear dimensionality reduction with a large POD basis.
Thus if a large POD basis is still insufficient for approximating the given problem, a nonlinear-augmenation manifold will be as well. 
Therefore, for sufficiently difficult problems, an alternative nonlinear dimensionality reduction technique, such as autencoders, may be preferable. 

Future work will focus on improving the attainable accuracy in the KM framework. 
In particular, we will explore other kernels, such as high-order polynomial kernels or sums and products of kernels, and examine their effect on the KM's accuracy.
We will also investigate if a natural nonlinear form for the map $\bn$ mapping low-order POD coefficients to high-order POD coefficients can be deduced, thus informing the selection of an ``optimal'' kernel for use in the KM approach. 
Lastly, we will incorporate KMs as the dimensionality reduction component in a ROM, such as the kernel-based ROMs of \cite{Diaz:2025}. 

%% file: appendix.tex
%!TEX root = main.tex
\section{RBF and regularization selection}\label{appendix}
Each of the nonlinear-correction manifolds that we compare is highly dependent on the regularization parameter $\lambda$. 
Additionally, the KM approach using an RBF kernel also depends on the choice of RBF kernel and shape parameter $\epsilon$. 
While in practice, one can use a hyper-parameter optimization package to tune these choices, as is done in \cite{deParga:2025}, we instead investigate the dependence of the compared approaches on the regularization, RBF kernel, and shape parameters for each of the numerical experiments in \Cref{sec:numerics}. 

\subsection{2D Advection-diffusion-reaction -- boundary layer}
\label{app:advdiff}

\begin{figure}[H]
    \centering
    \includegraphics[width=\textwidth]{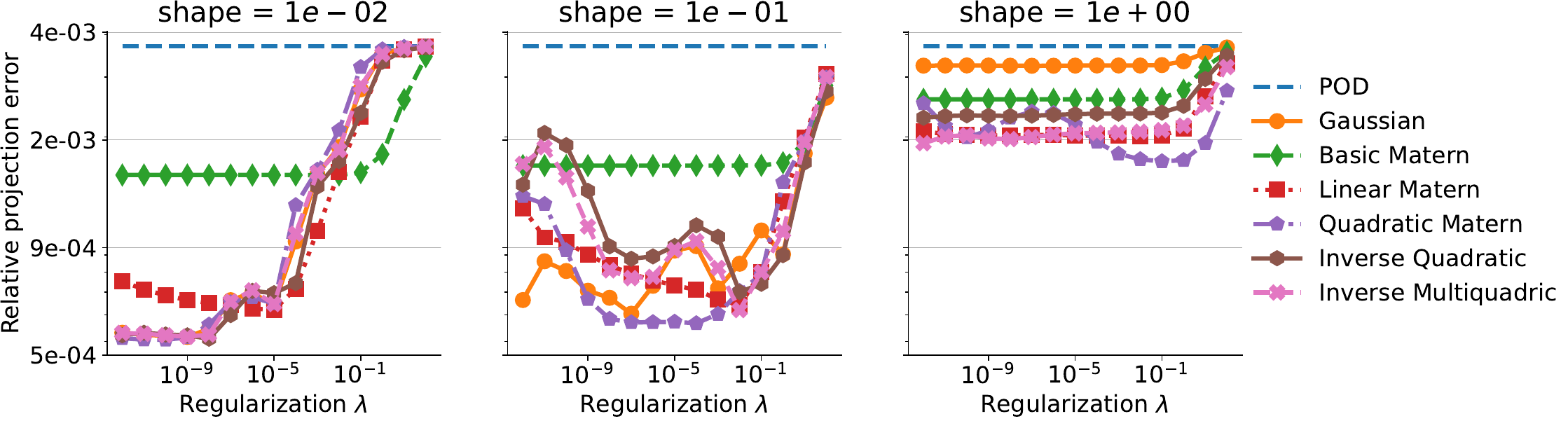}
    \caption{Projection error as a function of regularization $\lambda$ for different RBFs and shape parameters for the boundary layer example, $r=10$, $m=20$.}
    \label{fig:advdiff_error_vs_rbf}
\end{figure}

We first examine the effect of using different RBF kernels with varying shape parameters $\epsilon$ and regularization $\lambda$ on the testing error achieved by the KM approach.
For this test, we fix the latent size to $r=10$ and the number of augmenting modes to $m=20$.
\Cref{fig:advdiff_error_vs_rbf} shows that increasing the shape parameter shifts both the optimal regularization value and the lowest relative projection error. 
Among the combinations tested, the quadratic Mat\'{e}rn kernel with shape $\epsilon=10^{-2}$ achieved the lowest error of $E=5.07\times10^{-4}$ at regularization $\lambda=10^{-10}$.
Therefore, we use the quadratic Mat\'{e}rn kernel with shape parameter $\epsilon=10^{-2}$ in the following comparisons with the varying QM approaches.

\begin{figure}[H]
    \centering
    \includegraphics[width=\textwidth]{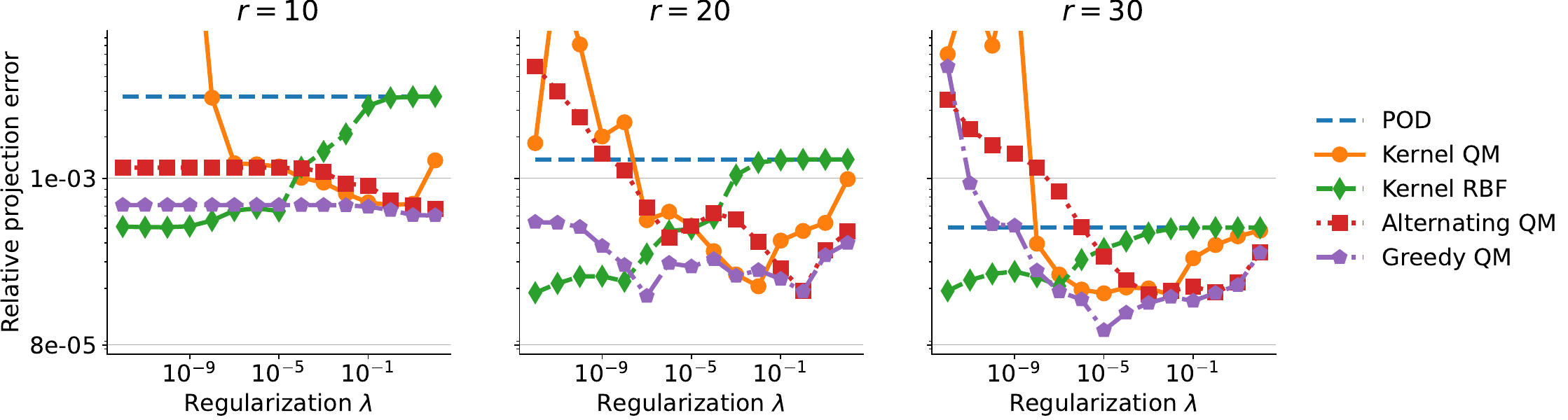}
    \caption{Projection error as a function of regularization $\lambda$ for different latent dimension sizes for the boundary layer example, $m=2r$.}
    \label{fig:advdiff_error_vs_reg}
\end{figure}

Next we examine the effect of the regularization parameter $\lambda$ on the different KM and QM approaches for different latent space dimensions.
See \Cref{fig:advdiff_error_vs_reg}.
For the Kernel RBF, Kernel QM, and Alternating QM approaches, the number of augmenting modes $m$ was set to $m=2r$.
From \Cref{fig:advdiff_error_vs_reg}, we see that for the three latent sizes examined, the Kernel QM and Alternating QM approaches benefit from larger regularization values, whereas the Kernel RBF achieves smaller errors for smaller regularization values.
Interestingly, the Greedy QM approach has a different optimal regularization range for each value of $r$. 
In the following tests where we vary the latent dimension, we fix the regularization for each dimension reduction approach to the values listed in \Cref{tbl:adv_diff_regs}.
However, for optimal results, tuning the regularization for each latent space dimension would be necessary.

\subsection{High-speed aerodynamics -- analytic model of surface heating}
\label{app:hifire}

\begin{figure}[H]
    \centering
    \includegraphics[width=\textwidth]{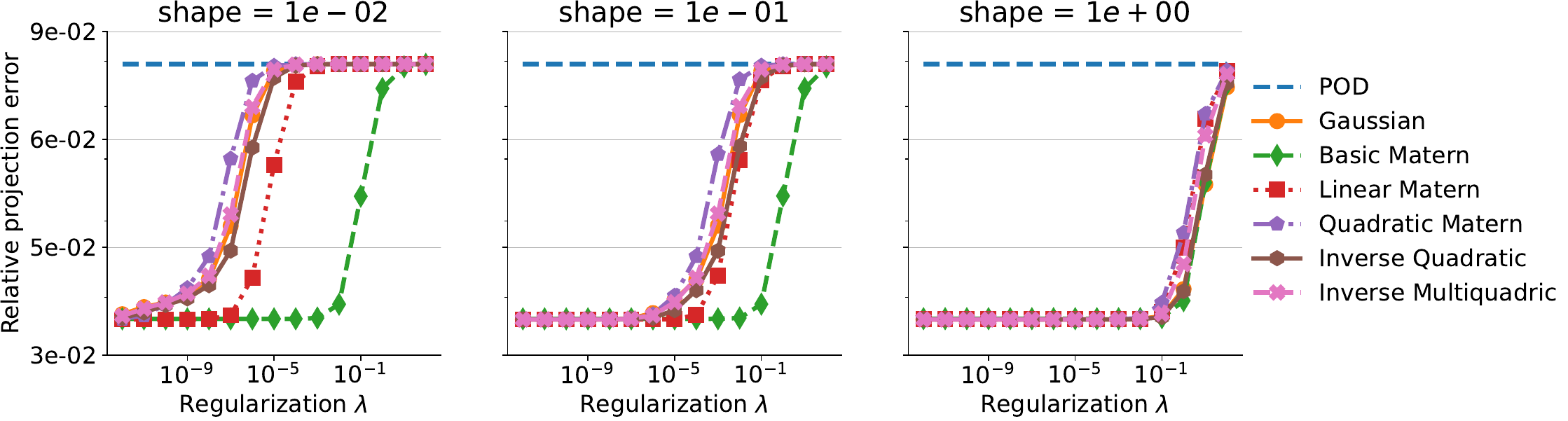}
    \caption{Mean projection error as a function of regularization $\lambda$ for different RBFs and shape parameters for the surface heating example, $r=10$, $m=100$.}
    \label{fig:hifire_error_vs_rbf}
\end{figure}

As in the previous numerical example, we first compare KMs with different RBF kernels and varying shape parameters $\epsilon$ and varying regularization $\lambda$.
For this example, each KM utilizes the input normalization discussed in \Cref{rmk:input_normalization}.
We fix the latent size to $r=10$ and the number of augmenting modes to $m=100$.
\Cref{fig:hifire_error_vs_rbf} shows that over the regularization range tested each KM attains a similar minimum error for each RBF and shape parameter combination.
For each shape value, as the regularization increases, the error also increases from a minimum value to the POD projection error. 
The regularization value for which this increase begins increases as the shape parameter increases.
Among the combinations tested, the Gaussian kernel with shape $\epsilon=10^{-1}$ achieved the lowest error of $E=3.75\times10^{-2}$ at regularization $\lambda=10^{-12}$.
Therefore, we use the Gaussian kernel with shape parameter $\epsilon=10^{-1}$ in the following comparisons with the varying QM approaches.

\begin{figure}[H]
    \centering
    \includegraphics[width=\textwidth]{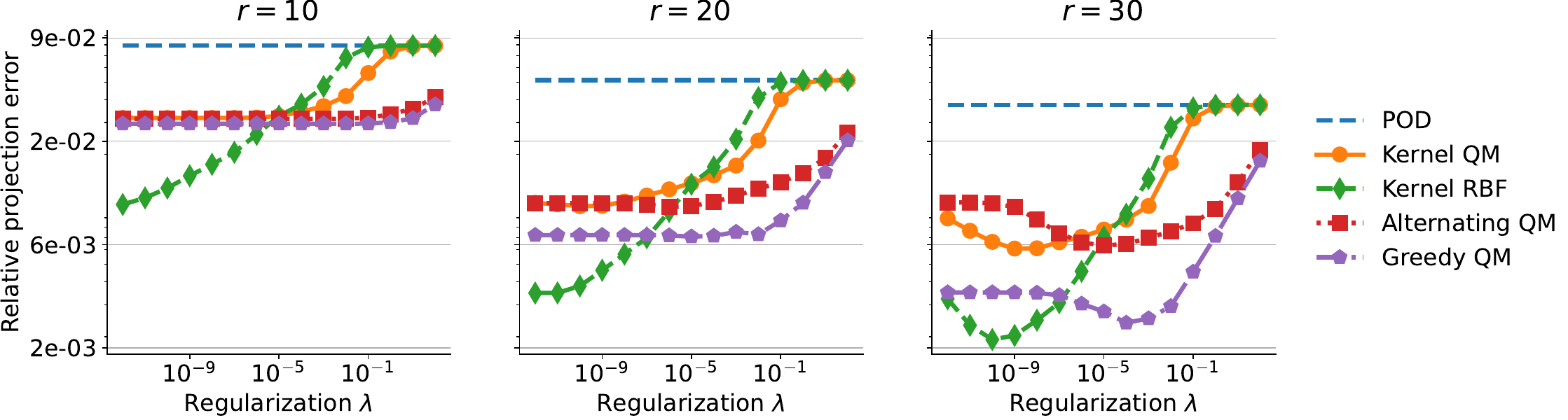}
    \caption{Mean projection error as a function of regularization $\lambda$ for different latent dimension sizes for the surface heating example, $m=10r$.}
    \label{fig:hifire_error_vs_reg}
\end{figure}

Next we examine the effect of the regularization parameter $\lambda$ on the different KM and QM approaches for different latent space dimensions.
See \Cref{fig:hifire_error_vs_reg}.
For both KM approaches and the Alternating QM approach, the number of augmenting modes $m$ was set to $m=10r$.
\Cref{fig:hifire_error_vs_reg} shows that for $r=10$ and $r=20$, the Kernel QM, Alternating QM, and Greedy QM are relatively insensitive to changes in regularization for smaller regularization values, while for $r=30$, there is a clear optimal regularization value in the tested range for each method.
The Kernel RBF approach increases in error as regularization increases for $r=10$ and $r=20$, but has an optimal regularization value at $\lambda=10^{-10}$ for $r=30$.
Based on the errors shown in \Cref{fig:hifire_error_vs_reg}, we fix the regularization for each dimension reduction approach to the values listed in \Cref{tbl:hifire_regs} for the remaining experiments.

\subsection{Solid mechanics - 3D flexible bracket}
\label{app:bracket}

\begin{figure}[H]
    \centering
    \includegraphics[width=\textwidth]{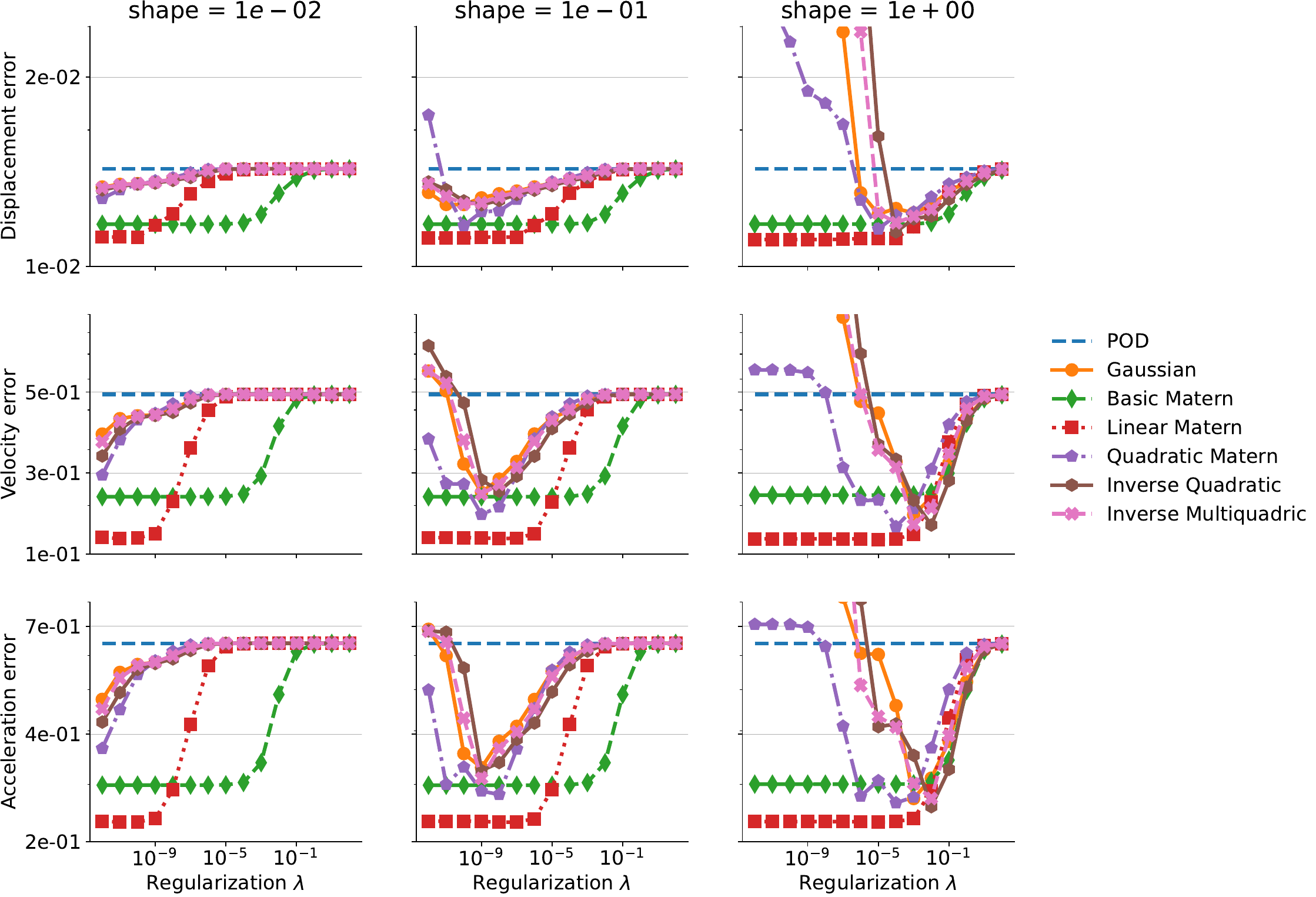}
    \caption{Displacement ($r=10$), velocity ($r=30$), and acceleration ($r=30$) projection errors as a function of regularization $\lambda$ for different RBFs and shape parameters for the flexible bracket example, $m=5r$.}
    \label{fig:bracket_error_vs_rbf}
\end{figure}

As in the previous examples, we first examine the effect of different RBFs and shape parameters on the KM projection error for varying regularization values $\lambda$.
We conduct this comparison for each of the displacement, velocity, and acceleration fields.
For displacement, we use $r=10$ and $m=50$, while for velocity and acceleration we use $r=30$ and $m=150$.
\Cref{fig:bracket_error_vs_rbf} shows across each field and shape parameter, the basic and linear Mat\'{e}rn kernels are the least sensitive to the regularization value $\lambda$, and the linear Mat\'{e}rn kernel yields the lowest projection error for each field.
We also observe that increasing the shape parameter $\epsilon$ also increases the optimal regularization value for the Gaussian, quadratic Mat\'{e}rn, inverse quadratic, and inverse multiquadric kernels.
Due to the insensitivity to regularization and yielding the lowest error, we use the linear Mat\'{e}rn kernel with shape $\epsilon=10^{-1}$ for the remaining comparisons on the flexible bracket example.

\begin{figure}[H]
    \centering
    \includegraphics[width=\textwidth]{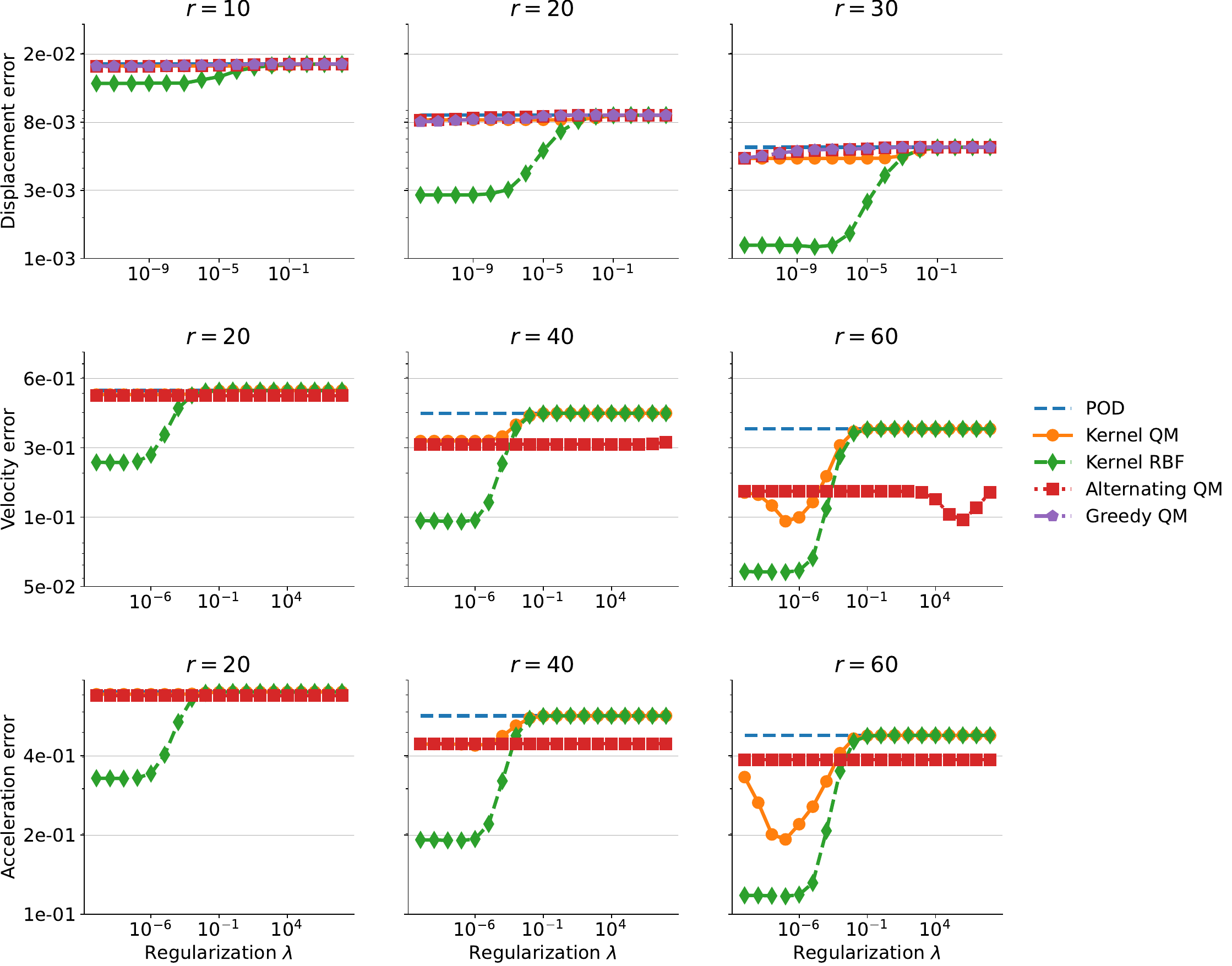}
    \caption{Displacement, velocity, and acceleration projection errors as a function of regularization $\lambda$ for different latent dimension sizes for the flexible bracket example, $m=5r$.}
    \label{fig:bracket_disp_error_vs_reg}
\end{figure}

Next we examine the effect of the regularization value $\lambda$ on the Kernel QM, Kernel RBF, Alternating QM, and Greedy QM approaches for the displacement, velocity, and acceleration fields. 
Since the velocity and acceleration fields require a significantly larger value of $r$ to obtain similar errors to displacement for smaller values of $r$, Greedy QM is omitted from the comparison for the velocity and acceleration fields. 
\Cref{fig:bracket_disp_error_vs_reg} shows that for displacement, each of the QM approaches are relatively insensitive to changes in the regularization and yield errors nearly identical to POD.
For the velocity field, Kernel QM and Alternating QM yield lower errors than POD for $r=40$ and $r=60$. 
Interestingly, for velocity at $r=60$, a large regularization value of $\lambda=10^6$ for Alternating QM yields the optimal error.
For the acceleration field, Alternating QM yields identical errors for each regularization value tested. 
Kernel QM yields identical acceleration errors as Alternating QM for $r=20$ for each regularization value, while for $r=40$, the two QMs have the same error for smaller regularization values and the Kernel QM error increases to match the POD error for larger regularization values. 
For $r=60$, Kernel QM yields optimal acceleration errors at $\lambda=10^{-7}$, slightly outperforming Alternating QM.
The Kernel RBF approach yields the lowest projection errors for each field, regularization value, and latent space dimension.

\subsection{2D Euler -- double Mach reflection}
\label{app:euler}
\begin{figure}[H]
    \centering
    \includegraphics[width=\textwidth]{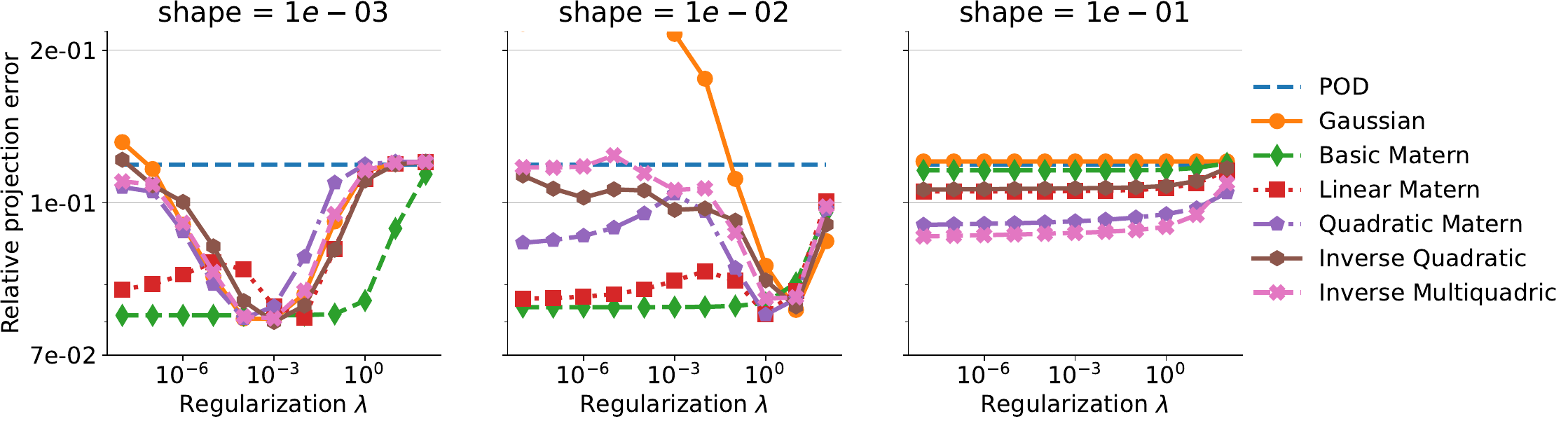}
    \caption{Projection error as a function of regularization $\lambda$ for different RBFs and shape parameters for the double Mach reflection example, $r=10$, $m=20$.}
    \label{fig:doublemach_error_vs_rbf}
\end{figure}

\Cref{fig:doublemach_error_vs_rbf} plots relative projection error as a function of regularization $\lambda$ for the Kernel RBF approach using several different RBF kernels and shape parameters $\epsilon$. 
For this comparison, we fix $r=10$ and $m=20$.
Notice that using a shape parameter $\epsilon = 10^{-3}$ yields the lowest errors. 
Increasing the shape parameter to $\epsilon = 10^{-2}$ increases the optimal regularization values for each RBF kernel except the basic Mat\'{e}rn kernel, which is relatively insensitive to changes in regularization. 
Taking the shape parameter $\epsilon = 10^{-1}$ degrades the accuracy of each of the KMs, but also reduces the sensitivity of the error with respect to regularization. 
For the remaining comparisons on the double Mach reflection example, we use the inverse quadratic kernel with shape parameter $10^{-3}$.

\begin{figure}[H]
    \centering
    \includegraphics[width=\textwidth]{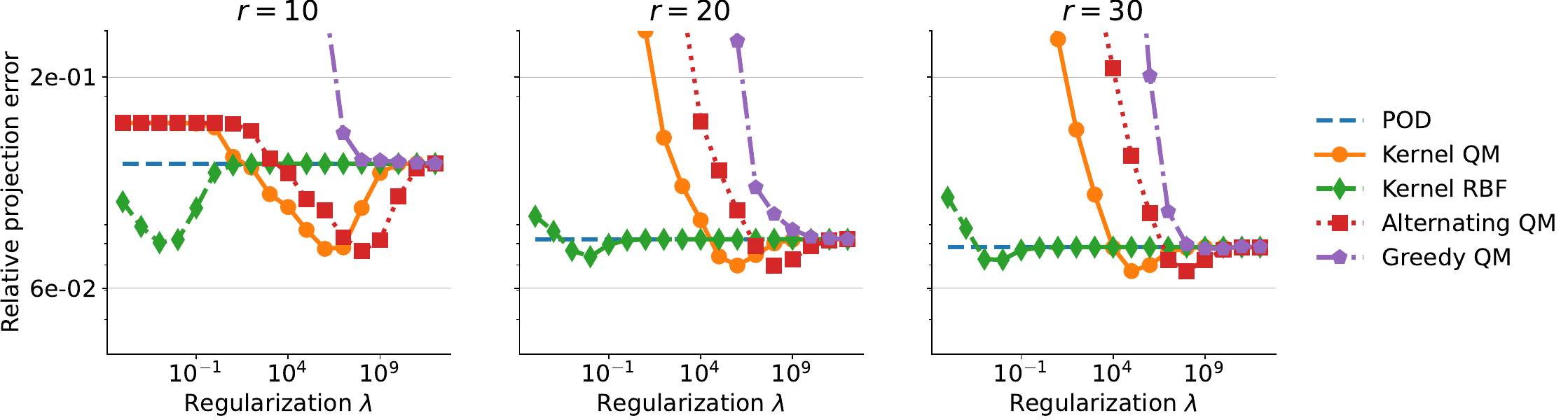}
    \caption{Projection error as a function of regularization $\lambda$ for different latent dimension sizes for the double Mach reflection example, $m=2r$.}
    \label{fig:doublemach_error_vs_reg}
\end{figure}

\Cref{fig:doublemach_error_vs_reg} plots relative projection error as a function of regularization $\lambda$ for the Kernel QM, Kernel RBF, Alternating QM, and Greedy QM approaches for several different values of $r$. 
For this comparison, we take $m=2r$ for the Kernel QM, Kernel RBF, and Alternating QM approaches. 
Observe that as $r$ increases, the best error achieved by each method gets closer to the POD error. 
Furthermore, as $r$ increases, the QM approaches require significantly larger regularization values; otherwise, the error greatly exceeds the POD error. 
The Kernel RBF approach is the least sensitive to regularization, but also yields larger errors compared to the Kernel QM and Alternating QM methods at their optimal regularization values. 
